\documentclass[
  12pt]{article}

\usepackage{amsthm,amsmath,amssymb}
\usepackage{iftex}
\usepackage[T1]{fontenc}
\usepackage[utf8]{inputenc}
\usepackage{textcomp}
\usepackage{lmodern}

\IfFileExists{upquote.sty}{\usepackage{upquote}}{}
\IfFileExists{microtype.sty}{
  \usepackage[]{microtype}
  \UseMicrotypeSet[protrusion]{basicmath}
}{}
\makeatletter
\@ifundefined{KOMAClassName}{
  \IfFileExists{parskip.sty}{%
    \usepackage{parskip}
  }{
    \setlength{\parindent}{0pt}
    \setlength{\parskip}{6pt plus 2pt minus 1pt}}
}{
  \KOMAoptions{parskip=half}}
\makeatother
\usepackage{xcolor}
\makeatletter
\ifx\paragraph\undefined\else
  \let\oldparagraph\paragraph
  \renewcommand{\paragraph}{
    \@ifstar
      \xxxParagraphStar
      \xxxParagraphNoStar
  }
  \newcommand{\xxxParagraphStar}[1]{\oldparagraph*{#1}\mbox{}}
  \newcommand{\xxxParagraphNoStar}[1]{\oldparagraph{#1}\mbox{}}
\fi
\ifx\subparagraph\undefined\else
  \let\oldsubparagraph\subparagraph
  \renewcommand{\subparagraph}{
    \@ifstar
      \xxxSubParagraphStar
      \xxxSubParagraphNoStar
  }
  \newcommand{\xxxSubParagraphStar}[1]{\oldsubparagraph*{#1}\mbox{}}
  \newcommand{\xxxSubParagraphNoStar}[1]{\oldsubparagraph{#1}\mbox{}}
\fi
\makeatother

\usepackage{longtable,booktabs,array}
\usepackage{calc}
\usepackage{etoolbox}
\makeatletter
\patchcmd\longtable{\par}{\if@noskipsec\mbox{}\fi\par}{}{}
\makeatother
\IfFileExists{footnotehyper.sty}{\usepackage{footnotehyper}}{\usepackage{footnote}}
\makesavenoteenv{longtable}
\usepackage{graphicx}
\makeatletter
\def\maxwidth{\ifdim\Gin@nat@width>\textwidth\textwidth\else\Gin@nat@width\fi}
\def\maxheight{\ifdim\Gin@nat@height>\textheight\textheight\else\Gin@nat@height\fi}
\makeatother

\setkeys{Gin}{width=\maxwidth,height=\maxheight,keepaspectratio}

\makeatletter
\def\fps@figure{htbp}
\makeatother

\makeatletter
\@ifpackageloaded{caption}{}{\usepackage{caption}}

\@ifpackageloaded{float}{}{\usepackage{float}}
\floatstyle{ruled}
\@ifundefined{c@chapter}{\newfloat{codelisting}{h}{lop}}{\newfloat{codelisting}{h}{lop}[chapter]}
\floatname{codelisting}{Listing}

\makeatother
\makeatletter
\@ifpackageloaded{caption}{}{\usepackage{caption}}
\@ifpackageloaded{subcaption}{}{\usepackage{subcaption}}
\makeatother

\usepackage[]{natbib}
\usepackage{bookmark}

\IfFileExists{xurl.sty}{\usepackage{xurl}}{}
\hypersetup{
  pdftitle={A Measure of Predictive Sharpness for Probabilistic Models},
  pdfauthor={Pekka Syrjänen},
  pdfkeywords={Predictive sharpness, Probability distributions, Probabilistic forecasting},
  colorlinks=true,
  linkcolor={blue},
  citecolor={blue},
  urlcolor={blue},
}

\numberwithin{equation}{section}
\theoremstyle{plain}

\newtheorem{proposition}{Proposition}[section]
\theoremstyle{definition}
\newtheorem{definition}{Definition}[section]
\newtheorem*{remark}{Remark}

\title{\bfseries A Measure of Predictive Sharpness for Probabilistic Models}
\author{%
  Pekka Syrjänen\thanks{pekka.syrjanen@helsinki.fi}\\
  University of Helsinki
}
\date{}

\begin{document}

\def\spacingset#1{\renewcommand{\baselinestretch}{#1}\small\normalsize}
\spacingset{1}

\maketitle

\bigskip
\begin{abstract}
We introduce a sharpness measure for probabilistic models that quantifies sharpness as an intrinsic property of the probability distribution. The measure is constructed based on a rank-based concentration principle that tracks upward transfers of probability mass along the rearranged profile of the predictive distribution. For finite outcome spaces, this yields a normalized sharpness measure with transparent mass--length representation and equivalent formulations as a Gini-type coefficient on the probability vector and a scaled 1-Wasserstein distance from the uniform distribution in rearranged space. We extend the functional to continuous and multidimensional domains and establish normalization, symmetry, continuity, and monotonicity properties. We investigate the relationship between the measure and uncertainty and dispersion measures, and illustrate its diagnostic applications using real and simulated data.
\end{abstract}

\noindent%
{\it Keywords: Predictive sharpness; Predictive distributions; Probabilistic forecasting}

\spacingset{1.8} 

\section{Introduction}
Probabilistic forecasts play an increasingly central role in statistical modeling, scientific inference, and applied decision-making \citep{buizza2008,collins2007,gneiting2014,timmermann2000,zhang2014}. A central objective of probabilistic forecasting is to maximize sharpness subject to calibration \citep{gneiting2007}. Calibration refers to the statistical consistency between predicted probabilities and the observed frequencies of outcomes and is therefore a joint property of forecasts and observations. Sharpness refers to the concentration of the predictive distribution and is a property of the forecast alone \citep{gneiting2007scoring}. For the evaluation and comparison of probabilistic forecasts, proper scoring rules are recommended, as they provide a numerical score that combines calibration and sharpness \citep{gneiting2007scoring}. At the same time, modelers and users are often interested in assessing distinct aspects of predictive performance \citep{arnold2024,gneiting2007}. In diagnostic settings, calibration and sharpness are therefore examined as separate properties of probabilistic predictions.

For such separate evaluation, the methodology for calibration is more widely discussed. Diagnostics such as the probability integral transform (PIT) are commonly used and broadly applicable \citep{gneiting2014}, and the theory of calibration, including multivariate extensions, continues to be actively developed \citep{allen2024,gneiting2023}. By comparison, the methodology for sharpness has been investigated less \citep{gneiting2008,gneiting2008b,jolliffe2008}, even though sharpness is regarded as a central property of probabilistic forecasts, at times on par with calibration \citep{winkler2008}. For example, \citet[p.~258]{gneiting2008b} referred to the "notorious notion of sharpness," acknowledging the multiplicity of possible approaches to its quantification. In practical settings, modelers may use summaries such as prediction interval widths or box plots \citep{gneiting2007}, and measures of distributional uncertainty or dispersion, namely, entropy and variance, are also sometimes applied \citep{bickel2008,tran2020}. In multivariate settings for ensemble forecasts, proposals for sharpness quantification include the volume of the convex hull or bounding box of the ensemble distribution \citep{weisheimer2005,judd2007}, the root mean squared Euclidean distance between the ensemble members and their mean vector \citep{stephenson2000}, and determinant sharpness, a covariance-based measure that extends the notion of standard deviation to the multivariate case \citep{gneiting2008}.

While these measures each reflect certain aspects of distributional spread, they also have limitations as measures of predictive concentration. Briefly, while interval width and box plots can offer useful summaries in univariate settings, they offer limited analytical capabilities and do not extend naturally to multivariate and higher-dimensional cases. Variance- and covariance-based measures depend on the spatial arrangement of probability mass and therefore do not always reflect probabilistic concentration per se (see Sections~3 and~5.3). Likewise, bounding-box measures are sensitive to extrema and primarily reflect geometric extent rather than the inherent concentration of the predictive distribution. Entropy, in contrast, is invariant to spatial rearrangements, but its nonlinearity and unit dependence (in continuous settings) make it less directly interpretable as a measure of predictive concentration. Moreover, entropy can remain similar across, or reverse the ordering of, distributions with comparable average uncertainty but differently concentrated probability mass (see Sections~3 and~5.2).

These limitations motivate the development of novel methods to sharpness quantification. In this article, we derive a sharpness functional for probabilistic forecasts by applying and extending the Gini methodology \citep{giorgi2017,yitzhaki2013}. The article makes the following main contributions. First, we derive a sharpness measure from a rank-based concentration principle that tracks upward transfers of probability mass along the ranked profile of the distribution. This yields a novel mass--length representation that provides a transparent and interpretable characterization of the distribution's concentration profile. Second, we show that the measure admits several equivalent scalar representations, including a Gini coefficient applied to the probability vector and a scaled 1-Wasserstein distance from the uniform distribution in rearranged space. Third, we show that the construction extends naturally to discrete, continuous, and multidimensional domains. Fourth, we develop graphical methods for examining multidimensional probabilistic forecasts by applying the mass--length expression. Fifth, we illustrate the diagnostic use of the measure using both real and simulated data, and discuss a complementary relationship with the multivariate energy score \citep{gneiting2007scoring}, particularly in sharpness analytics.

The primary application of the proposed measure is as a diagnostic tool for probabilistic models. The measure provides information beyond that available from proper scoring rules by offering an independent measure of sharpness that can be applied pre-observationally, for individual forecasts, and in aggregate. Together with calibration diagnostics (e.g., the PIT), it can be used to quantify sharpness alongside calibration. While functioning as a standalone sharpness measure, another use case is to apply the measure with entropy or variance for more detailed diagnostics, including resolving ambiguity about distributional shape. In the supplementary material, we illustrate the complementary information that these measures provide by further investigating and comparing their level sets (see Supplementary material, Section~S4).

Another potential area of application is Bayesian inference. In Bayesian inference, a canonical measure of learning is the Kullback--Leibler divergence,
\[
D_{\mathrm{KL}}(\pi(\theta \mid D) \| \pi(\theta)),
\]
which quantifies belief change (e.g., from the prior to the posterior). However, KL-divergence measures the difference between two distributions rather than capturing directly how concentrated the posterior is. Here, the proposed sharpness measure can function as a complementary index for posterior or predictive decisiveness: a measure of concentration that is invariant to geometric rearrangements and is interpretable as the degree of commitment over the outcome space.

We emphasize that the proposed measure does not constitute a part of a scoring rule nor does it evaluate properties such as the correctness of the dependence structure in predictions. It quantifies a property of the forecast itself: probabilistic concentration \citep{gneiting2007scoring}. Therefore, in diagnostic applications, it must be used alongside other measures that capture additional properties of interest. Given its tailored focus, the measure is nevertheless widely applicable: it is defined for discrete, continuous, and multidimensional distributions specified analytically (e.g., parametric, nonparametric, mixture, copula, or neural density models), expressed as weighted point-mass measures, approximated numerically through kernel density estimation, histograms, or grid-based discretizations, or obtained via Monte Carlo sampling from the predictive distribution.

The remainder of the article is organized as follows. In Section~2, we derive the sharpness measure, beginning with discrete cases and then extending the measure to continuous domains. Section~3 compares the measure with entropy and variance and presents numerical examples. Section~4 discusses graphical methods. Section~5 presents applications, focusing on model diagnostic use. Section~6 concludes. Technical details are presented in the supplementary material, including proofs of mathematical properties, derivations of forward and inverse domain transformations for domain comparisons, and an extended analysis of the relationships among the proposed measure, entropy, and variance. The accompanying material to the article includes Python scripts for reproducing the analyses in the paper and Python and R functions for implementing the proposed measure in practice.

\section{A Measure of Predictive Sharpness}

Throughout, let $(\Omega,\mathcal{F},\mu)$ be a measurable outcome space equipped with a finite reference measure, so that $\mu(\Omega)<\infty$. Let $\mathcal{P}(\Omega,\mathcal{F})$ denote the set of probability measures on $(\Omega,\mathcal{F})$, and let $P\in\mathcal{P}(\Omega,\mathcal{F})$ denote a predictive distribution on $\Omega$, so that $P(\Omega)=1$.

\subsection{Sharpness in Discrete Outcome Spaces}
We motivate our construction by examining concentration in a simple deterministic setting, where the objective is to predict a realized outcome from a limited set of possibilities. Consider that the outcome space $\Omega$ is comprised of $n = 4$ mutually exclusive and exhaustive elements $\{y_1, y_2, y_3, y_4\}$. In this setting, predictions take the form $y_{i_1} \vee \cdots \vee y_{i_k}$ for some subset $\{i_1, \ldots, i_k\} \subseteq \{1, 2, 3, 4\}$, and increasing concentration is reflected in the number of excluded possibilities, namely $n - k$. A simple measure of the concentration of such predictions is
\[
\frac{n - k}{n - 1},
\]
which equals $0$ when no outcomes are excluded and $1$ when all but one outcome are excluded. Thus, when $n = 4$, the exclusion of any one outcome yields $1/3$, and the exclusion of any two outcomes yields $2/3$.

We next extend this idea to the probabilistic setting, in which uncertainty over the outcomes is represented by a probability mass function $P=\{p_1,\dots,p_n\}$. In the probabilistic case, the most diffuse prediction is the uniform distribution $U$ on $\Omega$, and concentration increases as probability mass is shifted away from uniformity toward narrower subsets of the outcomes. As a first approximation to measuring such concentration, we consider the total variation distance (TVD) from $U$, which is defined as half the $L^1$ distance between the probability mass functions $P$ and $U$. Renormalizing this quantity to the unit interval (achieved by dividing by its maximum value $1-\frac{1}{n}$) gives
\begin{equation}\label{eq:tvdu}
\mathrm{TVD}_{\mathrm{u}} = \frac{\sum_{i=1}^{n} |p_i - \frac{1}{n}|}{2(1 - \frac{1}{n})},
\end{equation}
and a simple algebraic rearrangement shows that this expression reduces to the previous fractional formula whenever $n-k$ outcomes are fully excluded and the remaining outcomes have equal probability. Thus, when $n=4$, excluding one outcome yields $\mathrm{TVD}_{\mathrm{u}}=1/3$, and excluding two outcomes yields $\mathrm{TVD}_{\mathrm{u}}=2/3$. However, a fundamental problem with $\mathrm{TVD}_{\mathrm{u}}$ is that it remains insensitive to local redistributions of mass within the support of the distribution. For example, $\mathrm{TVD}_{\mathrm{u}}$ assigns the same value, $2/3$, to the distributions $\{0,0,\frac{1}{2},\frac{1}{2}\}$ and $\{0,0,\frac{1}{4},\frac{3}{4}\}$, even though the latter is more concentrated on $\Omega$. This motivates the consideration of a more refined measure sensitive to exclusion and concentration among the remaining possibilities.

To construct such a measure, we examine concentration on finite outcome spaces. First, concentration on finite domains ranges between two extremes: the uniform distribution and a degenerate distribution (point mass). Second, concentration increases when probability mass is transferred from lower- to higher-probability outcomes, while preserving the rank ordering of the probabilities. For example, concentration increases when shifting from the uniform distribution over four outcomes to $\{\frac{3}{24}, \frac{7}{24}, \frac{7}{24}, \frac{7}{24}\}$ and it increases further when additional mass is transferred upwards; for example, when the distribution is $\{\frac{3}{24}, \frac{6}{24}, \frac{6}{24}, \frac{9}{24}\}$. Third, inherent concentration within the distribution depends only on the probability values and not on their labeling or spatial arrangement. For example, distributions such as $\{\frac{3}{8}, \frac{2}{8}, \frac{2}{8}, \frac{1}{8}\}$ and $\{\frac{1}{8}, \frac{3}{8}, \frac{2}{8}, \frac{2}{8}\}$ are equally concentrated. Fourth, consider transfers that move probability mass from $p_{(j)}$ to $p_{(k)}$, with $k > j$, while preserving the ordering $p_{(1)} \le \cdots \le p_{(n)}$. For each $j$, the maximal possible transfer is determined by the uniform share of the total remaining probability mass distributed across $p_{(j)},\dots,p_{(n)}$. For example, when $n=4$ and the distribution is uniform, the maximal transfer for $p_{(1)}$ is $1/4$, for $p_{(2)}$ it is $1/3$ (possible when $\{0, \frac{1}{3}, \frac{1}{3}, \frac{1}{3}\}$), and similarly for the subsequent steps. Maximal \emph{concentration} over the entire outcome space is obtained by exhausting all such transfers, whereas partial reallocations represent intermediate concentration profiles. Finally, we may observe that no further transfers occur once the distribution is uniform on the remaining support (e.g., $\{\frac{3}{24}, \frac{7}{24}, \frac{7}{24}, \frac{7}{24}\}$). At each step, increased concentration is thus reflected by $p_{(j)}$ undercutting its local uniform share.

\begin{definition}
A \emph{sharpness functional} is a mapping
\[
S:\mathcal{P}(\Omega,\mathcal{F})\to\mathbb{R}
\]
that quantifies the concentration of a predictive distribution $P\in\mathcal{P}(\Omega,\mathcal{F})$ on $\Omega$.
\end{definition}

We consider the following as natural axiomatic requirements for a (positively oriented) sharpness measure $S$ that quantifies probabilistic concentration on finite outcome spaces:

(A1) (Normalization) The functional $S$ ranges between the uniform distribution and a point mass, namely, $S(P) = 0$ if and only if $P$ is the uniform distribution $U$ over $\Omega$, and $S(P) = 1$ if and only if $P$ is a point mass $\delta_y$ for some $y \in \Omega$.

(A2) (Symmetry) $S$ depends only on the probability values and is therefore invariant to relabelings and spatial rearrangements.

(A3) (Continuity) $S$ is continuous in the probability vector $P$.

(A4) (Monotonicity) If mass is transferred from a lower to a higher probability outcome without reversing the rank ordering of the probabilities, then $S$ strictly increases.

Axioms A1-A4 formalize the above observations. On finite domains, sharpness ranges between two extremes: the uniform distribution and a point mass. A sharpness measure should assign the lowest value to the uniform distribution and the highest to a point mass (or vice versa, depending on orientation), and normalization to the unit interval provides a natural scale for interpretation. (For further discussion on infinite outcome spaces, see Section~5.4.) Sharpness is defined as a property of the forecast only \citep{gneiting2007scoring}. Hence, A2 is natural. Notably, A2 distinguishes between measures that are used in sharpness diagnostics: dispersion measures such as standard deviation and variance do not satisfy A2, whereas entropy satisfies it. We consider A2 more natural for an intrinsic notion of sharpness, given that measures that reflect spatial dispersion may assign the same value to both highly sharp and unsharp distributions (see Section~3.2.1). At the same time, measures of sharpness and distributional dispersion are best viewed as complementary rather than competing measures of distributional structure (see Sections~3.1.2, 3.2.1, 5.1, and 5.3). Thus, A2 may be considered a natural requirement for measures of intrinsic sharpness, but not when considering measures of spatial concentration or spread. Concentration changes continuously with the probability vector, motivating A3. Finally, A4 captures the intuitive idea that transferring mass from lower- to higher-ranked outcomes increases concentration or inequality, corresponding to a so-called reverse "Robin Hood transfer" \citep{marshall1979}. In other words, a distribution becomes more concentrated as mass is transferred from the bottom towards the top.

We propose a construction that satisfies these axiomatic requirements under a transparent, linear scheme. The construction extends the linear exclusion principle, as observed in the fractional formula and \eqref{eq:tvdu}, where each exclusion of an outcome attains equal weight, to the full class $\mathcal{P}(\Omega,\mathcal{F})$, by also rewarding partial reallocations of mass linearly. Specifically, concentration is measured by tracking the upward transfer of mass along the ranked profile of the distribution. At each step, the amount transferred upward is given by the extent to which the probability value falls below its local uniform share, and this transfer is compared to the maximal feasible transfer at each step. Each such fraction of upward transfer is then rewarded linearly in proportion. A key benefit of this construction is that it preserves information about the distribution's full concentration profile, enabling concentration to be tracked in a transparent fashion across the distribution. In addition, the linearity of the construction ensures a consistent interpretation of the measure under changes in domain and scaling (see Section~5.4).

We proceed to formalize this idea. Let $\{p_{(j)}\}_{j=1}^{n}$ be the predicted probabilities sorted in ascending order, so that $p_{(1)} \leq \cdots \leq p_{(n)}$. Define $m_j = \sum_{k=j}^{n} p_{(k)}$ as the total remaining probability mass at step $j$, and $l_j = n - j + 1$ as the number of remaining outcomes at step $j$. We may observe that no upward transfer occurs when $p_{(j)} = m_j/l_j$, while the amount of mass transferred at $j$ is $(m_j/l_j) - p_{(j)}$. Relative to the maximal feasible transfer (1 / $l_j$), this corresponds to a normalized transfer that equals
\[
m_j - p_{(j)} l_j.
\]
Weighting each contribution by its share $1 / (n - 1)$ and summing over the steps yields the proposed sharpness measure:

\begin{equation}\label{eq:dml}
S(P) = \frac{1}{n - 1} \sum_{j=1}^{n - 1} \left( m_j - p_{(j)} l_j \right).
\end{equation}

\begin{remark}
The representation \eqref{eq:dml} quantifies sharpness by evaluating the rank-based concentration profile of the distribution. Given that $m_j - p_{(j)} l_j$ equals the fraction of maximum possible transfer at $j$, this expression enables a full stepwise reconstruction of the distribution's concentration profile. For any subset of the ranked probabilities $p_{(1)} \leq \cdots \leq p_{(n)}$, the average of $m_j - p_{(j)} l_j$ expresses the fraction of the maximal transfer realized across those steps. Moreover, the linear rewarding scheme ensures that the measure matches the fractional formula whenever $n - k$ outcomes are fully excluded and the remaining have equal probability. In Section~4, we further exploit this property in the continuous setting to demonstrate novel graphical methods for representing probabilistic concentration.
\end{remark}

We next establish useful identities for the sharpness functional. First, collecting the coefficients on each $p_{(j)}$, the total coefficient on $p_{(j)}$ is:
\[
\frac{1}{n - 1} \left( j - (n - j + 1) \right) = \frac{2j - n - 1}{n - 1}.
\]
This yields a (rescaled) Gini coefficient formulation \citep{gastwirth2017} as 
\begin{equation}\label{eq:dgin}
S(P) = \sum_{j = 1}^n \left( \frac{2j - n - 1}{n - 1} \right) p_{(j)},
\end{equation}
with the formula applied to the sorted probability vector. While this form sacrifices the transparent decomposition of the concentration profile given by the mass--length expression \eqref{eq:dml}, it offers a more efficient weighted-sum representation. Alternatively, let $J$ be a random variable on $\{1,\dots,n\}$ with
\[
\mathbb P(J=j)=p_{(j)}, \qquad j=1,\dots,n.
\]
Then
\[
\mathbb E[J]=\sum_{j=1}^{n} j\,p_{(j)},
\]
and substituting into \eqref{eq:dgin} yields
\begin{equation}\label{eq:dex}
S(P)=\frac{2}{n-1}\left(\mathbb{E}[J]-\frac{n+1}{2}\right),
\end{equation}
which provides an alternative expression for $S(P)$ as the expected rank of probability mass in the rearranged space, relative to the uniform midpoint $(n+1)/2$. This also yields
\begin{equation}
S(P) = \frac{2}{n-1}\, W_1(P^\uparrow, U),
\end{equation}
where $W_1$ denotes the $1$-Wasserstein (Earth Mover's) distance on $\{1,\dots,n\}$ with ground metric $d(i,j)=|i-j|$, $U$ is the uniform distribution on $\{1,\dots,n\}$, and $P^\uparrow$ denotes the probability measure on $\{1,\dots,n\}$ defined by $P^\uparrow(\{j\})=p_{(j)}$. These equivalent representations link concentration, inequality, and optimal transport. Specifically, $S(P)$ is defined so as to assign a linear reward to upward transfers of mass within the ranked profile of the distribution. This corresponds to the "inequality" of the probability values, as measured by the Gini coefficient. Up to rescaling, this is equal to the optimal transport cost of moving mass from the uniform distribution to the rearranged distribution $P^\uparrow$ on the ordered space $\{1,\dots,n\}$. In addition, these identities make it easy to show that:

\begin{proposition}
S(P) satisfies the properties A1, A2, A3, and A4.
\end{proposition}

\begin{proof}
(A1) By \eqref{eq:dex}, if $P$ is uniform, then $\mathbb{E}[J]=(n+1)/2$ and $S(P)=0$. If $P$ is a point mass at rank $n$, then $\mathbb{E}[J]=n$ and $S(P)=1$. Since the sorted probabilities satisfy $p_{(1)} \leq \cdots \leq p_{(n)}$, the expected rank satisfies $\mathbb{E}[J]\in[(n+1)/2,n]$, with equality at the lower and upper endpoints attained only by the uniform distribution and point masses, respectively. Hence, $S(P)\in[0,1]$ with the required normalization. (A2) Since $S(P)$ depends only on the sorted probability vector, it is invariant under relabeling. (A3) Using \eqref{eq:dgin}, $S(P)$ is a finite weighted sum of the components of the sorted probability vector. Since the sorted probabilities are continuous functions of the probability vector $P$, $S(P)$ is continuous. (A4) Using \eqref{eq:dgin}, a transfer $\delta>0$ from rank $j$ to rank $k>j$ that preserves the ordering yields
\[
S(P')-S(P)=\frac{2k-n-1}{n-1}\delta-\frac{2j-n-1}{n-1}\delta=\frac{2(k-j)}{n-1}\delta>0.
\]
\end{proof}

\subsection{Sharpness in Continuous Outcome Spaces}
We next extend the sharpness measure to continuous outcome spaces. Let $\Omega \subset \mathbb{R}^d$, $d \geq 1$, be bounded and measurable with finite Lebesgue measure $|\Omega| := \lambda(\Omega) < \infty$, and let $f : \Omega \to \mathbb{R}_{\geq 0}$ be a probability density satisfying $\int_{\Omega} f(y)\,dy = 1$. The continuous extension follows an analogous transfer principle to the discrete case but is formulated in terms of the distribution's rearranged density. Specifically, let
\[
f^{\uparrow} : [0,|\Omega|] \to \mathbb{R}_{\geq 0}
\]
denote the nondecreasing rearrangement of $f$, given by
\[
f^{\uparrow}(t) = \inf \left\{ s \ge 0 \mid \left|\{y \in \Omega : f(y) \le s\}\right| \ge t \right\}.
\]
Define $m(t) = \int_t^{|\Omega|} f^{\uparrow}(s)\,ds$ as the remaining probability mass at $t$, and $l(t) = |\Omega| - t$ as the remaining measure of the rearranged domain at $t$. The continuous extension of the measure is then given by
\begin{equation}\label{eq:cml}
S(f) = \frac{1}{|\Omega|}\int_0^{|\Omega|} \left(m(t) - f^{\uparrow}(t)\,l(t)\right)\,dt.
\end{equation}

This expression quantifies how much the distribution deviates from uniformity, tracking the concentration of probability mass over regions of the domain. This construction depends only on $f^{\uparrow}$ and $|\Omega|$, so it is invariant to rearrangements of the density values in $\Omega$ and therefore applies to bounded domains in arbitrary finite dimension. In the supplementary material, we further prove that:

\begin{proposition}
$S(f)$ satisfies continuous analogues of A1--A4: it is normalized on $[0,1)$, with $S(f)=0$ if and only if $f=1/|\Omega|$ a.e.\ on $\Omega$, and $S(f)\to 1$ along sequences of densities converging weakly to a Dirac point mass; invariant under measure-preserving rearrangements; Lipschitz continuous under $L^{1}$-convergence of rearranged densities; and strictly monotone under first-order stochastic dominance in the rearranged space.
\end{proposition}

$S(f)$ also preserves a continuous analogue of the exclusion principle discussed above: the exclusion of a fraction of the outcome space contributes proportionally to the sharpness score. In the supplementary material, we prove that:

\begin{proposition}
\label{prop:exclusion}
If the prediction is uniform on a measurable subset $A \subset \Omega$ and zero elsewhere, then
\[
S(f)=1-\frac{|A|}{|\Omega|}.
\]
\end{proposition}

Similarly to $S(P)$, $S(f)$ can be simplified to a more computationally efficient expression. Using Fubini’s theorem:
\begin{align*}
\int_0^{|\Omega|} m(t)\, dt
&= \int_0^{|\Omega|} \left( \int_t^{|\Omega|} f^{\uparrow}(s)\, ds \right) dt \\
&= \int_0^{|\Omega|} f^{\uparrow}(s) \left( \int_0^s dt \right) ds \\
&= \int_0^{|\Omega|} s\, f^{\uparrow}(s)\, ds.
\end{align*}

\noindent Next, rewrite the second term:
\begin{align*}
\int_0^{|\Omega|} f^{\uparrow}(t)\, l(t)\, dt 
&= \int_0^{|\Omega|} f^{\uparrow}(t)\, (|\Omega| - t)\, dt \\
&= |\Omega| \int_0^{|\Omega|} f^{\uparrow}(t)\, dt - \int_0^{|\Omega|} t\, f^{\uparrow}(t)\, dt \\
&= |\Omega| - \int_0^{|\Omega|} t\, f^{\uparrow}(t)\, dt.
\end{align*}
Substituting both into the definition yields
\begin{equation}\label{eq:cet}
S(f) = \frac{2}{|\Omega|} \int_0^{|\Omega|} t\, f^{\uparrow}(t)\, dt - 1.
\end{equation}

\noindent Let $T$ be a random variable on $[0,|\Omega|]$ with density $f^{\uparrow}$, so that $\mathbb E[T] = \int_0^{|\Omega|} t\, f^{\uparrow}(t)\,dt$. This gives
\begin{equation}\label{eq:cex}
S(f) = \frac{2}{|\Omega|}\, \mathbb E[T] - 1.
\end{equation}

\noindent Another form is thus,
\begin{equation}
S(f)=\frac{2}{|\Omega|}W_1(\mu_{f^{\uparrow}},\mu_U),
\end{equation}
where $\mu_{f^{\uparrow}}(dt)=f^{\uparrow}(t)\,dt$ and $\mu_U(dt)=\frac{1}{|\Omega|}\,dt$ are probability measures on $[0,|\Omega|]$, and $W_1$ denotes the $1$-Wasserstein distance on $[0,|\Omega|]$, with ground metric $d(s,t)=|s-t|$. We postpone the derivation of the Gini coefficient formulation to Section~4.2, where the interpretation and graphical representation of the resulting expression are also discussed. $S(f)$ provides a continuous analogue of the discrete sharpness measure, where it reflects the concentration of mass over regions of the outcome space.

\section{Comparisons and Numerical Examples}

We next examine the behavior of the sharpness measure with comparisons to entropy and variance. These comparisons show that each measure reflects distinct aspects of distributional structure, allowing the measures to provide complementary information about distributional shape.

\subsection{Comparison with Other Measures}

\subsubsection{Relationship to Shannon entropy}
Entropy and the sharpness measure are functionals of the probability values themselves, rather than of the labels or spatial arrangement of outcomes. The sharpness measure $S(P)$ admits the rank-weighted representation
\[
S(P)=\sum_{j=1}^n w_j\,p_{(j)}, \qquad w_j=\frac{2j-n-1}{n-1}, \qquad w_1<\cdots<w_n,
\]
and depends only on the ordered probability vector $p_{(1)} \leq \cdots \leq p_{(n)}$. Likewise, entropy
\[
H(P) = -\sum_{i=1}^n p_i \log p_i
\]
is symmetric in its arguments and therefore depends only on the multiset $\{p_i\}$. In the continuous case, let $\Omega \subset \mathbb{R}^d$, $d\geq1$, be measurable with finite Lebesgue measure $|\Omega|:=\lambda(\Omega)<\infty$, and let $f:\Omega\to\mathbb{R}_{\geq0}$ be a probability density satisfying $\int_{\Omega} f(y)\,dy=1$. If $f^{\uparrow}$ denotes the nondecreasing rearrangement of $f$ on $[0,|\Omega|]$, then equimeasurability gives
\[
\int_{\Omega} \phi(f(y))\,dy=\int_0^{|\Omega|}\phi(f^{\uparrow}(t))\,dt.
\]
Thus also
\[
H(f)=-\int_0^{|\Omega|}f^{\uparrow}(t)\log f^{\uparrow}(t)\,dt,
\]
and
\[
S(f)=\frac{2}{|\Omega|}\int_0^{|\Omega|}t\,f^{\uparrow}(t)\,dt-1.
\]

A more precise relationship is established via majorization. Let $P$ and $Q$ be discrete probability distributions on the same finite outcome space and write $P \succ Q$ if $P$ majorizes $Q$. Then

\begin{proposition}
If $P \succ Q$, then
\[
H(P) \le H(Q)
\qquad\text{and}\qquad
S(P) \ge S(Q).
\]
\end{proposition}

Proposition~3.1 follows immediately from the established properties of $H(P)$ and $S(P)$. The first inequality follows from the Schur-concavity of entropy. From \eqref{eq:dgin}, the sharpness functional is linear with strictly increasing weights, so it is Schur-convex, yielding the latter inequality \citep{marshall1979}. Therefore, whenever two predictive distributions are unambiguously ordered by the majorization order, entropy and sharpness move in opposite directions: sharper predictions have lower entropy. However, majorization is only a partial order, and thus entropy and sharpness may also order distributions differently. This divergence is structural, where entropy quantifies uncertainty through a strictly concave transform of probabilities, while sharpness quantifies the displacement of mass toward higher ranks. Consequently, the two measures capture distinct aspects of distributional structure, as further illustrated through numerical examples in Section~3.2.

Given these structural differences, an extremal connection between these measures can be further shown to exist, where we find the maximum value of either entropy or sharpness while the other measure is fixed. In the continuous case, by \eqref{eq:cex} fixing sharpness is equivalent to fixing the mean of $t$ under the rearranged density $f^{\uparrow}$. The density of maximum differential entropy subject to normalization and a linear moment constraint belongs to the exponential family \citep[Section~11.1]{cover1991}. Applying this known result with the constraint
\[
\int_0^{|\Omega|} t\, f^{\uparrow}(t)\, dt = \mu
\]
on the bounded interval $[0,|\Omega|]$ yields
\[
f^{\uparrow}(t) = \frac{e^{\lambda t}}{\int_0^{|\Omega|} e^{\lambda s}\, ds},
\qquad t \in [0,|\Omega|],
\]
where $\lambda$ is chosen so that $S$ equals the fixed value. An analogous argument applies in the discrete case. Since $S(P)$ is linear in the ranked probabilities, maximizing entropy under normalization and fixed sharpness produces the exponential tilt form
\[
p_{(j)} = \frac{e^{\beta w_j}}{\sum_{k=1}^n e^{\beta w_k}},
\]
with $\beta$ determined by the sharpness constraint \citep[Section~11.2]{cover1991}. These distributions therefore represent maximum-entropy predictive distributions within a given sharpness level set. The same exponential families maximize sharpness at fixed entropy, given that maximizing a linear functional under an entropy constraint is dual to the maximum-entropy problem with a linear moment constraint \citep{csiszar1975}. Further analysis of this relationship is presented in the supplementary material, where we also delve into the opposite case of sharpness or entropy minimization while fixing the other measure.

\subsubsection{Relationship to variance}
Sharpness and variance, while each reflecting a certain type of distributional concentration or spread, measure different aspects of distributional structure. Sharpness measures the extent to which the predictive distribution is concentrated, independently of where the mass is located in the outcome space. Variance measures dispersion around the mean and therefore depends inherently on how the mass is positioned. Consequently, there is no universal monotone relationship between sharpness and variance. These measures can vary significantly while the other measure is held fixed, which is illustrated further below through numerical examples and visualizations.

\subsection{Numerical Examples}

\subsubsection{Discrete distributions}
We next examine numerical examples. We begin with the discrete case, considering an outcome space that consists of $n = 4$ outcomes. Table~\ref{tbl-one} reports the values of $S(P)$, Shannon entropy (in bits), the corresponding KL divergence from the uniform distribution (also in bits), and variance for illustrative distributions.

\begin{longtable}{@{}lcccc@{}}
\caption{Sharpness scores for discrete probability distributions over a domain where $n = 4$. Entropy and KL-divergence are expressed in bits. Variance is reported as $\operatorname{Var}(Y)$, where $Y\sim P$ on the numerically coded support $\{0,1,2,3\}$.}
\label{tbl-one}\\
\toprule
\multicolumn{1}{l}{\textbf{Distribution}} &
\multicolumn{1}{c}{\textbf{S(P)}} &
\multicolumn{1}{c}{\textbf{Shannon Entropy}} &
\multicolumn{1}{c}{\textbf{KL-Divergence}} &
\multicolumn{1}{c}{\textbf{Variance}} \\
\midrule
\endfirsthead

\toprule
\multicolumn{1}{l}{\textbf{Distribution}} &
\multicolumn{1}{c}{\textbf{S(P)}} &
\multicolumn{1}{c}{\textbf{Shannon Entropy}} &
\multicolumn{1}{c}{\textbf{KL-Divergence}} &
\multicolumn{1}{c}{\textbf{Variance}} \\
\midrule
\endhead

\bottomrule
\endfoot

\bottomrule
\endlastfoot

\{0.25, 0.25, 0.25, 0.25\} & 0.000 & 2.000 & 0.000 & 1.250 \\
\{0.24, 0.24, 0.28, 0.24\} & 0.040 & 1.997 & 0.003 & 1.201 \\
\{0, 1/3, 1/3, 1/3\}       & 0.333 & 1.585 & 0.415 & 0.667 \\
\{0.25, 0.5, 0.25, 0\}     & 0.500 & 1.500 & 0.500 & 0.500 \\
\{0, 0, 0.5, 0.5\}         & 0.667 & 1.000 & 1.000 & 0.250 \\
\{0.0, 0.05, 0.27, 0.68\}         & 0.753 & 1.105 & 0.895 & 0.333 \\
\{0.16, 0, 0, 0.84\}         & 0.893 & 0.634 & 1.366 & 1.201 \\
\{0.1, 0, 0, 0.9\}         & 0.933 & 0.469 & 1.531 & 0.810 \\
\{0, 0, 0.01, 0.99\}       & 0.993 & 0.081 & 1.919 & 0.010 \\
\{0, 0, 0, 1.0\}           & 1.000 & 0.000 & 2.000 & 0.000 \\
\end{longtable}

Under majorization, sharpness and entropy order distributions in opposite directions, meaning sharpness increases as entropy decreases. However, the two measures may change in the same direction if uncertainty and concentration toward higher ranks increase or decrease together. Specifically, entropy is more sensitive to the inclusion of low-probability outcomes. Consequently, sharpness and entropy can both increase if a low-probability tail is retained but probability mass becomes more concentrated among the higher-ranked outcomes, as illustrated in Table~\ref{tbl-one}. Conversely, both measures may decrease if probability mass is shifted away from lower-ranked outcomes while the remaining mass becomes more evenly distributed. Further differences between the measures arise with respect to scaling, where sharpness increases linearly with rank-based concentration, whereas entropy decreases nonlinearly as uncertainty is reduced. Maximum entropy grows as $\log n$, whereas $S(P)$ is normalized to lie within $[0,1]$.

\begin{figure}
\centering{
\includegraphics[width=14cm,height=\textheight]{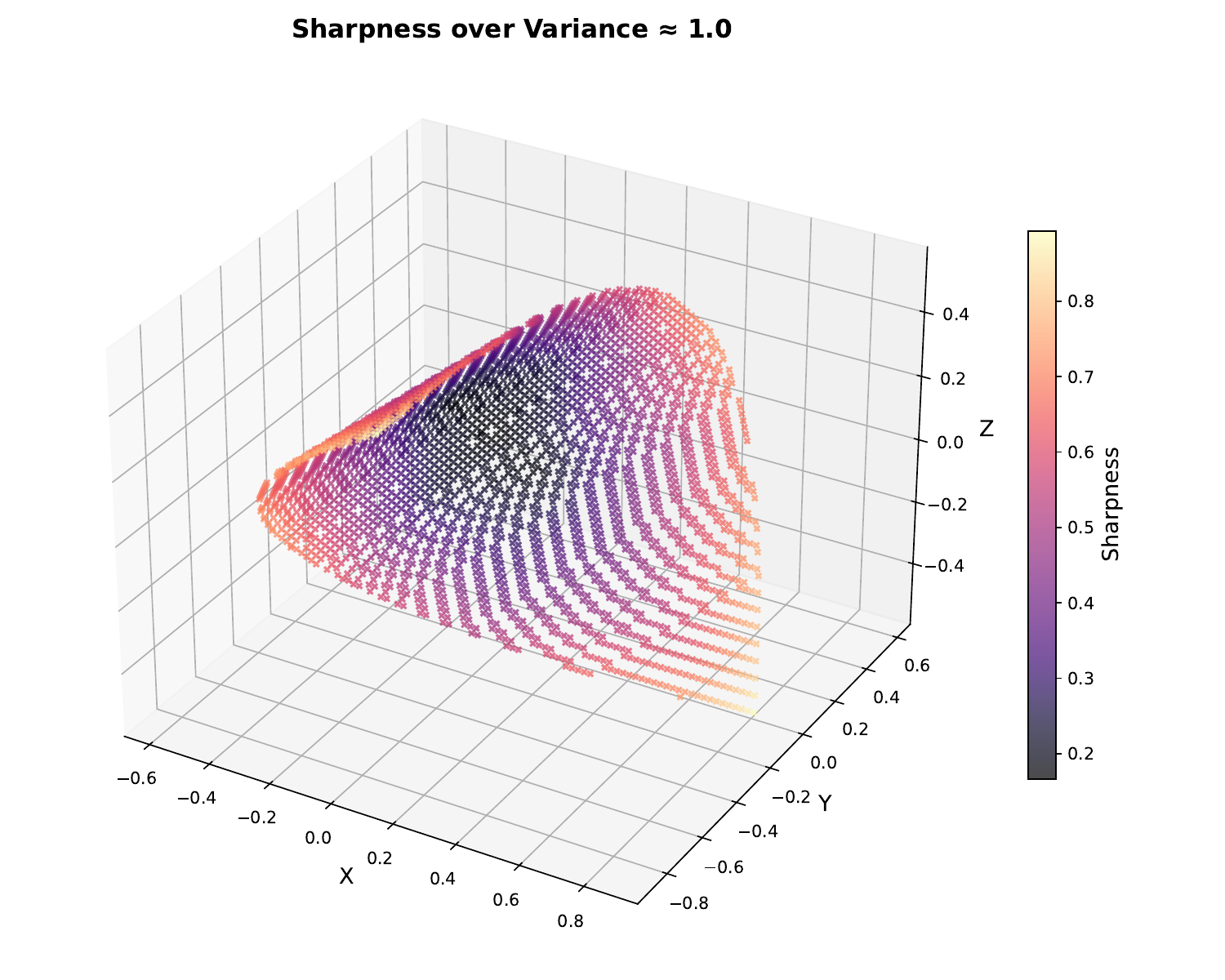}
}

\caption{\label{fig:variance}Distributions with approximately equal variance, $\mathrm{Var}(Y) \approx 1.0$, on the 3-simplex (n=4), where each point represents a discrete distribution over four outcomes. A distribution such as $\{0.19,\ 0.32,\ 0.3,\ 0.19\}$ obtains a low value of $\mathrm{S}(P) \approx 0.17$, while a high value of $\mathrm{S}(P) \approx 0.89$ is obtained by $\{0.86,\ 0,\ 0.02,\ 0.12\}$.}
\end{figure}

Sharpness and variance formalize distinct notions, so they naturally provide complementary information. Variance distinguishes between distributions with similar concentration but different spatial arrangement of probability mass, whereas sharpness distinguishes distributions with similar variance but different levels of concentration. This is illustrated in Figure~\ref{fig:variance}, where we visualize the sharpness scores of distributions having approximately equal variance $\mathrm{Var}(Y) \approx 1.0$.

\subsubsection{Continuous distributions}
In the continuous setting, we examine a set of representative probability density functions defined on the interval $\Omega = [0,4]$. These are shown in Table~\ref{tbl-two}.

\begin{longtable}{@{}lcccc@{}}
\caption{Sharpness, differential entropy (nats), and variance $\operatorname{Var}(Y)$ for continuous probability distributions on $[0,4]$, where $Y$ has density $f$. The distributions are truncated and renormalized to $[0,4]$.}
\label{tbl-two}\\
\toprule
\multicolumn{1}{l}{\textbf{Distribution}} &
\multicolumn{1}{c}{$\mathbf{   S(f)}$} &
\multicolumn{1}{c}{\textbf{   Entropy}} &
\multicolumn{1}{c}{\textbf{   Variance}} \\
\midrule
\endfirsthead

\toprule
\multicolumn{1}{l}{\textbf{Distribution}} &
\multicolumn{1}{c}{$\mathbf{   S(f)}$} &
\multicolumn{1}{c}{\textbf{   Entropy}} &
\multicolumn{1}{c}{\textbf{   Variance}} \\
\midrule
\endhead

\bottomrule
\endfoot

\bottomrule
\endlastfoot

$f(y) = 1/|\Omega|$ & 0.000 & 1.386 & 1.333 \\
Gaussian: $\mu = 2.8, \sigma = 1$ & 0.354 & 1.149 & 0.666 \\
$f(y) = \frac{1}{Z} [0.5\, \varphi(y; 1.2, 0.3^2) + 0.5\, \varphi(y; 3.0, 0.4^2)]$ & 0.459 & 1.023 & 0.925 \\
$f(y) = \frac{1}{Z} [0.6\, \varphi(y; 1.2, 0.3^2) + 0.4\, \varphi(y; 3.0, 0.4^2)]$ & 0.492 & 0.977 & 0.886 \\
Piecewise: 0 on $[0,2)$; 0.5 on $[2,4]$ & 0.500 & 0.693 & 0.333 \\
Gaussian: $\mu = 2.8, \sigma = 0.5$ & 0.610 & 0.690 & 0.236 \\
Piecewise: 0.15 on $[0,1)$; 0 on $[1,3)$; 0.85 on $[3,4]$ & 0.675 & 0.423 & 1.231 \\
Gaussian: $\mu = 2.8, \sigma = 0.1$ & 0.920 & -0.884 & 0.010 \\
Gaussian: $\mu = 2.8, \sigma = 0.01$ & 0.992 & -3.186 & 0.000 \\
$\delta(y - 2.8)$ (Dirac delta) & 1.000 & --- & 0.000 \\
\end{longtable}

Similarly to $S(P)$, the continuous sharpness measure $S(f)$ reflects proportionally increases in the concentration of probability mass. It rewards concentration anywhere within the distribution, favoring greater concentration over narrower regions. $S(f)$ thus distinguishes between distributions that may be ranked similarly by entropy, as illustrated in Table~\ref{tbl-two} by a piecewise and a Gaussian distribution that mirror the discrete examples examined in the previous section. A reciprocal relationship with variance can again be observed, where variance reflects spatial spread and sharpness captures inherent concentration. Thus, for example, more spread-out distributions, such as multimodal Gaussian mixtures, may have relatively high variance even when their mass is more locally concentrated, as captured by $S(f)$. Similarly to $S(P)$, $S(f)$ provides an interpretable metric of concentration, where the scale of the measure ranges between maximal diffusion ($0$) and maximal concentration ($1$).

\subsection{Sharpness in Multidimensional Outcome Spaces}

Finally, to demonstrate the behavior of $S(f)$ in a multidimensional setting, we provide a simple numerical example by using a distribution defined over the three-dimensional unit cube $\Omega = [0,1]^3$. Namely, we construct a density that places 99\% of the probability mass uniformly within one octant---specifically, the subcube $[0, 0.5]^3$---and distributes the remaining 1\% evenly across the other seven octants. This yields the following piecewise-constant probability density function:
\[
f_{0.99}(x, y, z) =
\begin{cases}
\displaystyle \frac{0.99}{(0.5)^3} = 7.92, & \text{if } x, y, z \in [0, 0.5], \\
\displaystyle \frac{0.01}{(0.5)^3 \cdot 7} \approx 0.011429, & \text{otherwise}.
\end{cases}
\]

This density thus strongly concentrates mass in a localized region of the outcome space, suppressing the majority of the domain. Using $S(f)$, we obtain
\[
S(f_{0.99}) = 0.865,
\]
which reflects the proximity of the distribution to excluding 7/8 of the possible outcome space. Further analysis of multidimensional distributions is presented in the following section.

\section{Graphical Representations}

The sharpness functional admits several equivalent expressions that further enable constructing different graphical representations of probabilistic concentration. These representations make it possible to examine not only the overall sharpness score, but also how probability mass is concentrated across the ranked profile of the distribution. These representations may be particularly useful in multidimensional settings. In this section, we develop graphical representations based on the mass--length formulation and the equivalent Lorenz-curve representation.

\subsection{The Mass--Length Representation and Concentration Plots}

\begin{figure}
\centering{
\includegraphics[width=9.5cm,height=\textheight]{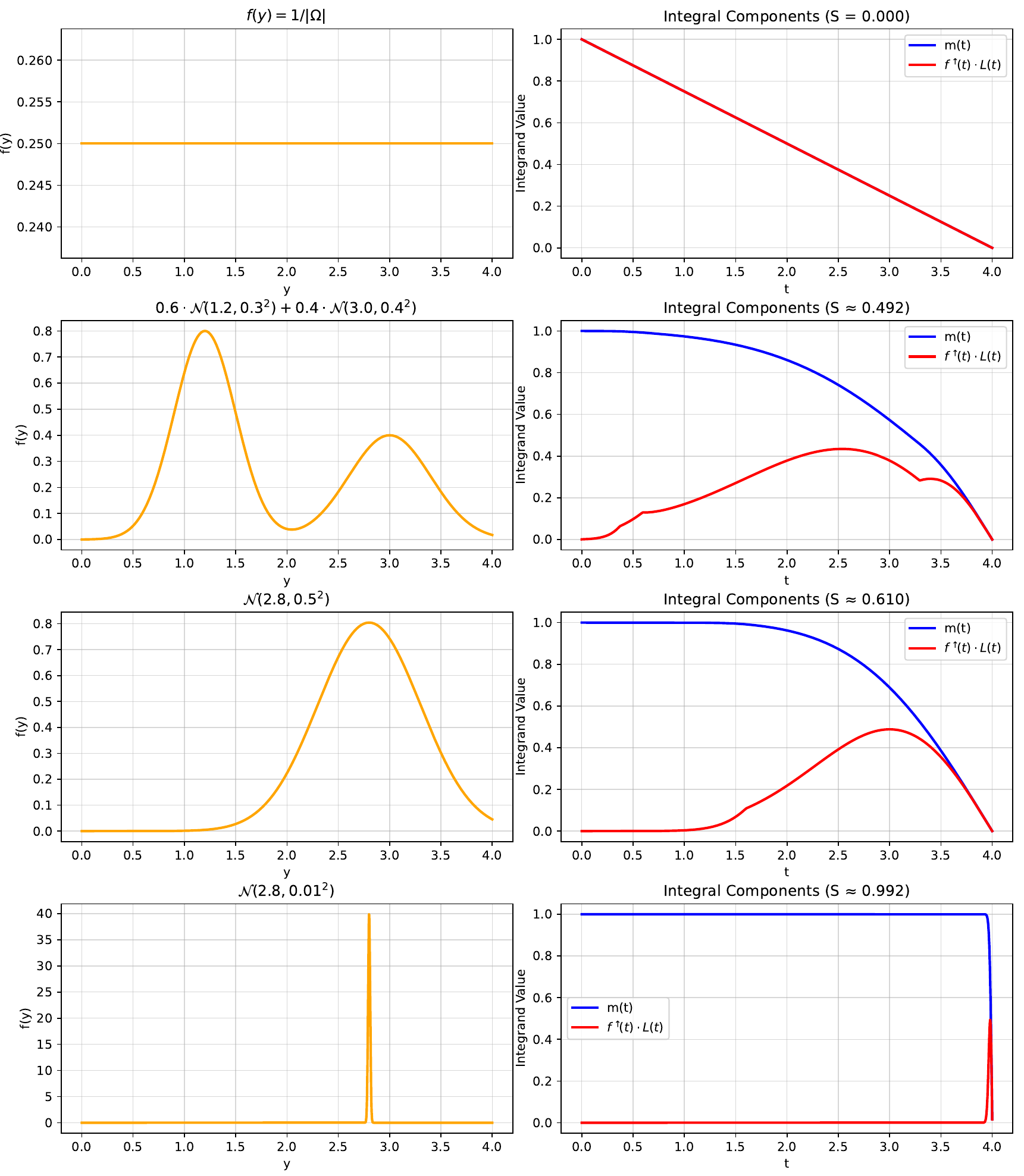}
}

\caption{\label{fig:integrand}Plots for select pdfs in Table~\ref{tbl-two} and the integral components of $S(f)$.}
\end{figure}

We first examine the mass--length representation of $S(f)$. This expression represents sharpness by the area between two curves, displayed in Figure~\ref{fig:integrand}.

\clearpage

The mass--length representations encode substantial information about how concentration across the distribution is structured: the two curves remain separated when regions of the outcome space are suppressed, descend in parallel over piecewise constant segments, and approach one another more sharply when mass concentrates in regions of the outcome space.

\begin{figure}
\centering{
\includegraphics[width=13cm,height=\textheight]{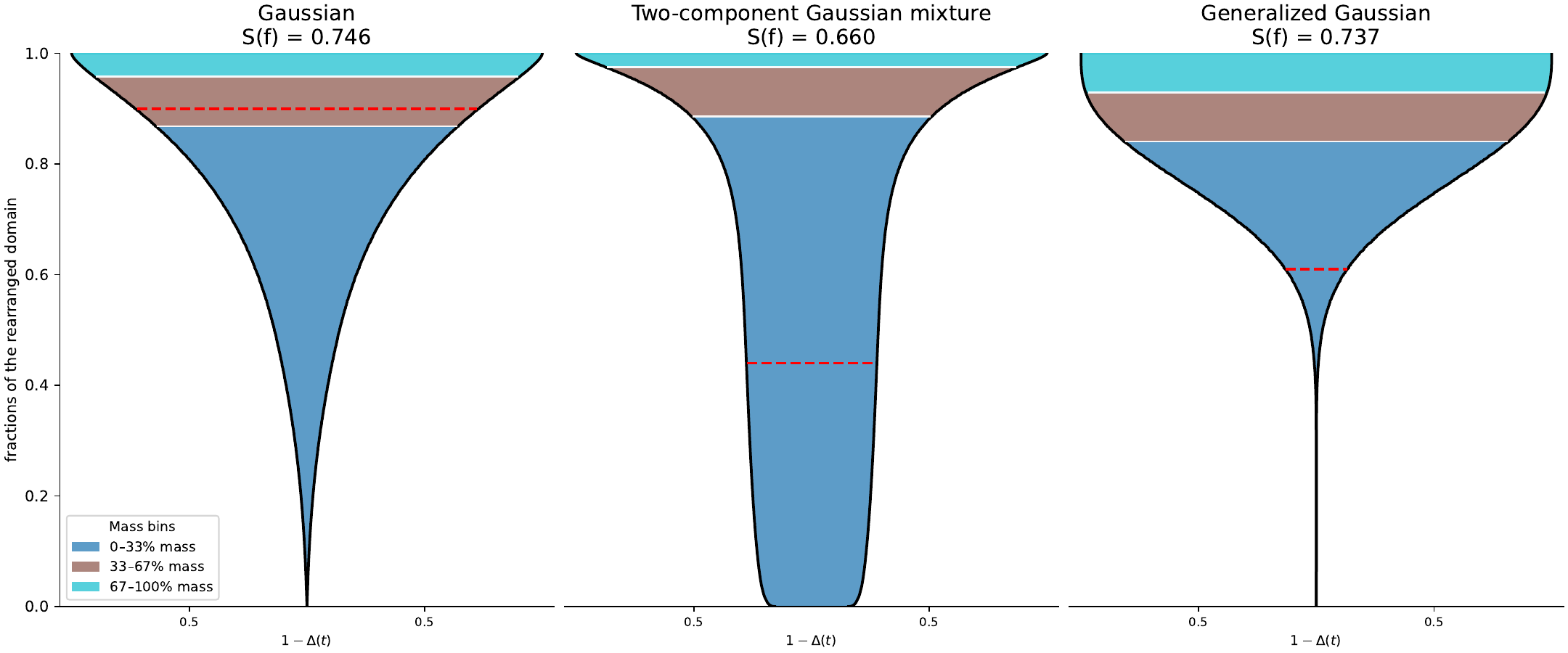}
}

\caption{\label{fig:3D}Concentration plots for three-dimensional probability density functions defined on $\Omega = [0,2]^3$, each with entropy $H(f) \approx 1.0$. The x-axis measures contribution to the sharpness score, while the 0.5 intercept marks when the contribution falls below half of the total possible (obtained by full exclusion). The plots display fractions of the domain suppressed (or fully excluded), constant-density regions, the rate at which concentration increases across the distribution's ranked profile, as well as the amount of probability mass packed into fractions of the domain. An observed value is also shown on each plot (by the dashed red line).}
\end{figure}

This transparent behavior enables us to convert the mass--length plots into more visually intuitive concentration plots that capture several attributes of interest concerning the distribution's concentration. The plot is constructed as an inverse of the mass--length plot, so that more of the probability mass along the rearranged space is retained as we move from the bottom of the figure to the top. The slope of the figure is determined by the gap between the mass--length curves, thus depicting the rate at which concentration grows across the distribution's ranked profile. The figures are further colored by 33\% mass increments, displaying how much probability mass is packed into fractions of the domain, and are mirrored about the vertical axis for cleaner display. Figure~\ref{fig:3D} illustrates such concentration plots for three 3D probability density functions defined on  $\Omega = [0,2]^3$.

Figure~\ref{fig:3D} shows three probability density functions with differential entropy $H(f) \approx 1.0$. Despite having approximately equal entropy and, in two cases, approximately equal sharpness, the concentration plots show differences in how concentration is structured across the distributions' ranked profiles. Thus, the plots provide information about distributional structure that is not captured by either of the scalar summaries alone. The plots can be used for further analysis, for example, to identify differences in concentration patterns across distributions. The plots can be especially useful in multidimensional cases, where visualization is otherwise more limited. For example, large collections of distributions may be screened efficiently for differences in concentration profiles, revealing different underlying distributional shapes.

The nondecreasing rearrangement discards information about the spatial placement of mass in the original domain, so the plots do not automatically show spatial organization of concentration, such as multimodality (although the mass-length representation exhibits characteristic patterns for certain multimodal shapes as shown, for example, in Figure~\ref{fig:integrand}). However, spatial information can be incorporated by applying the mass--length expression. Specifically, we can choose points or regions $y_{\mathrm{p}} \in \Omega$ from the original distribution and represent these meaningfully within the mass--length functional space. For a given point $y_{\mathrm{p}}$ (represented by a small neighborhood around $y_{\mathrm{p}}$), its rearranged position $t_{\mathrm{p}} \in [0, |\Omega|]$ can be approximated by matching its density value to the rearranged profile
\[
t_p = t_j \quad \text{where} \quad j = \arg\min_i \left| f^{\uparrow}(t_i) - f(y_p) \right|,
\]
using a fine-grained discretization $\{y_i\}_{i=1}^N$ of the original domain. Using this approach, one may choose regions of interest and examine them on either the mass--length plots or the concentration plots. It is then possible to examine several meaningful attributes, such as the relative placement of the region within the distribution's ranked profile and the type of concentration pattern in the region (e.g., whether it falls in an excluded region, a piecewise constant region, or a region of increasing concentration), as well as other properties such as relative rank and mass placed in higher-density regions. The most direct use case is to plot the observed value on the concentration plots, displaying its relative position within the distribution, shown also in Figure~\ref{fig:3D}. In the supplementary material, we provide formulas for these mappings, including calculating the sharpness score contributions of specific regions of the original outcome space, which may be used for detecting differences in concentration across different parts of the original outcome space (see Supplementary material, Section~S3).

\subsection{Lorenz Curve Representation}

$S(f)$ admits a reinterpretation as a Gini-type coefficient applied to the predictive distribution. Thus, sharpness can be further visualized using Lorenz curves. We proceed to derive the Gini expression next. The classical Gini coefficient is given by $G = 1 - 2 \int_0^1 L(p)\, dp$, where the Lorenz curve is defined as $L(p) = \mu^{-1} \int_0^p F^{-1}(t)\, dt$, $\mu$ is the population mean, and $F^{-1}(t)$ denotes the quantile function of the income distribution  \citep{gastwirth1972,kakwani1980}. In this context, probability mass replaces income, so that the Lorenz curve measures the cumulative share of total probability mass accumulated in fractions of the rearranged domain.

Let $f^{\uparrow} \colon [0, |\Omega|] \to \mathbb{R}_{\geq 0}$ be the nondecreasing rearrangement of a predictive density $f$ defined over a bounded domain $\Omega \subset \mathbb{R}^d$, and let $u = t / |\Omega| \in [0,1]$ denote the normalized domain. The Lorenz-type curve over this rearranged space is given by
\begin{equation}
L(u) = \int_0^{u|\Omega|} f^{\uparrow}(s)\, ds,
\end{equation}
which gives the cumulative probability mass over the fraction $u$ of the rearranged outcome space. Using the simplified expression \eqref{eq:cet} and applying the change of variables $t = u|\Omega|$, with $dt = |\Omega|\, du$, we obtain:
\[
S(f) = 2|\Omega| \int_0^1 u\, f^{\uparrow}(u|\Omega|)\, du - 1.
\]
Observing that
\[
L'(u) = \frac{d}{du} \int_0^{u|\Omega|} f^{\uparrow}(s)\, ds = f^{\uparrow}(u|\Omega|)\cdot |\Omega|
\quad\text{for a.e. } u\in[0,1],
\]
it follows that $f^{\uparrow}(u|\Omega|) = \frac{1}{|\Omega|} L'(u)$, and thus:
\[
S(f) = 2 \int_0^1 u\, L'(u)\, du - 1.
\]
Applying integration by parts,
\[
\int_0^1 u\, L'(u)\, du = \left[ u L(u) \right]_0^1 - \int_0^1 L(u)\, du = 1 - \int_0^1 L(u)\, du,
\]
since $L(1) = \int_0^{|\Omega|} f^{\uparrow}(s)\, ds = 1$. Substituting gives
\[
S(f) = 2(1 - \int_0^1 L(u)\, du) - 1
\]
\begin{equation}
= 1 - 2 \int_0^1 L(u)\, du.
\end{equation}

\begin{figure}
\centering{
\includegraphics[width=11.5cm,height=\textheight]{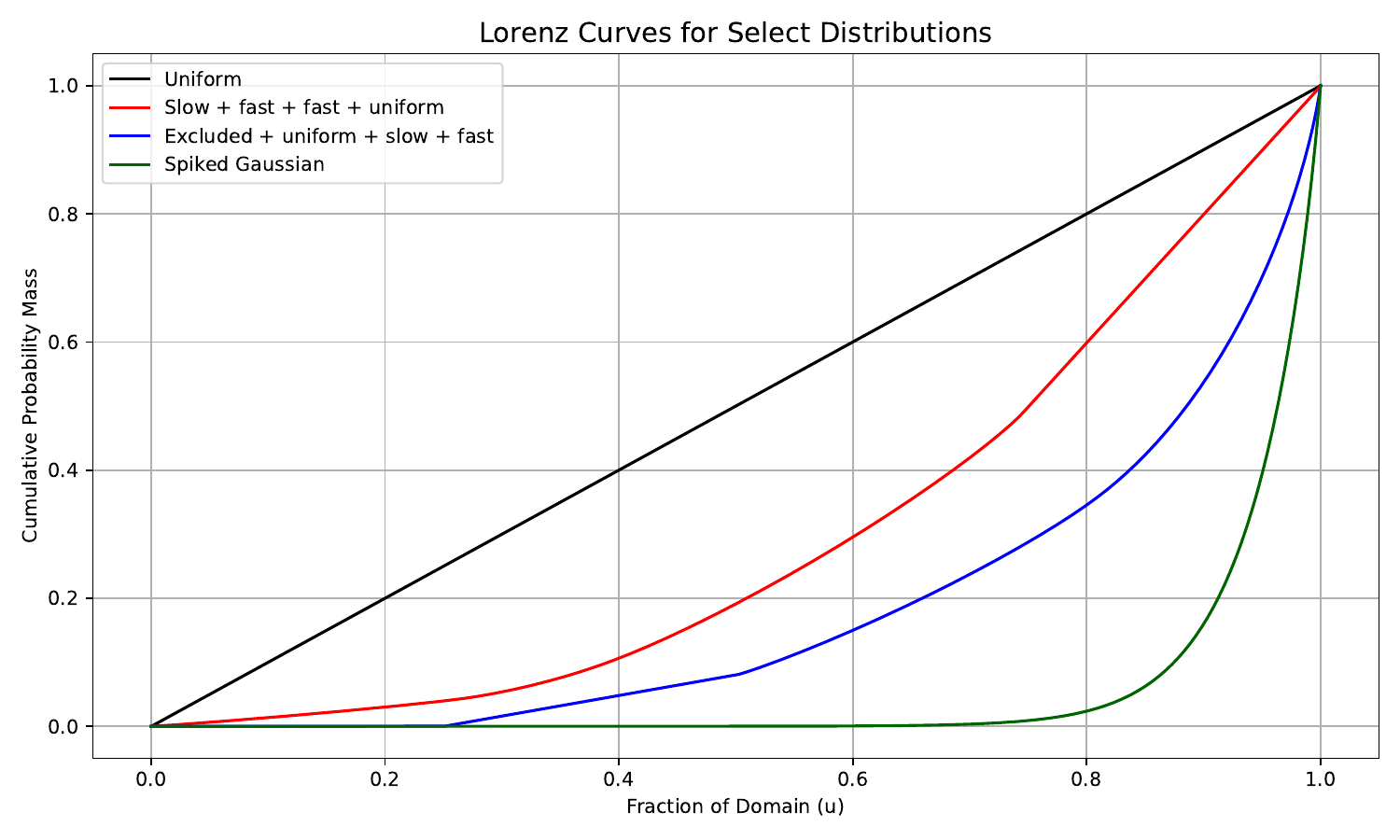}
}
\caption{\label{fig:gini}Lorenz curve representation of three 3D PDFs defined on $\Omega = [0,2]^3$, defined based on different types of concentration patterns. For comparison, a concentration plot representation of the same PDFs is provided in the supplementary material (Section~S3.3).}
\end{figure}

The Lorenz curve representation---illustrated in Figure~\ref{fig:gini}---provides another visually intuitive means of comparing the sharpness of probability distributions. However, compared with the mass--length expression, this formulation includes less information about the distribution's concentration profile, as the cumulative nature of the Lorenz curve collapses the ranked concentration profile into a single cumulative trajectory, thereby obscuring local concentration patterns. The formulation is also computationally somewhat less direct than the simplified formula \eqref{eq:cet}. As such, within the probabilistic context, it may be most useful when visualizing relative sharpness rankings across distributions. Nevertheless, we may observe that given the mathematical equivalence of the formulas, the traditional Gini coefficient can likewise be expressed in the mass--length form. In this setting, the total remaining share of wealth again replaces the total remaining probability mass, and $f^{\uparrow}$ is replaced by the mean-normalized quantile function $F^{-1}/\mu$ of the income distribution. This provides access to the graphical methods provided by the mass--length formulation, for example, examining specific quantiles of the population on the concentration plots, and constitutes an area for further exploration.

\section{Case Studies and Applications}
In this section, we examine applications of the sharpness measure in model diagnostics. In Section~5.1, we demonstrate the behavior of the measure using real data from standard weather forecasting settings. In Section~5.2, we illustrate multivariate forecast evaluation using simulated data. In Section~5.3, we compare the measure with the multivariate energy score \citep{gneiting2007scoring} and determinant sharpness \citep{gneiting2008}. In Section~5.4, we discuss computational methods and domain conventions.

\subsection{WeatherBench2}

Given that the sharpness measure was constructed from a theoretical rank-based concentration principle, we first demonstrate the behavior of the measure using real forecasting data. We use publicly available WeatherBench2 global medium-range (1--14 days) weather forecasting data \citep{rasp2024}, choosing ensemble forecasts from the WeatherBench2 IFS ensemble dataset together with ERA5 reanalysis as the verifying reference (for CRPS and the energy score). Dispersion measures, including prediction interval width, standard deviation, and multivariate extensions such as determinant sharpness, are the standard comparison measures in these settings, as outcomes fall naturally on low-dimensional scales in natural units and spatial dispersion around the mean is of primary interest. However, the sharpness measure should closely align with these measures in these settings, and provide complementary information when probabilistic concentration and spatial dispersion differ.

Two examples are considered. First, for 24-hour accumulated precipitation, we analyze global forecasts initialized at the beginning of January 2021 at a lead time of 72 hours and compare the sharpness score with 90\% prediction interval width, ensemble standard deviation, and CRPS. For this demonstration, we use only the raw histogram data. Second, for 10 m wind, we analyze 24-hour and 120-hour forecasts initialized between January and February 2021 at four locations, representing each bivariate $(u, v)$ ensemble forecast by a rolling-trained Gaussian-mixture density and comparing sharpness with entropy and determinant sharpness, the multivariate extension of standard deviation previously applied in 10 m wind ensemble forecast evaluation \citep{gneiting2008}. We also reproduced the results discussed below with the raw histogram data, verifying that the relationships between the measures were not produced by our particular smoothing approach.

\clearpage

\begin{figure}
\centering{
\includegraphics[width=15cm,height=\textheight]{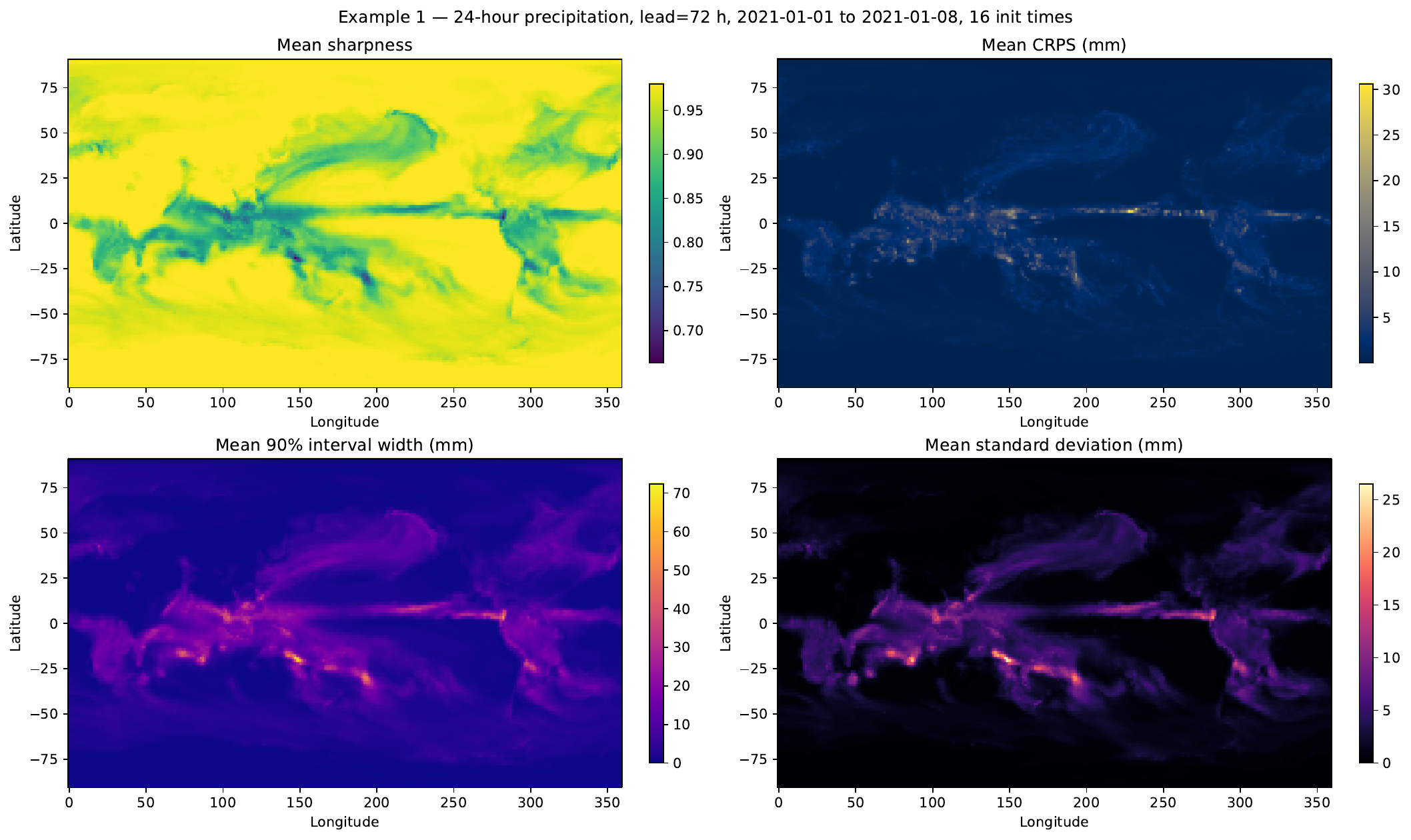}
}
\caption{\label{fig:precip}Global 24-hour precipitation forecasts from the IFS ensemble dataset, initialized 1--8 January 2021. Sharpness is shown alongside CRPS, 90\% prediction interval width, and ensemble standard deviation.}
\end{figure}

\begin{figure}
\centering{
\includegraphics[width=16cm,height=\textheight]{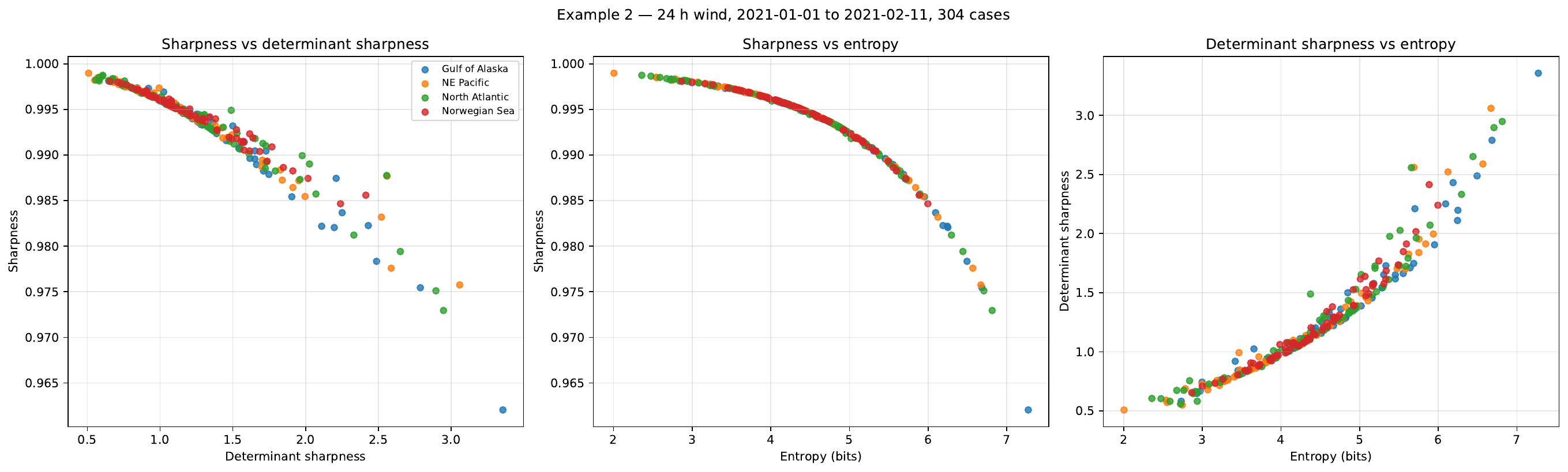}
}
\caption{\label{fig:wind}Crossplots for 24-hour $(u, v)$ wind forecasts at four locations, comparing sharpness, entropy, and determinant sharpness. In this setting, sharpness and entropy remain tightly coupled, as forecasts broadly share the same shape, with some top concentration and occasional outlying ensemble members.}
\end{figure}

In these settings, the proposed sharpness measure shows the consistent behavior expected of a measure of forecast concentration. For precipitation, the spatial distribution of mean sharpness is closely aligned with conventional measures: regions with broader predictive distributions exhibit lower sharpness, whereas more concentrated forecasts exhibit higher sharpness, as shown in Figure~\ref{fig:precip}. For bivariate wind forecasts, sharpness, entropy, and determinant sharpness are closely aligned at a short 24-hour lead time, as ensemble members tend to cluster closely together in the outcome space. Each measure quantifies, in a distinctive fashion, the spread or concentration of this small cluster, resulting in the tight coupling shown in Figure~\ref{fig:wind}.

\clearpage

\begin{figure}
\centering{
\includegraphics[width=16cm,height=\textheight]{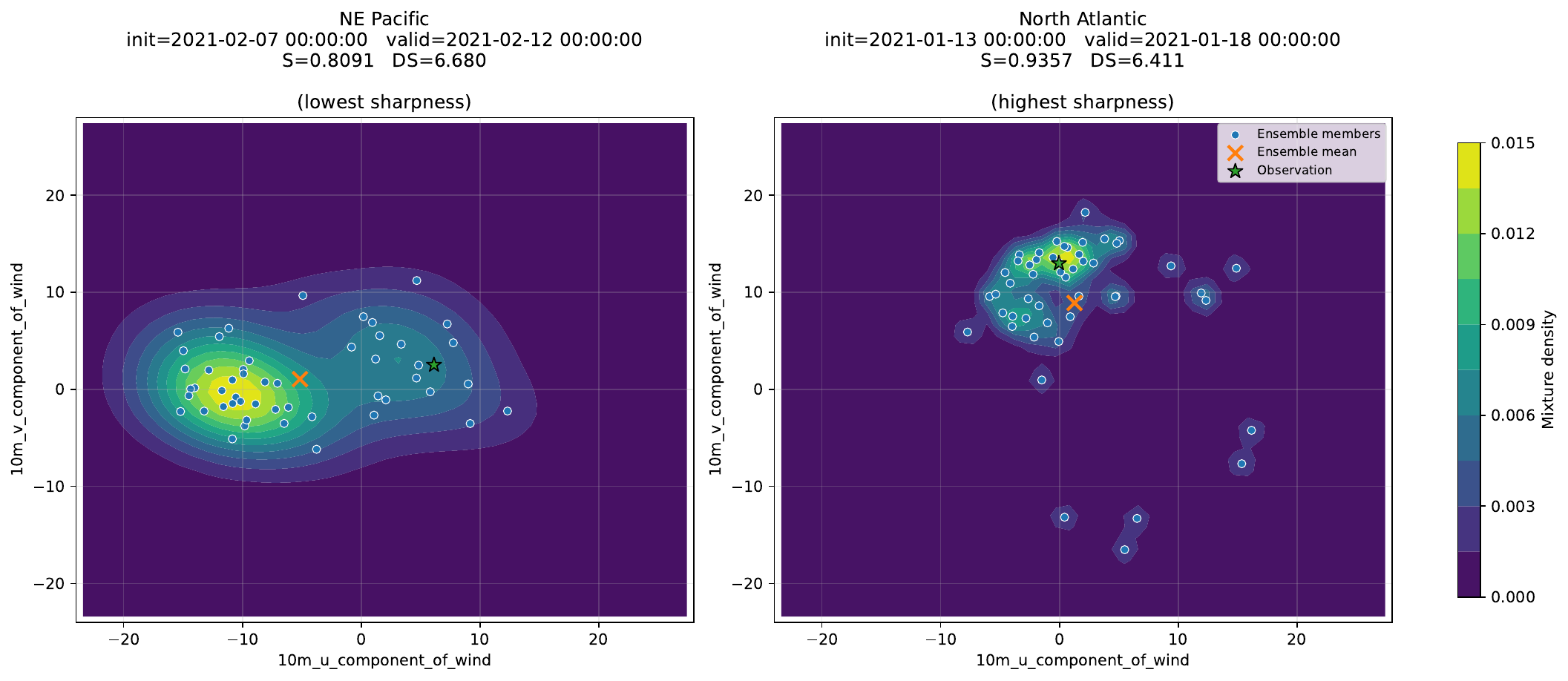}
}
\caption{\label{fig:wind2}Two 10 m wind forecasts with 120-h lead time with approximately equal determinant sharpness, distinguished by a low ($S(f) \approx 0.81$) and a high ($S(f) \approx 0.94$) sharpness score. A high sharpness score reflects forecast concentration over a smaller region of the outcome space.}
\end{figure}

However, more substantial differences between sharpness and determinant sharpness emerge at greater lead times, where sharpness provides an interpretable companion to determinant sharpness in diagnosing ensemble forecast shape. Under this forecasting setting, the sharpness measure reflects closely the exclusion of possible regions of the outcome space, given that ensemble members tend to cluster. With our setup, each $10\,\mathrm{m\,s^{-1}} \times 10\,\mathrm{m\,s^{-1}}$ area in $u$- and $v$-components represents around $3.7\%$ of the total possible outcome space. Thus, a sharpness score of $0.96$ reflects the ensemble forecast clustering almost entirely within such an area, and lower scores indicate that the ensemble forecast is spreading over correspondingly larger areas. Determinant sharpness reflects overall spread, exhibiting sensitivity to outliers. This difference enables the measures to provide complementary information, with sharpness distinguishing more concentrated distributions at equal determinant sharpness, and determinant sharpness distinguishing between more and less spatially spread-out distributions at equal sharpness (here, generally between unimodal and bimodal shapes). This complementary behavior is illustrated in Figure~\ref{fig:wind2}, while more detailed analysis of the complementary information that these measures provide is also provided in Section~5.3.

\subsection{Simulated multivariate forecasts}

We next illustrate the sharpness measure in a controlled multivariate setting, which is constructed as a simple simulation case study based on macroeconomic forecasting. We consider predictive densities on the bounded cube
\[
\Omega=[-3,3]^3,
\]
where the three coordinates represent GDP growth, inflation, and the change in the unemployment rate, all measured in percentage points. Forecasts are issued monthly over a one-year horizon, and a single realized outcome is observed at the end of the period. The forecasts are not necessarily designed to mimic genuine forecasters in detail, but to generate interpretable joint predictive densities whose shape changes over time in distinctive ways. At each issue date $t=1,\dots,12$, forecasters observe a noisy public signal about the expected outcome. Specifically, the signal is centered at the realized outcome and its covariance decreases over time, so that information gradually improves as the time horizon shortens. This creates a setting in which all forecasters are eventually pulled toward the same realized outcome, while retaining freedom to use different forecasting strategies.

Four stylized forecasters are considered. Forecaster A represents a forecaster with a fixed macroeconomic background view (aligned with the realized outcome) who gradually concentrates their forecasts on their favored scenario. Forecasts are represented by a single Gaussian density informed by the fixed view and the evolving public signal. Forecaster B is a scenario-based forecaster with some tail scenario uncertainty and a central predicted region. However, the forecast model begins to exhibit some pathological behavior with a sharpening core region while the tail scenarios retain non-negligible mass. Forecaster C is a regime-switcher represented by a two-component Gaussian mixture. Early in the forecast cycle the forecaster places most predictive mass on a "wrong" regime that is too pessimistic about growth; as the horizon shortens, the weight shifts toward the "right" regime centered near the evolving public signal. Finally, Forecaster D issues a forecast that increasingly concentrates over a central region while also exhibiting small tail risk.

\begin{figure}[t]
\centering
\includegraphics[width=16cm]{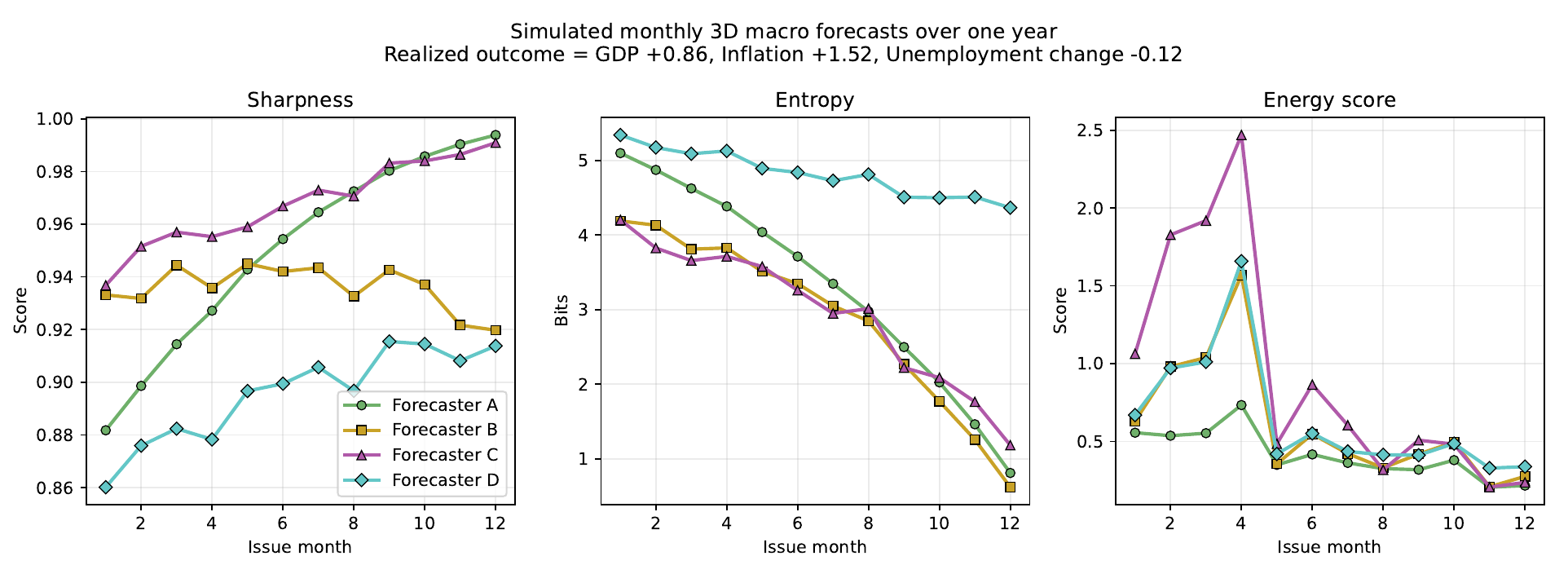}
\caption{\label{fig:macro_sim}Monthly simulated macroeconomic forecasts on $\Omega=[-3,3]^3$, comparing sharpness, differential entropy, and the energy score for four stylized forecasters.}
\end{figure}

Figure~\ref{fig:macro_sim} shows comparison metrics for the four stylized forecasters. Sharpness, entropy, and the energy score are calculated for each forecast. The comparisons illustrate the joint information provided and the differences in interpretation between the respective measures. For example, entropy and sharpness can decouple if tail risks increase while forecasts are also becoming more concentrated as the realization horizon nears (Forecast D). The linear weights for sharpness are insensitive to local spikes, so sharpness reflects that the pathological forecast is not genuinely becoming sharper (Forecast B). In contrast to entropy and sharpness, the energy score captures the greater spread of the two-regime forecast (Forecast C), as its dispersion term is specifically sensitive to spatial dispersion (see Section~5.3). In this simulation, all forecasts trace the "real" signal, and hence each ends up with a similar energy score.

A benefit of the sharpness measure is that it admits a reinterpretation with respect to the original outcome variables. By the exclusion principle (Proposition~\ref{prop:exclusion}), as an approximate heuristic it can be interpreted based on the exclusion of fractions of the total outcome space. In any dimension $d$, the difference between maximum sharpness and the given sharpness score corresponds to $(1 - S(f))^{1/d}$ per-variable (side-length) retained possibility space. This can be related back to the original variables, so that the remaining fraction reflects (to a certain approximation) the range of each variable still considered by the model within the more probable range. (The predictive density may also be more concentrated for some variables, entailing that the others must be more diffuse. Such asymmetries are naturally captured by spatially sensitive dispersion measures, discussed further below.) For example, in this setup, the sharpness score of Forecaster A's final forecast corresponds to an equivalent retained side length of about 18\% of the original range in each variable, or roughly 1.1 percentage points. Thus, sharpness scores can also sometimes be reported with the respective side-length equivalents to aid interpretation and, if needed, also with the corresponding concentration plot visualizations for further diagnosis.

\subsection{Sharpness and the energy score}
In this section, we further clarify the relationship of the sharpness measure with proper scoring rules, discussing in particular the multivariate energy score \citep{gneiting2007scoring}. Proper scoring rules provide a principled way to evaluate probabilistic forecasts by encouraging forecasters to report their true beliefs \citep{gneiting2007scoring,gneiting2007,gneiting2014}. In particular, the continuous ranked probability score (CRPS) \citep{matheson1976}, and its multivariate extension the energy score \citep{gneiting2007scoring}, offer an attractive method for overall forecast evaluation \citep{gneiting2008}. At the same time, scoring rules combine both calibration and sharpness into a single scalar value, while decompositions require observations and do not isolate an intrinsic sharpness component of the predictive distribution \citep{arnold2024}. Thus, the proposed sharpness measure can provide complementary information, focusing specifically on the forecast's intrinsic probabilistic concentration.

We first establish a key consistency property implied by proper scoring rules. In particular, for an autocalibrated random forecast $F$, proper scoring rules imply a certain information-ordering requirement on sharpness measures: a better-informed forecast $F$ should be at least as sharp on average as its less-informed, unconditional version $F_0$ \citep{jordan2026sharpness}. The requirement for this is that Jensen's inequality holds,
\[
\mathbb{E}[S(F)] \geq S(F_0) = S(\mathbb{E}[F]).
\]
which, for a positively oriented sharpness measure, requires convexity. \citet{jordan2026sharpness} recently showed that certain common sharpness measures, such as central prediction interval width, fail to satisfy this requirement (in the negative orientation). By its Gini equivalence shown above, $S$ is convex, and hence the proposed sharpness measure satisfies this requirement. In other words, incorporating information cannot reduce expected sharpness: the informed (autocalibrated) forecast is at least as sharp on average as its less-informed version obtained by averaging over the available information, thus ensuring consistency with the best-informed forecast principle underlying proper scoring rule evaluation.

In addition to satisfying this fundamental requirement, the proposed sharpness measure exhibits a certain structural similarity with the multivariate energy score, specifically in relation to its dispersion component. First, we may observe that by its Gini identity, the sharpness measure can be written also in an unsorted, pairwise form. In the discrete case,
\begin{align*}
S(P)
&=\sum_{j=1}^n \frac{2j-n-1}{n-1}\,p_{(j)} \\
&=\frac{1}{n-1}\sum_{i<j}\bigl(p_{(j)}-p_{(i)}\bigr) \\
&=\frac{1}{2(n-1)}\sum_{i=1}^n\sum_{j=1}^n |p_i-p_j|.
\end{align*}
Similarly, in the continuous case,
\begin{align*}
S(f)
&=\frac{2}{|\Omega|}\int_0^{|\Omega|} t\,f^{\uparrow}(t)\,dt-1 \\
&=\frac{1}{2|\Omega|}\int_0^{|\Omega|}\int_0^{|\Omega|}\bigl|f^{\uparrow}(s)-f^{\uparrow}(t)\bigr|\,ds\,dt \\
&=\frac{1}{2|\Omega|}\int_\Omega\int_\Omega |f(y)-f(z)|\,dy\,dz,
\end{align*}

\noindent where the last equality follows from equimeasurability. For a predictive distribution $P$ on $\mathbb{R}^d$, the energy score is
\[
\mathrm{ES}(P,x)=\mathbb{E}_{P}\|X-x\|^\beta-\frac12\,\mathbb{E}_{P}\|X-X'\|^\beta,
\]
where $X,X' \sim P$ are independent, and $\beta\in(0,2)$ controls how strongly the score responds to forecast–outcome and within-forecast distances. With the standard choice $\beta = 1$, this yields the CRPS in the univariate case \citep{gneiting2007scoring}. The second term is interpreted as the forecast's internal spread or dispersion component. If $P$ admits a density, the second term can be written in full pairwise form as
\[
\frac12\,\mathbb{E}_{P}\|X-X'\| = \frac12\int_{\mathbb{R}^d}\int_{\mathbb{R}^d}\|u-v\|\,dP(u)\,dP(v).
\]
If the predictive distribution is approximated by $m$ draws $X_1,\dots,X_m \sim P$, the corresponding empirical approximation is
\[
\widehat{\frac12\,\mathbb{E}_{P}\|X-X'\|} = \frac{1}{2m^2}\sum_{i=1}^m\sum_{j=1}^m \|X_i-X_j\|.
\]

We observe that with weight $\beta = 1$, this term too is proportional to the multivariate distance-Gini mean difference \citep{koshevoy1997}, with the exact constant depending on the normalization convention (under Koshevoy-Mosler's convention, the constant is $d$, where $d$ is the dimension). In the univariate case, the dispersion term of the CRPS thus equals the distance-Gini mean difference (GMD), and differs from the Gini coefficient only by normalization with respect to the mean \citep[p.~256]{koshevoy1997}:
\[
GMD= \frac12\int_{\mathbb R}\int_{\mathbb R}|u-v|\,dP(u)\,dP(v).
\]

Thus, both the energy score dispersion term and $S$ admit pairwise Gini-type representations. Nevertheless, they measure different forms of concentration: the energy-score term measures pairwise spatial distances between forecasted outcomes, whereas $S$ measures pairwise differences in the predictive densities over the outcome space. This enables the measures to capture different aspects of predictive concentration, where the energy score reflects spatial differences and the proposed sharpness measure reflects intrinsic probabilistic concentration.

\begin{figure}[t]
\centering
\includegraphics[width=10.0cm]{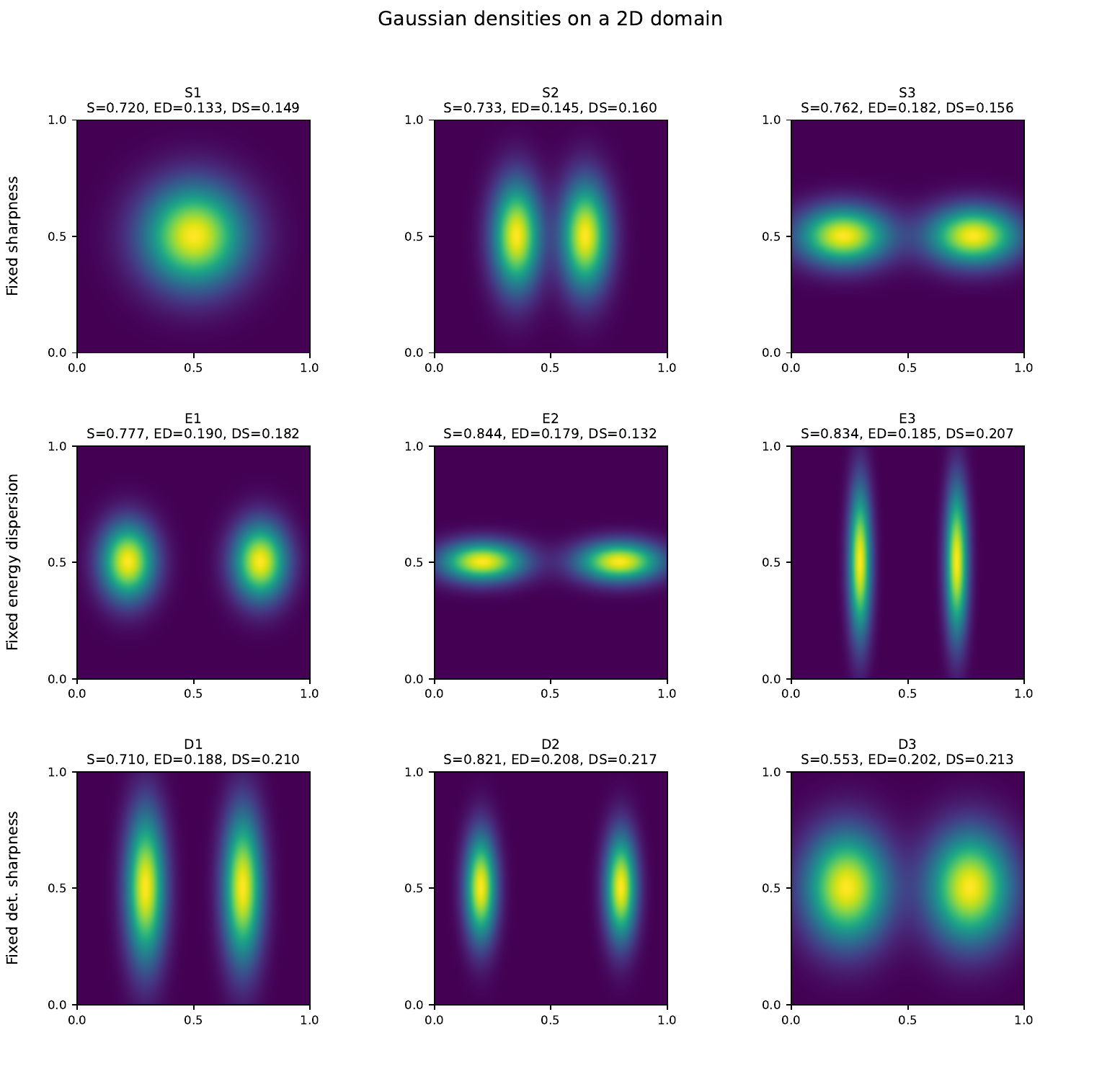}
\caption{\label{fig:gauss}Truncated and renormalized Gaussian mixture densities on $[0,1]^2$. Each row keeps one summary metric approximately fixed, displaying variation in the others: the top row fixes sharpness (S), the middle row fixes the energy score dispersion term (ED), and the bottom row fixes determinant sharpness (DS). In the second and third columns, the other remaining measure is fixed while the final one captures variation in the third column.}
\end{figure}

Both of these measures can yet be contrasted with determinant sharpness, which measures spatial spread in a yet different way. Determinant sharpness,
\[
DS=(\det \Sigma)^{1/(2d)},
\]
summarizes the overall spread encoded by the covariance matrix \citep[p.~221]{gneiting2008}, and thus captures a different aspect of forecast spread than does the energy score dispersion term. A simple two-dimensional case example can make these distinctions concrete. Figure~\ref{fig:gauss} shows predictive densities on $\Omega=[0,1]^2$, each defined as anisotropic Gaussian mixtures. In each case, two of the measures may be fixed while the third captures variation in either distributional spread or concentration, showing sensitivity to different aspects of distributional structure.

The proposed sharpness measure thus adds a quantity that is not provided by determinant sharpness and the energy score dispersion term: a measure of the forecast's inherent concentration. In this sense, determinant sharpness captures covariance volume, the energy score captures pairwise spread in the forecasted outcomes, and $S$ captures inherent probabilistic concentration.

\subsection{Computation and domain conventions}
Finally, we discuss computation and practical domain conventions. In practical applications, the sharpness score is scaled by the domain choice. When an outcome variable lies on a well-defined, bounded scale, domain choice is straightforward. However, in many applications, the domain choice is not immediately given. For example, in many weather forecasting settings, extreme values are physically possible but exceedingly improbable, and the range of natural variability varies by the setting. Furthermore, even though many measures, including scoring rules such as the Brier score, implicitly measure performance relative to the uniform distribution, the uniform distribution is not always a natural baseline of comparison \citep{winkler2008}. For example, for precipitation, the uniform distribution over the maximum range of natural variability is not a starting point that appropriately reflects prior knowledge.

We observe that by the linearity and scaling properties of the sharpness measure, the domain choice affects only the relative scaling of sharpness score distances, rather than altering the underlying information conveyed about the predictive distribution. The relative differences between sharpness scores remain unaltered under domain expansion and restriction, provided that the full predictive distribution is evaluated under each convention. To give an example, if forecasts only assigned densities to outcomes ranging between $0$ and $10$ in the given units but a domain of size $|\Omega| = 20$ was used, the differences between sharpness scores would be more compressed in the upper range. However, the sharpness scores for the distributions defined under domain convention $|\Omega| = 10$, $|\Omega| = 20$, or larger remain directly comparable, and can be transformed between conventions using the domain transformation formulas provided in Table~\ref{tbl-three}. Thus, alternative domain conventions can also be compared retrospectively, provided that the full predictive distribution is evaluated under each convention. The derivation of the domain change formulas is provided in the supplementary material, Section~S2.

Sharpness score differences can be examined equivalently as a normalized gain from $S_1$ to $S_2$,
\[
\Delta_{\mathrm{rel}}=\frac{S_2-S_1}{1-S_1},
\]
which becomes convenient if the scores are compressed in the upper range. In either case, a change in sharpness from $S_1$ to $S_2$ remains tied to the original variables through the side-length equivalent discussed above,
\[
(1-S_1)^{1/d}-(1-S_2)^{1/d}.
\]
Using the normalized gain, this can be written as
\[
(1-S_1)^{1/d}\Bigl[1-(1-\Delta_{\mathrm{rel}})^{1/d}\Bigr],
\]
since
\[
1-S_2=(1-\Delta_{\mathrm{rel}})(1-S_1).
\]

The measure can thus be rescaled to focus only on the support of a distribution for arbitrary, bounded domains, and alternative truncation schemes of zero-probability regions yield, via transformation formulas, equivalent sharpness scores for any valid probability distributions. In practical applications, we have found that a convenient approach is to define the domain empirically by examining the maximum range of variability in the outcome variables of interest, and if needed, to focus on the effective support by examining the relative sharpness change. An alternative is to define the domain based on a theoretically relevant range. For comparison to sharpness benchmarks other than the uniform distribution, the relative measure (taken to that alternative benchmark) presents a viable alternative.

A limitation of the sharpness measure is that it is not defined for distributions with infinite support. However, the measure can nonetheless be applied when using a common truncation scheme. For example, one way to define the relevant domain is to choose a common bounded range that contains at least 99\% of the probability mass of all the distributions of interest. The distributions may then be truncated and renormalized to this range, and the sharpness measure applied in the usual fashion. The resulting scores quantify concentration relative to the chosen bounded range, but the truncated distributions may nonetheless approximate the original distributions arbitrarily closely by retaining an increasingly large share of their probability mass.

With respect to interpretation, what counts as "sharp" is relative to the given practical domain and the performance of current models \citep{thompson2023}. For example, consider that 10-m wind forecasts are able to reliably predict the wind vector within about $4\, \mathrm{m\,s^{-1}}$ in the $u$- and $v$-components. The ability to issue forecasts within about $2\, \mathrm{m\,s^{-1}}$ in these components is "sharper," but the value of this increase depends on practical needs. Similarly, what specific numerical value is assigned to such an increase by the sharpness measure depends on what specific domain convention is used. However, given the linearity properties of the measure, each alternative convention reflects the same underlying shift with respect to the original variables. For example, on a domain of size $50\,\mathrm{m\,s^{-1}} \times 50\,\mathrm{m\,s^{-1}}$, an improvement from about $4\, \mathrm{m\,s^{-1}}$ to about $2\, \mathrm{m\,s^{-1}}$ in each component corresponds approximately to sharpness score change from $S(f) = 0.9936$ to $S(f) = 0.9984$ (in other words, the respective sharpness scores lie closer to these uniform-on-support values). If the domain is defined instead as $10\,\mathrm{m\,s^{-1}} \times 10\,\mathrm{m\,s^{-1}}$, the absolute sharpness change is from $S(f) = 0.84$ to $S(f) = 0.96$. However, both changes reflect the same underlying change, and can be confirmed to count equally in the side-length form (a reduction of $2\, \mathrm{m\,s^{-1}}$ when measured in the side-length equivalents). Thus, the interpretation of the sharpness measure remains tied to the original variables, and numerical differences between sharpness scores are interpreted based on relative benchmarks in the given application domain.

\begin{longtable}{@{}ll@{}}
\caption{Sharpness score (S) transformations across domains.}
\label{tbl-three}\\
\toprule
\multicolumn{1}{l}{\textbf{Transformation}} &
\multicolumn{1}{l}{\textbf{Formula}} \\
\midrule
\endfirsthead

\toprule
\multicolumn{1}{l}{\textbf{Transformation}} &
\multicolumn{1}{l}{\textbf{Formula}} \\
\midrule
\endhead

\bottomrule
\endfoot

\bottomrule
\endlastfoot

Forward (Discrete): embed from size $m$ to size $n > m$
& $S_n = 1 + \dfrac{m - 1}{n - 1} (S_m - 1)$ \\
Inverse (Discrete): restrict from size $n$ to size $m < n$ 
& $S_m = 1 + \dfrac{n - 1}{m - 1} (S_n - 1)$ \\
Forward (Continuous): extend from measure $\ell$ to $L > \ell$ 
& $S_L = 1 + \left( \dfrac{\ell}{L} \right) (S_\ell - 1)$ \\
Inverse (Continuous): restrict from measure $L$ to $\ell < L$ 
& $S_\ell = 1 + \left( \dfrac{L}{\ell} \right) (S_L - 1)$ \\
\end{longtable}

For computation, it is recommended to use the simplified expressions of both the discrete and continuous sharpness formulas, \eqref{eq:dgin} and \eqref{eq:cet}, respectively. These use sorting and therefore have overall complexity $O(n\log n)$ in the discrete case, and likewise $O(N\log N)$ on a fixed discretization with N cells in the continuous case, with linear-time evaluation once the values are ordered. The pairwise formulas introduced above avoid sorting but are quadratic, $O(n^2)$ (or $O(N^2)$), and are therefore not the fastest exact implementation. In high-dimensional settings (e.g., $d > 4$), the measure remains practical when the predictive density is directly evaluable or otherwise available in a tractable form; the main difficulty arises only when one must first construct a suitable mass representation from forecast draws. In high-dimensional cases, the most efficient practical approach is typically to avoid full tensor discretization and instead evaluate the density at Monte Carlo or quasi-Monte Carlo points drawn over the bounded domain, then estimate the score using the sorted formula on those sampled values, allowing more efficient evaluation. Python and R implementations of the sharpness formula for different practical settings are provided in the accompanying material.

\section{Discussion}

In this article, we have defined a sharpness functional for probabilistic models by applying a rank-based concentration principle that tracks upward transfers of probability mass along the rearranged profile of the distribution. The measure focuses on the structure of the predictive distribution itself---specifically, the extent to which it concentrates probability mass over a domain. The measure offers a distinct perspective from conventional tools such as scoring rules, entropy, or variance-based measures, focusing specifically on inherent probabilistic concentration.

The construction extends naturally to finite discrete, continuous, and multidimensional outcome spaces. It is consistent with a deterministic notion of concentration based on the exclusion of alternative possibilities within an outcome space. The measure also admits equivalent formulations as a Gini-type coefficient and as a scaled 1-Wasserstein distance from the uniform distribution in rearranged space. These connections demonstrate a robust mathematical foundation for the proposed construction.

The introduced framework provides several graphical and analytical tools for analyzing the concentration of probability distributions and complements many existing measures in forecast diagnostics. A complementary relationship between entropy and dispersion-based measures was demonstrated. The relationships among the sharpness measure, entropy, and dispersion-based measures warrant further investigation. In particular, these measures enable distinguishing different types of distributional shapes, which may provide additional information for model comparison as well as selection, especially when considered alongside calibration and predictive performance measures (see Supplementary material, Section~S4).

Further work may also investigate the graphical methods provided by the sharpness functional in greater detail. In particular, the geometric properties of the representations developed in Section~4 may be further investigated. The rearrangement-based formulation separates the intrinsic concentration profile of a distribution from its spatial organization, thus creating a framework for studying probabilistic concentration. Future research may investigate how these representations characterize features such as multimodality, as well as studying how spatial information could be further incorporated within the rearranged space (see Section~4 and Supplementary material, Section~S3).

In summary, we view the current approach as complementary to existing tools, and as an interpretable approach to studying predictive concentration. We hope that the preceding analyses and investigations may contribute to the development of novel approaches to forecast evaluation and refinement.

\section*{Reproducibility Statement}\label{reproducibility-statement}

The reproducibility material for the article is available from the following open GitHub repository: https://github.com/psyrjane/Predictive-sharpness. The repository also contains Python and R scripts for implementing the sharpness measure in practical settings.

\section*{AI Statement}\label{ai-statement}

ChatGPT (commercial version, versions between 2025-2026) was used for coding assistance and language improvement. In particular, AI was used to assist with figure formatting and to produce Python scripts for extracting and preparing the data needed for the analyses in Section 5. The desired analyses and visualizations were specified by the author, after which AI generated scripts that were reviewed and manually corrected where necessary. All AI-generated code and the resulting outputs were checked by the author. AI was used to check the text for grammatical errors and for more idiomatic phrasing, with all suggestions reviewed and manually edited by the author.

\bibliography{bibliography.bib}

@article{gneiting2007scoring,
  author    = {Tilmann Gneiting and Adrian E. Raftery},
  title     = {Strictly Proper Scoring Rules, Prediction, and Estimation},
  journal   = {J. Am. Stat. Assoc.},
  year      = {2007},
  volume    = {102},
  pages     = {359--378}
}

@article{gneiting2014,
  author  = {Tilmann Gneiting and Matthew Katzfuss},
  title   = {Probabilistic Forecasting},
  journal = {Annu. Rev. Stat. Appl.},
  year    = {2014},
  volume  = {1},
  pages   = {125--151}
}

@article{gneiting2007,
  author  = {Gneiting, Tilmann and Balabdaoui, Faouzi and Raftery, Adrian E.},
  title   = {Probabilistic Forecasts, Calibration and Sharpness},
  journal = {J. R. Stat. Soc. Ser. B Stat. Methodol.},
  year    = {2007},
  volume  = {69},
  pages   = {243--268}
}

@article{arnold2024,
  author  = {Arnold, S. and Walz, E. M. and Ziegel, J. and Gneiting, T.},
  title   = {Decompositions of the Mean Continuous Ranked Probability Score},
  journal = {Electron. J. Stat.},
  year    = {2024},
  volume  = {18},
  pages   = {4992--5044}
}

@book{kakwani1980,
  author       = {Kakwani, Nanak Chand},
  title        = {Income Inequality and Poverty: Methods of Estimation and Policy Applications},
  year         = {1980},
  publisher    = {World Bank},
  address = {New York}
}

@article{gastwirth1972,
  author  = {Gastwirth, Joseph L.},
  title   = {The Estimation of the Lorenz Curve and Gini Index},
  journal = {Rev. Econ. Stat.},
  year    = {1972},
  volume  = {54},
  pages   = {306--316}
}

@article{thompson2023,
  author = {Thompson, W. H. and Skau, S.},
  title = {On the scope of scientific hypotheses},
  journal = {R. Soc. Open Sci.},
  year = {2023},
  volume = {10},
  pages = {230607}
}

@article{matheson1976,
  author  = {Matheson, James E. and Winkler, Robert L.},
  title   = {Scoring Rules for Continuous Probability Distributions},
  journal = {Management Sci.},
  year    = {1976},
  volume  = {22},
  pages   = {1087--1096}
}

@article{bickel2008,
  author  = {Bickel, J. Eric and Kim, Seong Dae},
  title   = {Verification of The Weather Channel Probability of Precipitation Forecasts},
  journal = {Mon. Wea. Rev.},
  year    = {2008},
  volume  = {136},
  pages   = {4867--4881}
}

@article{buizza2008,
  author  = {Buizza, Roberto},
  title   = {The Value of Probabilistic Prediction},
  journal = {Atmos. Sci. Lett.},
  year    = {2008},
  volume  = {9},
  pages   = {36--42}
}

@article{zhang2014,
  author  = {Zhang, Yao and Wang, Jianxue and Wang, Xifan},
  title   = {Review on Probabilistic Forecasting of Wind Power Generation},
  journal = {Renew. Sustain. Energy Rev.},
  year    = {2014},
  volume  = {32},
  pages   = {255--270}
}

@article{timmermann2000,
  author  = {Timmermann, Allan},
  title   = {Density Forecasting in Economics and Finance},
  journal = {J. Forecast.},
  year    = {2000},
  volume  = {19},
  pages   = {231--234}
}

@article{collins2007,
  author  = {Collins, Mat},
  title   = {Ensembles and Probabilities: A New Era in the Prediction of Climate Change},
  journal = {Philos. Trans. R. Soc. A},
  year    = {2007},
  volume  = {365},
  pages   = {1957--1970}
}

@book{yitzhaki2013,
  author    = {Yitzhaki, Shlomo and Schechtman, Edna},
  title     = {The Gini Methodology: A Primer on a Statistical Methodology},
  year      = {2013},
  publisher = {Springer},
  address   = {New York}
}

@article{giorgi2017,
  author  = {Giorgi, Giovanni Maria and Gigliarano, Chiara},
  title   = {The Gini Concentration Index: A Review of the Inference Literature},
  journal = {J. Econ. Surv.},
  year    = {2017},
  volume  = {31},
  pages   = {1130--1148}
}

@article{csiszar1975,
  author  = {Csisz{\'a}r, Imre},
  title   = {I-Divergence Geometry of Probability Distributions and Minimization Problems},
  journal = {Ann. Probab.},
  year    = {1975},
  volume  = {3},
  pages   = {146--158}
}

@book{marshall1979,
  author    = {Marshall, Albert W. and Olkin, Ingram and Arnold, Barry C.},
  title     = {Inequalities: Theory of Majorization and Its Applications},
  year      = {1979},
  publisher = {Academic Press},
  address   = {New York}
}

@book{cover1991,
  author    = {Cover, Thomas M. and Thomas, Joy A.},
  title     = {Elements of Information Theory},
  year      = {1991},
  publisher = {Wiley},
  address   = {New York}
}

@article{gastwirth2017,
  author  = {Gastwirth, Joseph L.},
  title   = {Is the Gini Index of Inequality Overly Sensitive to Changes in the Middle of the Income Distribution?},
  journal = {Stat. Public Policy},
  year    = {2017},
  volume  = {4},
  pages   = {1--11}
}

@article{weisheimer2005,
  author  = {Weisheimer, Antje and Smith, Leonard A. and Judd, Kevin},
  title   = {A New View of Seasonal Forecast Skill: Bounding Boxes from the DEMETER Ensemble Forecasts},
  journal = {Tellus A Dyn. Meteorol. Oceanogr.},
  year    = {2005},
  volume  = {57},
  pages   = {265--279}
}

@article{judd2007,
  author  = {Judd, Kevin and Smith, Leonard A. and Weisheimer, Antje},
  title   = {How Good Is an Ensemble at Capturing Truth? Using Bounding Boxes for Forecast Evaluation},
  journal = {Q. J. R. Meteorol. Soc.},
  year    = {2007},
  volume  = {133},
  pages   = {1309--1325},
}

@article{stephenson2000,
  author  = {Stephenson, David B. and Doblas-Reyes, Francisco J.},
  title   = {Statistical Methods for Interpreting Monte Carlo Ensemble Forecasts},
  journal = {Tellus A Dyn. Meteorol. Oceanogr.},
  year    = {2000},
  volume  = {52},
  pages   = {300--322}
}

@article{gneiting2008,
  author  = {Gneiting, Tilmann and Stanberry, Larissa I. and Grimit, Eric P. and Held, Leonhard and Johnson, Nicholas A.},
  title   = {Assessing Probabilistic Forecasts of Multivariate Quantities, with an Application to Ensemble Predictions of Surface Winds},
  journal = {TEST},
  year    = {2008},
  volume  = {17},
  pages   = {211--235}
}

@article{gneiting2008b,
  author  = {Gneiting, Tilmann and Stanberry, Larissa I. and Grimit, Eric P. and Held, Leonhard and Johnson, Nicholas A.},
  title   = {Rejoinder on: Assessing probabilistic forecasts of multivariate quantities, with an application to ensemble predictions of surface winds},
  journal = {TEST},
  year    = {2008},
  volume  = {17},
  pages   = {256--264}
}

@article{jolliffe2008,
  author  = {Jolliffe, Ian T.},
  title   = {Comments on: Assessing probabilistic forecasts of multivariate quantities, with an application to ensemble predictions of surface winds},
  journal = {TEST},
  year    = {2008},
  volume  = {17},
  pages   = {249--250}
}

@article{winkler2008,
  author  = {Winkler, Robert L. and Jose, Victor R. R.},
  title   = {Comments on: Assessing probabilistic forecasts of multivariate quantities, with an application to ensemble predictions of surface winds},
  journal = {TEST},
  year    = {2008},
  volume  = {17},
  pages   = {251--255}
}

@article{allen2024,
  author  = {Allen, Sam and Ziegel, Johanna and Ginsbourger, David},
  title   = {Assessing the calibration of multivariate probabilistic forecasts},
  journal = {Q. J. R. Meteorol. Soc.},
  year    = {2024},
  volume  = {150},
  pages   = {1315--1335}
}

@article{gneiting2023,
  author  = {Gneiting, Tilmann and Resin, Johannes},
  title   = {Regression diagnostics meets forecast evaluation: Conditional calibration, reliability diagrams, and coefficient of determination},
  journal = {Electron. J. Stat.},
  year    = {2023},
  volume  = {17},
  pages   = {3226--3286}
}

@article{koshevoy1997,
  author  = {Koshevoy, Gleb A. and Mosler, Karl},
  title   = {Multivariate Gini Indices},
  journal = {J. Multivariate Anal.},
  volume  = {60},
  pages   = {252--276},
  year    = {1997}
}

@article{rasp2024,
  author  = {Rasp, Stephan and Hoyer, Stephan and Merose, Alexander and Langmore, Ian and Battaglia, Peter and Russell, Tyler and Sanchez-Gonzalez, Alvaro and Yang, Vivian and Carver, Rob and Agrawal, Shreya and Chantry, Matthew and Ben Bouallegue, Zied and Dueben, Peter and Bromberg, Carla and Sisk, Jared and Barrington, Luke and Bell, Aaron and Sha, Fei},
  title   = {WeatherBench 2: A Benchmark for the Next Generation of Data-Driven Global Weather Models},
  journal = {J. Adv. Model. Earth Syst.},
  year    = {2024},
  volume  = {16},
  pages   = {e2023MS004019}
}

@article{tran2020,
  author  = {Tran, Kevin and Neiswanger, Willie and Yoon, Junwoong and Zhang, Qingyang and Xing, Eric and Ulissi, Zachary W.},
  title   = {Methods for comparing uncertainty quantifications for material property predictions},
  journal = {Machine Learning: Science and Technology},
  year    = {2020},
  volume  = {1},
  pages   = {025006},
  doi     = {10.1088/2632-2153/ab7e1a}
}

@book{luenberger2008,
  author    = {Luenberger, David G. and Ye, Yinyu},
  title     = {Linear and Nonlinear Programming},
  edition   = {4},
  year      = {2008},
  publisher = {Springer},
  address   = {New York}
}

@inproceedings{jordan2026sharpness,
  author    = {Jordan, Alexander I.},
  title     = {On Sharpness Diagrams},
  booktitle = {Workshop on Towards Trustworthy Predictions: Theory and Applications of Calibration for Modern AI at AISTATS},
  year      = {2026},
  address   = {Tangier}
}

\newpage
\bigskip
\begin{center}
{\LARGE\bfseries Supplementary Material}
\end{center}

\setcounter{page}{1}
\setcounter{equation}{0}
\setcounter{section}{0}
\setcounter{figure}{0}
\setcounter{table}{0}
\renewcommand {\thepage} {S\arabic{page}}
\renewcommand {\theequation} {S\arabic{equation}}
\renewcommand {\thesection} {S\arabic{section}}
\renewcommand{\thefigure}{S\arabic{figure}}
\renewcommand{\thetable}{S\arabic{table}}

\makeatletter
\renewcommand{\theHsection}{supp.\arabic{section}}
\renewcommand{\theHsubsection}{supp.\arabic{section}.\arabic{subsection}}
\renewcommand{\theHsubsubsection}{supp.\arabic{section}.\arabic{subsection}.\arabic{subsubsection}}
\renewcommand{\theHequation}{supp.\arabic{equation}}
\renewcommand{\theHfigure}{supp.\arabic{figure}}
\renewcommand{\theHtable}{supp.\arabic{table}}
\makeatother

\bigskip

Section \ref{sec::proofs} provides proofs of mathematical properties.

Section \ref{sec::transformations} derives formulas for sharpness score transformations across discrete and continuous domains.

Section \ref{sec::formulas} derives formulas for analysis in the rearranged space, including local contributions to sharpness score and recovery of key points.

Section \ref{sec::levelsets} investigates relationships among sharpness, entropy, and variance.

\spacingset{1.8} 

\bigskip
\bigskip
\bigskip
\bigskip

\section{Proofs}\label{sec::proofs}
In this section, we prove formal properties of the continuous sharpness measure $S(f)$.

\subsection{Proof of Proposition~2.2}

Let $\Omega\subset\mathbb{R}^d$ be bounded and measurable with $0<|\Omega|<\infty$, and let $f:\Omega\to\mathbb{R}_{\ge 0}$ satisfy
\[
\int_\Omega f(y)\,dy=1.
\]
Let $f^\uparrow:[0,|\Omega|]\to\mathbb{R}_{\ge 0}$ denote the nondecreasing rearrangement of $f$, and recall that
\[
S(f)=\frac{2}{|\Omega|}\int_0^{|\Omega|} t\,f^\uparrow(t)\,dt-1.
\]

Then:

\begin{itemize}
\item[(A1)] $S(f)$ satisfies
\[
0\le S(f)<1.
\]
Moreover, $S(f)=0$ if and only if $f=1/|\Omega|$ a. e. on $\Omega$, and $S(f)\to 1$ along sequences of densities converging weakly to a Dirac point mass.

\item[(A2)] $S(f)$ is invariant under measure-preserving rearrangements of the domain.

\item[(A3)] $S(f)$ is Lipschitz continuous with respect to the $L^1$-distance of rearrangements:
\[
|S(f_1)-S(f_2)|\le 2\|f_1^\uparrow-f_2^\uparrow\|_{L^1(0,|\Omega|)}.
\]

\item[(A4)] If $f_1,f_2$ satisfy
\[
\int_{0}^{t} f_1^\uparrow(s)\,ds \ge \int_{0}^{t} f_2^\uparrow(s)\,ds
\qquad\text{for all } t\in[0,|\Omega|],
\]
then $S(f_1)\le S(f_2)$, with strict inequality if the above holds strictly for
some $t\in(0,|\Omega|)$.
\end{itemize}

\begin{proof}
We verify A1--A4 using the equivalent representations of $S(f)$ established in the main text.

\medskip
\noindent\textit{(A1) Normalization.}
We begin with the simplified formula,
\[
S(f)=\frac{2}{|\Omega|}\int_0^{|\Omega|} t\,f^\uparrow(t)\,dt-1.
\]
Given that $f^\uparrow$ is a probability density on $[0,|\Omega|]$, the quantity
\[
\int_0^{|\Omega|} t\,f^\uparrow(t)\,dt
\]
represents the expectation of $t$ under $f^\uparrow$. Because $0\le t\le |\Omega|$ on this interval,
\[
\int_0^{|\Omega|} t\,f^\uparrow(t)\,dt \le |\Omega|\int_0^{|\Omega|} f^\uparrow(t)\,dt = |\Omega|.
\]
Moreover, since $t<|\Omega|$ for almost every $t\in[0,|\Omega|]$, equality cannot hold for an absolutely continuous density. Hence
\[
\int_0^{|\Omega|} t\,f^\uparrow(t)\,dt<|\Omega|,
\qquad\text{so}\qquad
S(f)<1.
\]

To obtain the lower bound, note that both $t\mapsto t$ and $f^\uparrow$ are nondecreasing on $[0,|\Omega|]$. Hence, by Chebyshev's integral inequality,
\[
\frac{1}{|\Omega|}\int_0^{|\Omega|} t\,f^\uparrow(t)\,dt
\ge
\left(\frac{1}{|\Omega|}\int_0^{|\Omega|} t\,dt\right)
\left(\frac{1}{|\Omega|}\int_0^{|\Omega|} f^\uparrow(t)\,dt\right)
=
\frac{|\Omega|}{2}\cdot \frac{1}{|\Omega|}.
\]
Therefore
\[
\int_0^{|\Omega|} t\,f^\uparrow(t)\,dt \ge \frac{|\Omega|}{2},
\]
and thus
\[
S(f)\ge \frac{2}{|\Omega|}\cdot \frac{|\Omega|}{2}-1=0.
\]
Equality holds only if $f^\uparrow$ is constant almost everywhere on $[0,|\Omega|]$, and since
\[
\int_0^{|\Omega|} f^\uparrow(t)\,dt=1,
\]
this constant must be $1/|\Omega|$. By equimeasurability of $f$ and $f^\uparrow$, this is equivalent to
\[
f(y)=\frac{1}{|\Omega|}
\qquad\text{for almost every }y\in\Omega.
\]

Finally, the upper endpoint is attained only in the limit of concentrating sequences. Let $y_0\in\Omega$ be a Lebesgue density point of $\Omega$, and for $r>0$ define
\[
A_r:=\Omega\cap B(y_0,r),
\qquad
f_r(y):=\frac{\mathbf 1_{A_r}(y)}{|A_r|}.
\]
Then each $f_r$ is a probability density on $\Omega$, and $|A_r|\downarrow 0$ as $r\downarrow 0$. Its nondecreasing rearrangement is
\[
f_r^\uparrow(t)=
\begin{cases}
0, & 0\le t<|\Omega|-|A_r|,\\[1mm]
|A_r|^{-1}, & |\Omega|-|A_r|\le t\le |\Omega|.
\end{cases}
\]
Therefore
\[
S(f_r)
=
\frac{2}{|\Omega|}\int_{|\Omega|-|A_r|}^{|\Omega|}\frac{t}{|A_r|}\,dt-1
=
1-\frac{|A_r|}{|\Omega|}
\longrightarrow 1
\qquad\text{as }r\downarrow 0.
\]

Moreover, $f_r(y)\,dy \Rightarrow \delta_{y_0}$ weakly. Indeed, for every bounded continuous $\varphi$,
\[
\left|\int_\Omega \varphi(y)f_r(y)\,dy-\varphi(y_0)\right|
=
\left|\frac{1}{|A_r|}\int_{A_r}\bigl(\varphi(y)-\varphi(y_0)\bigr)\,dy\right|
\le
\sup_{y\in A_r}|\varphi(y)-\varphi(y_0)|
\to 0
\]
by continuity of $\varphi$ at $y_0$. Thus $1$ is the supremum of $S(f)$ on absolutely continuous densities, and it is approached along sequences of densities converging weakly to a Dirac point mass. Accordingly, we extend $S$ to Dirac point masses by continuous extension, defining
\[
S(\delta_{y_0}) := \lim_{r\downarrow 0} S(f_r) = 1,
\]
where $f_r(y)\,dy \Rightarrow \delta_{y_0}$ as above.

\medskip
\noindent\textit{(A2) Symmetry.}
Since $S(f)$ depends on $f$ only through its nondecreasing rearrangement $f^\uparrow$, and $f^\uparrow$ is determined entirely by the distribution of values of $f$, $S(f)$ is invariant under all measure-preserving rearrangements of the domain.

\medskip
\noindent\textit{(A3) Continuity.}
Using the simplified expression for $S(f)$ and noting that $t/|\Omega|\le 1$ for $t\in[0,|\Omega|]$,
\[
\begin{aligned}
|S(f_1)-S(f_2)|
&=\frac{2}{|\Omega|}\left|\int_0^{|\Omega|} t\,(f_1^\uparrow(t)-f_2^\uparrow(t))\,dt\right| \\
&\le \frac{2}{|\Omega|}\int_0^{|\Omega|} t\,|f_1^\uparrow(t)-f_2^\uparrow(t)|\,dt \\
&\le 2\|f_1^\uparrow-f_2^\uparrow\|_{L^1(0,|\Omega|)}.
\end{aligned}
\]
Hence $S$ is Lipschitz continuous with constant $2$ with respect to the $L^1$-distance
of rearrangements.

\medskip
\noindent\textit{(A4) Monotonicity.}
Write
\[
S(f)=\frac{2}{|\Omega|}\,\mathbb E[T]-1,
\]
where \(T\) has density \(f^\uparrow\) on \([0,|\Omega|]\). Let \(T_i\) have density \(f_i^\uparrow\) and cdf
\[
F_i(t)=\int_0^t f_i^\uparrow(s)\,ds.
\]
If
\[
F_1(t)\ge F_2(t)\qquad \text{for all } t\in[0,|\Omega|],
\]
then \(T_2\) first-order stochastically dominates \(T_1\), so
\[
\mathbb E[T_2]\ge \mathbb E[T_1],
\]
with strict inequality if the dominance is strict. Given that
\[
S(f_i)=\frac{2}{|\Omega|}\mathbb E[T_i]-1,
\]
it follows that
\[
S(f_1)\le S(f_2),
\]
with strict inequality if the dominance is strict.
\end{proof}

\subsection{Proof of Proposition~2.3}

We prove that $S(f)$ satisfies the exclusion principle: the exclusion of a region of the outcome space contributes its fractional share to the sharpness score.

\begin{proof}
Let $a := |A|$. Since
\[
f(y)=
\begin{cases}
\frac{1}{a}, & y \in A,\\
0, & y \in \Omega \setminus A,
\end{cases}
\]

\noindent the non-decreasing rearrangement $f^\uparrow(t)$ over $[0, |\Omega|]$ is given by
\[
f^\uparrow(t) = 
\begin{cases}
0, & 0 \leq t < |\Omega| - a, \\
\frac{1}{a}, & |\Omega| - a \leq t \leq |\Omega|,
\end{cases}
\]

\noindent We now evaluate the sharpness measure:
\[
S(f) = \frac{1}{|\Omega|} \left( \int_0^{|\Omega|} m(t) \, dt - \int_0^{|\Omega|} f^\uparrow(t) l(t) \, dt \right),
\]

\noindent where $m(t) = \int_t^{|\Omega|} f^\uparrow(s) \, ds$ and $l(t) = |\Omega| - t$.

\vspace{0.5em}
\noindent First, we evaluate $\int_0^{|\Omega|} m(t) \, dt$.

\noindent For $0 \leq t < |\Omega| - a$, we have
\[
m(t) = \int_t^{|\Omega| - a} 0 \, ds + \int_{|\Omega| - a}^{|\Omega|} \frac{1}{a} \, ds = 1.
\]
For $|\Omega| - a \leq t \leq |\Omega|$, we have
\[
m(t) = \int_t^{|\Omega|} \frac{1}{a} \, ds = \frac{|\Omega| - t}{a}.
\]
Thus,
\[
\int_0^{|\Omega|} m(t) \, dt = \int_0^{|\Omega| - a} 1 \, dt + \int_{|\Omega| - a}^{|\Omega|} \frac{|\Omega| - t}{a} \, dt = |\Omega| - a + \frac{1}{a} \cdot \frac{a^2}{2} = |\Omega| - \frac{a}{2}.
\]

\vspace{0.5em}
\noindent We next evaluate $\int_0^{|\Omega|} f^\uparrow(t) l(t) \, dt$. Since $f^\uparrow(t) = 0$ on $[0, |\Omega| - a)$, we have
\vspace{0.5em}
\[
\int_0^{|\Omega|} f^\uparrow(t) l(t) \, dt = \int_{|\Omega| - a}^{|\Omega|} \frac{1}{a} (|\Omega| - t) \, dt = \frac{1}{a} \cdot \frac{a^2}{2} = \frac{a}{2}.
\]
\vspace{0.5em}
\noindent Combining:
\[
S(f) = \frac{1}{|\Omega|} \left( |\Omega| - \frac{a}{2} - \frac{a}{2} \right) = \frac{1}{|\Omega|} (|\Omega| - a) = 1 - \frac{a}{|\Omega|}.
\]
\end{proof}

\section{Domain Transformations}\label{sec::transformations}

\subsection{Discrete Case}

Let $P=(p_1,\dots,p_m)$ be a probability distribution on a domain of size $m$, and let $S_m(P)$ denote its sharpness score. In the forward direction, we seek the sharpness score obtained when $P$ is embedded in a larger domain of size $n>m$ by assigning probability zero to the remaining $n-m$ outcomes. In the inverse direction, we seek to recover the sharpness score $S_m(P)$ on the smaller domain of size $m<n$ by removing $n-m$ outcomes of probability zero. Define the embedding $\tilde P$ on a domain of size $n$ by
\[
\tilde P=(\underbrace{0,\dots,0}_{n-m},p_1,\dots,p_m),
\]
and let $S_n(\tilde P)$ denote its sharpness score on the size-$n$ domain.

\begin{proposition}
The sharpness score of $\tilde P$ on the size-$n$ domain is
\[
S_n(\tilde P)=1+\frac{m-1}{n-1}\bigl(S_m(P)-1\bigr).
\]
Equivalently,
\[
S_m(P)=1+\frac{n-1}{m-1}\bigl(S_n(\tilde P)-1\bigr).
\]
\end{proposition}

\begin{proof}
From Eq.~(2.2) of the main text,
\[
S_n(\tilde P)=\frac{1}{n-1}\sum_{j=1}^{n-1}\bigl(m_{(j)}-\tilde p_{(j)}l_{(j)}\bigr).
\]
Each appended zero in $\tilde P$ contributes $1$ to the unnormalized sum, since for such terms $\tilde p_{(j)}=0$ and $m_{(j)}=1$. Hence the $n-m$ added outcomes contribute $n-m$ in total. The remaining $m-1$ terms coincide with those of the original distribution $P$, whose unnormalized contribution is $(m-1)S_m(P)$. Therefore,
\[
S_n(\tilde P)=\frac{(n-m)+(m-1)S_m(P)}{n-1},
\]
and rearranging gives the stated formula. The inverse transformation follows immediately.
\end{proof}

\subsection{Continuous Case}

Let $f$ be a probability density on a measurable domain $A \subset \mathbb{R}^d$ with Lebesgue measure $\ell=|A|$, and let $S_\ell(f)$ denote its sharpness score. In the forward direction, we seek the sharpness score obtained when $f$ is extended to a larger domain $B \supset A$ with $|B|=L>\ell$ by assigning density zero on $B \setminus A$. In the inverse direction, we seek to recover the sharpness score $S_\ell(f)$ on the smaller domain of measure $\ell<L$ by removing a zero-density region of measure $L-\ell$. Define the extension $\tilde f$ on $B$ by
\[
\tilde f(y)=
\begin{cases}
f(y), & y \in A,\\
0, & y \in B \setminus A,
\end{cases}
\]
and let $S_L(\tilde f)$ denote its sharpness score on the domain of measure $L$.

\begin{proposition}
The sharpness score of $\tilde f$ on the domain of measure $L$ is
\[
S_L(\tilde f)=1+\frac{\ell}{L}\bigl(S_\ell(f)-1\bigr).
\]
Equivalently,
\[
S_\ell(f)=1+\frac{L}{\ell}\bigl(S_L(\tilde f)-1\bigr).
\]
\end{proposition}

\begin{proof}
From Eq.~(2.7) of the main text,
\[
S_L(\tilde f)=\frac{2}{L}\int_0^L t\,\tilde f^\uparrow(t)\,dt-1,
\]
where $\tilde f^\uparrow$ denotes the nondecreasing rearrangement of $\tilde f$. Since $\tilde f=0$ on a set of measure $L-\ell$, its rearrangement satisfies
\[
\tilde f^\uparrow(t)=
\begin{cases}
0, & t \in [0,L-\ell],\\
f^\uparrow(t-(L-\ell)), & t \in (L-\ell,L],
\end{cases}
\]
where $f^\uparrow$ is the nondecreasing rearrangement of $f$ on $[0,\ell]$. Therefore,
\[
S_L(\tilde f)=\frac{2}{L}\int_{L-\ell}^L t\,f^\uparrow(t-(L-\ell))\,dt-1.
\]
Changing variables by $u=t-(L-\ell)$ gives
\[
S_L(\tilde f)=\frac{2}{L}\int_0^\ell \bigl(u+(L-\ell)\bigr)f^\uparrow(u)\,du-1.
\]
Using $\int_0^\ell f^\uparrow(u)\,du=1$, this becomes
\[
S_L(\tilde f)=\frac{2}{L}\int_0^\ell u f^\uparrow(u)\,du+\frac{2(L-\ell)}{L}-1.
\]
From Eq.~(2.7) again,
\[
S_\ell(f)=\frac{2}{\ell}\int_0^\ell u f^\uparrow(u)\,du-1,
\]
so
\[
\int_0^\ell u f^\uparrow(u)\,du=\frac{\ell}{2}\bigl(S_\ell(f)+1\bigr).
\]
Substituting yields
\[
S_L(\tilde f)=1+\frac{\ell}{L}\bigl(S_\ell(f)-1\bigr).
\]
The inverse transformation follows immediately.
\end{proof}

\section{Formulas for Analysis in the Rearranged Space}\label{sec::formulas}

The mass-length functional representation of the sharpness measure $S(f)$ provides a structured space in which aspects of the original probability distribution $f$ can be represented alongside the global sharpness measure. This section demonstrates how to recover points or subsets of interest from the original distribution, assess local contributions to sharpness score, and evaluate diagnostic quantities such as relative rank and mass above a density threshold in the rearranged space. In visualization, these quantities can be further mapped onto the mass-length representation or the concentration plots introduced in the main text. A Python implementation for both types of plots is provided in the accompanying material.

\subsection{Mapping from Original Domain to Rearranged Space}

\begin{figure}
\centering{
\includegraphics[width=10cm,height=\textheight]{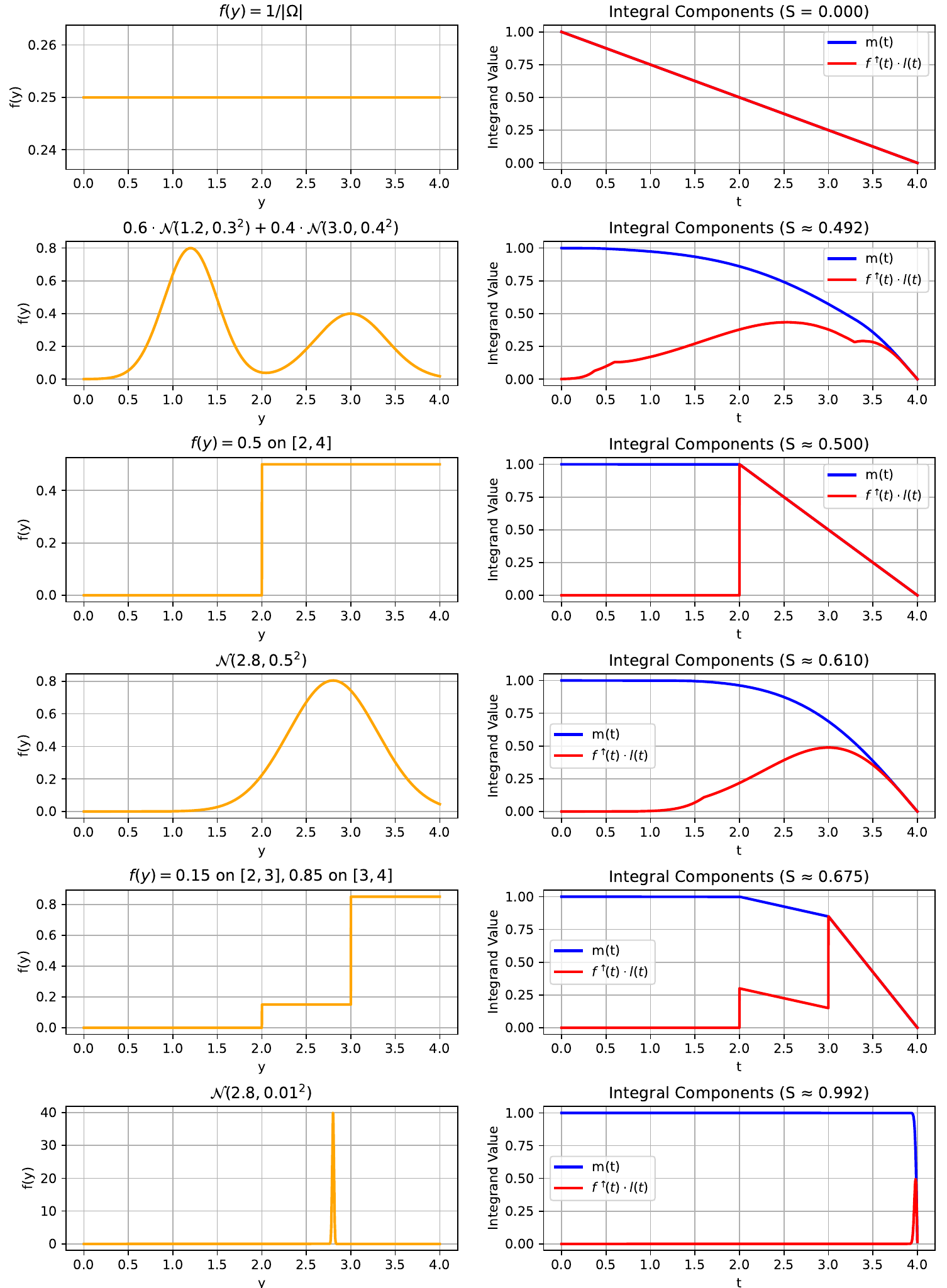}
}

\caption{\label{fig:integrands}Plots for some of the probability distribution functions in Table~2 of the article and the integral components of $S(f)$.}
\end{figure}

Let $y_p \in \Omega$ be a reference location in the original outcome space (in numerical implementations, represented by a small neighborhood around $y_p$), and let $f(y_p)$ denote the density value at $y_p$. To locate an approximate rearranged position $t_p \in [0, |\Omega|]$ of $y_p$, we match $f(y_p)$ to the value of $f^\uparrow(t)$ over a uniform discretization $\{t_i\}_{i=1}^N$:
\[
t_p = t_j \quad \text{where} \quad j = \arg\min_i \left| f^\uparrow(t_i) - f(y_p) \right|.
\]
This mapping identifies a corresponding point in the rearranged domain whose density is closest to that of $y_p$. If multiple points in the original domain share the same density, their mapping to $t$ is non-unique. However, in that case, the rearranged function $f^\uparrow(t)$ is constant over a subinterval $I \subset [0, |\Omega|]$, that is, $f^\uparrow(t) = \tau$ for all $t \in I$, and the mass-length functional representation exhibits linear structure: the remaining mass function $m(t) = \int_t^{|\Omega|} f^\uparrow(s)\, ds$ decreases linearly when $\tau > 0$, and remains constant when $\tau = 0$; likewise, the term $f^\uparrow(t)\,l(t)$ decreases linearly when $\tau > 0$ and remains constant when $\tau = 0$. As a result, the sharpness integrand $m(t) - f^\uparrow(t)\,l(t)$ is constant on $I$. This geometric behavior encodes the rank and measure of the equal-density region in the rearranged space, even though the relative spatial locations of the values in the original outcome space are not preserved. Figure~\ref{fig:integrands} illustrates the behavior of the mass-length functional over several representative probability distributions.

\clearpage

\subsection{Recovering Key Points}

Key reference points of the original probability distribution $f$—such as the mean, median, or an observed value—are located approximately in the rearranged domain by matching their density values to the sorted profile $f^\uparrow(t)$. Let $f^\uparrow(t)$ be evaluated on a uniform grid $\{t_1, \dots, t_N\}$ over $[0, |\Omega|]$, with spacing $\Delta t = |\Omega| / N$.

\paragraph{Mode.} Let $y_{\text{mode}} \in \Omega$ be a mode of the original distribution, such that
\[
f(y_{\text{mode}}) = \max_{y \in \Omega} f(y).
\]
To find its position in rearranged space, identify the index $j$ such that:
\[
t_{\text{mode}} = t_j \quad \text{where} \quad j = \arg\min_i \left| f^\uparrow(t_i) - f(y_{\text{mode}}) \right|.
\]

\paragraph{Median.} In the univariate case, let $y_{\text{med}} \in \Omega$ be defined as:
\[
y_{\text{med}} = \inf \left\{ y \in \Omega \,\middle|\, \int_{a}^{y} f(y')\,dy' \geq \frac{1}{2} \right\},
\]
where $a = \inf \Omega$ is the lower bound of the domain.
Then, identify its location in the rearranged space by:
\[
t_{\text{med}} = t_j \quad \text{where} \quad j = \arg\min_i \left| f^\uparrow(t_i) - f(y_{\text{med}}) \right|.
\]

\paragraph{Mean.} The mean of the distribution is given by
\[
y_{\text{mean}} = \int_\Omega y \, f(y)\, dy.
\]
Let $f(y_{\text{mean}})$ denote its density under $f$. The corresponding position in the rearranged space is identified by:
\[
t_{\text{mean}} = t_j \quad \text{where} \quad j = \arg\min_i \left| f^\uparrow(t_i) - f(y_{\text{mean}}) \right|.
\]

\paragraph{Specific point (e.g., an observed value).} For a specific reference location $y_{\mathrm{p}} \in \Omega$, such as a realized outcome, identify the index $j$ such that:
\[
t_{\mathrm{p}} = t_j \quad \text{where} \quad j = \arg\min_i \left| f^\uparrow(t_i) - f(y_{\mathrm{p}}) \right|.
\]

These mappings allow us to recover salient points from the original distribution $f$. When a reference location $y_p \in \Omega$ lies in a constant-density region---that is, when $f(y_p)=\tau$ and the set of points in $\Omega$ having density $\tau$ has positive measure---the non-decreasing rearrangement $f^\uparrow(t)$ is likewise constant over an interval $I_t=[t_a,t_b]\subset[0,|\Omega|]$ whose length equals the measure of that set. Thus, an alternative approach is not to map $y_p$ to a single location, but instead to identify the constant region in the rearranged space. In practice, this is done by scanning $f^\uparrow(t)$ for all indices $t_i$ such that
\[
|f^\uparrow(t_i) - \tau| \leq \varepsilon,
\]
for a small numerical tolerance $\varepsilon$. The first and last such indices define the endpoints $t_a$ and $t_b$ of the constant plateau region in the rearranged space:
\[
I_t = [t_a, t_b].
\]

\subsection{Local Contribution to Sharpness Score}

To compute the local contribution of a specific region of the rearranged space to the total sharpness score, we use the mass-length decomposition. Let $[t_a, t_b] \subseteq [0, |\Omega|]$ be a subinterval of the rearranged domain. The normalized contribution of this region to the overall sharpness score is given by:
\[
\Delta([t_a, t_b]) = \frac{1}{|\Omega|} \int_{t_a}^{t_b} \left( m(t) - f^\uparrow(t) l(t) \right) dt,
\]

\noindent where $m(t) = \int_t^{|\Omega|} f^\uparrow(s)\, ds$ is the remaining mass, and $l(t) = |\Omega| - t$ is the remaining length. The contribution of a narrow sliver (e.g., one bin in a discretization) is given by:
\[
\Delta(t_i) \approx \frac{1}{|\Omega|} \left( m(t_i) - f^\uparrow(t_i) l(t_i) \right) \cdot \Delta t,
\]

\noindent where $t_i$ is the placement of the sliver and $\Delta t$ is its width.

\begin{figure}
\centering{
\includegraphics[width=15.5cm,height=\textheight]{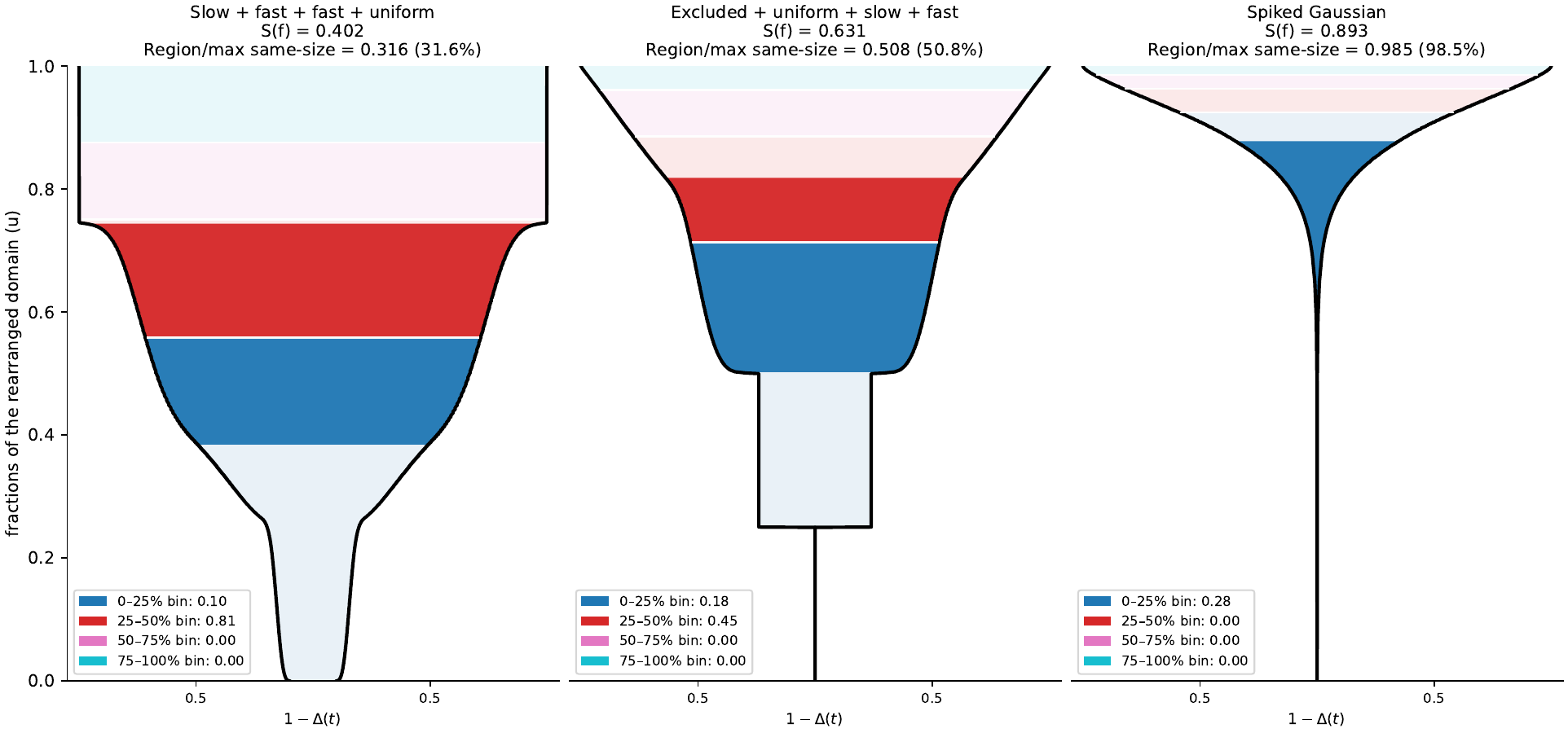}
}

\caption{\label{fig:cplots}Concentration plots for 3D probability distribution functions defined on $\Omega = [0,2]^3$, highlighting the region $[0,1]\times[1,2]\times[0,2]$ from the original space (covering 25 \% of the total volume of the 3D outcome space). The highlighted regions show where the density values from this region are located in the rearranged space. The plots illustrate the fractions of the outcome space containing higher and lower density values than the selected region, and how sharply concentration increases across different fractions of the outcome space. From the left: in the first plot, the highlighted region falls in the middle of the distribution, below a flat piecewise constant region at the top; in the second plot, the highlighted region is again in the middle, while around 25 \% of the domain is fully suppressed, and another 25 \% of the domain falls below the highlighted region but attains positive probability values in a piecewise constant region; in the third plot, most of the domain is almost fully suppressed, but the highlighted region contains some probability mass in the lowest-density part of the distribution.}
\end{figure}

In addition to computing contributions in the rearranged domain, we can also compute the contribution of regions defined in the original outcome space. Let $A \subseteq \Omega$ be a measurable region, for example an interval or a hyper-rectangular region (in multiple dimensions). Under a discretization of $A$, each grid cell in $A$ is mapped to a rearranged index according to its density value, using the approach described above. Denote by $T(A)$ the resulting collection of rearranged indices associated with the cells in $A$ (with repeated indices allowed when multiple cells have the same density). Then the local contribution of $A$ to the sharpness score is computed as
\[
\Delta(A) \approx \frac{1}{|\Omega|} \sum_{t_i \in T(A)} \left( m(t_i) - f^\uparrow(t_i) l(t_i) \right) \Delta t,
\]
where $\Delta t = |\Omega|/N$ is the uniform grid resolution. Figure~\ref{fig:cplots} illustrates this mapping for three 3-dimensional probability distribution functions defined on $\Omega = [0,2]^3$. To further illustrate how different types of concentration patterns affect the shape of the plots, the first two distributions on the left are defined based on different concentration patterns occurring in four equal-sized regions of the domain (full exclusion, slower concentration, faster concentration, and uniformity). The rightmost shape represents a standard Gaussian-type distribution.

\clearpage

\subsection{Mapping Support for a Region in the Rearranged Space}

Consider we are given a measurable subregion $A \subseteq \Omega$ and we wish to examine its support in the rearranged space. To assess whether any part of the region is supported, we evaluate whether $f(y) > 0$ for some $y \in A$. In the rearranged representation, the corresponding positive densities from $A$ can then be located using the approach defined above. The indices $t_i$ associated with the positive densities identify the part of the rearranged domain linked to $A$. Visualizing these positions in the rearranged space allows simultaneously displaying multiple properties of the chosen region, including relative rank and mass allocated to higher density regions.

\subsection{Mass Above a Density Threshold}

Given a point $y_{\mathrm{p}} \in \Omega$, let $t_{\mathrm{p}} \in [0, |\Omega|]$ be its corresponding location in the rearranged space, identified via
\[
t_{\mathrm{p}} = t_j \quad \text{where} \quad j = \arg\min_i \left| f^\uparrow(t_i) - f(y_{\mathrm{p}}) \right|.
\]
Define the density level at this location as $\tau = f^\uparrow(t_{\mathrm{p}})$. Then, the mass allocated to all higher-density outcomes is:
\[
M_*(\tau) = \int_{\{ t \in [0, |\Omega|] : f^\uparrow(t) > \tau \}} f^\uparrow(t)\, dt.
\]

This is implemented numerically by locating the first point in the rearranged space with density strictly greater than $\tau$ and computing the cumulative remaining mass $m(t)$ from that point onward.

\subsection{Relative Likelihood and Rank}

Let $y_{\text{p}} \in \Omega$ be a point in the original space, with density $f(y_{\text{p}})$.  
The relative likelihood of $y_{\text{p}}$ is defined as its density relative to the maximum density of the distribution:
\[
\text{RL}(y_{\text{p}}) = \frac{f(y_{\text{p}})}{\max_{y \in \Omega} f(y)}.
\]
Since each density value corresponds to a location in the rearranged space, the relative likelihood can also be examined in the rearranged space by mapping $f(y_{\text{p}})$ to its corresponding density in $f^\uparrow(t)$ and examining it in relation to a mode of the distribution.

The relative rank of $y_{\text{p}}$ can be represented in the rearranged space by examining the proportion of rearranged domain points with lower density than $f(y_{\text{p}})$. Specifically,
\[
R(y_{\text{p}}) = \frac{1}{N} \cdot \left| \left\{ i \,\middle|\, f^\uparrow(t_i) < f(y_{\text{p}}) \right\} \right|,
\]
where $\{t_i\}_{i=1}^N$ denotes the discretization of the rearranged space.

\subsection{Summary Table}

Table~\ref{tbl-supp1} collects the formulas discussed in this section.

\begin{table}[H]
\centering
\begin{tabular}{ll}
\toprule
\textbf{Quantity} & \textbf{Recovery Formula} \\
\midrule
Mode & $t_{\mathrm{mode}} = t_j,\ j = \arg\min_i |f^\uparrow(t_i) - f(y_{\mathrm{mode}})|$ \\
Median & $t_{\mathrm{med}} = t_j,\ j = \arg\min_i |f^\uparrow(t_i) - f(y_{\mathrm{med}})|$ \\
Mean & $t_{\mathrm{mean}} = t_j,\ j = \arg\min_i |f^\uparrow(t_i) - f(y_{\mathrm{mean}})|$ \\
Specific point & $t_{\mathrm{p}} = t_j,\ j = \arg\min_i |f^\uparrow(t_i) - f(y_{\mathrm{p}})|$ \\
Mass above & $\tau = f^\uparrow(t_{\mathrm{p}}),\quad M_*(\tau) = \int_{\{ t \in [0, |\Omega|] : f^\uparrow(t) > \tau \}} f^\uparrow(t)\, dt$ \\
Local contribution (rearranged) & $\Delta([t_a, t_b]) = \frac{1}{|\Omega|} \int_{t_a}^{t_b} \left( m(t) - f^\uparrow(t) l(t) \right) dt$ \\
Local contribution (original) & $\Delta(A) \approx \frac{1}{|\Omega|} \sum_{t_i \in T(A)} \left( m(t_i) - f^\uparrow(t_i) l(t_i) \right) \Delta t$ \\
Relative likelihood & $\displaystyle RL(y_{\mathrm{p}}) \approx \frac{f^\uparrow(t_{\mathrm{p}})}{f^\uparrow(t_{\mathrm{mode}})}$ \\
Relative rank & $\displaystyle R(y_{\mathrm{p}}) \approx \frac{1}{N} \cdot \left| \left\{ i \,\middle|\, f^\uparrow(t_i) < f(y_{\text{p}}) \right\} \right|$ \\
\bottomrule
\end{tabular}
\caption{Key quantities approximated in the rearranged space $f^\uparrow(t)$, derived from a distribution $f$ on $\Omega$.}\label{tbl-supp1}
\end{table}

\section{Analysis of Sharpness, Entropy, and Variance}\label{sec::levelsets}

This section investigates the relationships among sharpness, entropy, and variance, three complementary measures that characterize the shape and spread of probability distributions. To demonstrate the complementary diagnostic information that these measures provide, we evaluate level sets of these measures: collections of distributions that share approximately equal values of sharpness, entropy, or variance. We begin by exploring the mathematical relationships between entropy and sharpness, in particular, how we can identify the maximum or minimum sharpness or entropy while the other measure is fixed. We then investigate these relationships numerically and through visualizations. In the visualizations, we restrict attention to discrete distributions.

\subsection{Min/Max Entropy and Sharpness}

We characterize the distributions that maximize and minimize entropy within sharpness level sets, and vice versa. Since both entropy and sharpness are invariant under permutations, the optimization problems reduce to the ordered probability profile $P^\uparrow$ (discrete) and the nondecreasing rearrangement $f^\uparrow$ (continuous), respectively.

\begin{proposition}[Maximum entropy at fixed sharpness]
Let $s\in[0,1)$.

\smallskip
\noindent
\emph{Discrete case.}
Let $P$ be a probability distribution on $\{1,\dots,n\}$, and write
\[
P^\uparrow=(p_{(1)},\dots,p_{(n)}), \qquad p_{(1)}\le \cdots \le p_{(n)}.
\]
Then the maximizers of $H(P)$ subject to $S(P)=s$ are exactly those distributions whose ordered profile satisfies
\[
p_{(j)}=\frac{e^{\beta w_j}}{\sum_{k=1}^n e^{\beta w_k}},
\qquad
w_j=\frac{2j-n-1}{n-1},
\]
where $\beta\ge 0$ is chosen so that $S(P)=s$.

\smallskip
\noindent
\emph{Continuous case.}
Let $f$ be a probability density on a bounded measurable set $\Omega\subset\mathbb R^d$, and let $f^\uparrow$ denote its nondecreasing rearrangement on $[0,|\Omega|]$. Then the maximizers of $H(f)$ subject to $S(f)=s$ are exactly those densities whose rearrangement satisfies
\[
f^\uparrow(t)=\frac{e^{\lambda t}}{\int_0^{|\Omega|} e^{\lambda u}\,du},
\qquad t\in[0,|\Omega|],
\]
where $\lambda\ge 0$ is chosen so that $S(f)=s$.
\end{proposition}

\begin{proof}[Proof sketch]
In the discrete case, by Eq.~(2.3),
\[
S(P)=\sum_{j=1}^n w_j p_{(j)},
\qquad
w_1<\cdots<w_n.
\]
Since $H(P)=H(P^\uparrow)$, maximizing entropy at fixed sharpness is equivalent to maximizing Shannon entropy over ordered profiles $P^\uparrow$ subject to normalization and the single linear constraint $\sum_j w_j p_{(j)}=s$. By the maximum-entropy theorem \citep[Theorem~11.1.1]{cover1991}, the unique maximizer under these constraints has the exponential form $p_j\propto e^{\beta w_j}$. Because $w_j$ is strictly increasing and $s\ge 0$, the corresponding parameter satisfies $\beta\ge 0$, so the resulting profile is nondecreasing and thus of the required form. Enforcing the normalization constraint $\sum_{j=1}^n p_{(j)}=1$ gives
\[
p_{(j)}=\frac{e^{\beta w_j}}{\sum_{k=1}^n e^{\beta w_k}}.
\]

In the continuous case, by equimeasurability,
\[
H(f)=H(f^\uparrow)
\quad\text{and}\quad
S(f)=\frac{2}{|\Omega|}\int_0^{|\Omega|} t\,f^\uparrow(t)\,dt-1.
\]
Thus maximizing entropy at fixed sharpness is equivalent to maximizing differential entropy over densities $g$ on $[0,|\Omega|]$ subject to normalization and the linear moment constraint
\[
\int_0^{|\Omega|} t\,g(t)\,dt=\frac{|\Omega|}{2}(1+s).
\]
By \citet[Theorem~11.1.1]{cover1991}, the unique maximizer is $g(t)\propto e^{\lambda t}$. Since $s\in[0,1)$, the corresponding mean lies in $[|\Omega|/2,|\Omega|)$, hence $\lambda\ge 0$, so the maximizer is nondecreasing and so has the required nondecreasing form. Enforcing the normalization constraint $\int_0^{|\Omega|} g(t)\,dt=1$ gives
\[
f^\uparrow(t)=\frac{e^{\lambda t}}{\int_0^{|\Omega|} e^{\lambda u}\,du},
\qquad t\in[0,|\Omega|].
\]
\end{proof}

\begin{proposition}[Maximum sharpness at fixed entropy]
Let $h_0$ lie in the range of the entropy functional.

\smallskip
\noindent
\emph{Discrete case.}
Let $P$ be a probability distribution on $\{1,\dots,n\}$, and write
\[
P^\uparrow=(p_{(1)},\dots,p_{(n)}), \qquad p_{(1)}\le \cdots \le p_{(n)}.
\]
Then the maximizers of $S(P)$ subject to $H(P)=h_0$ are exactly those distributions whose ordered profile satisfies
\[
p_{(j)}=\frac{e^{\beta w_j}}{\sum_{k=1}^n e^{\beta w_k}},
\qquad
w_j=\frac{2j-n-1}{n-1},
\]
where $\beta\ge 0$ is chosen so that $H(P)=h_0$.

\smallskip
\noindent
\emph{Continuous case.}
Let $f$ be a probability density on a bounded measurable set $\Omega\subset\mathbb R^d$, and let $f^\uparrow$ denote its nondecreasing rearrangement on $[0,|\Omega|]$. Then the maximizers of $S(f)$ subject to $H(f)=h_0$ are exactly those densities whose rearrangement satisfies
\[
f^\uparrow(t)=\frac{e^{\lambda t}}{\int_0^{|\Omega|} e^{\lambda u}\,du},
\qquad t\in[0,|\Omega|],
\]
where $\lambda\ge 0$ is chosen so that $H(f)=h_0$.
\end{proposition}

\begin{proof}[Proof sketch]
In both settings, entropy depends only on the ordered profile or rearrangement, while sharpness is a linear functional thereof; see Eq.~(2.3) in the discrete case and Eq.~(2.7) in the continuous case. The characterization of maximizers of a linear functional under a fixed-entropy constraint is therefore the dual problem to maximizing entropy under a fixed linear constraint, and the extremizers are the same exponential family \citep{csiszar1975}. Applying this result yields the forms displayed above.
\end{proof}

Thus, in both the discrete and continuous settings, the same exponential family (i) maximizes entropy at fixed sharpness, and (ii) maximizes sharpness at fixed entropy. As $\beta$ (respectively $\lambda$) increases from $0$ to $+\infty$, these profiles interpolate from the uniform distribution ($s=0$) to the degenerate limit ($s\to 1$).

We next examine the corresponding minimum problems. In contrast to the maximum case, the minima are boundary-driven. For fixed sharpness, minimum entropy is attained on boundary profiles in the discrete case, whereas in the continuous case no minimizer exists. For fixed entropy, in the discrete case the minimizer is obtained by comparing boundary families, and in the continuous case by comparing their piecewise-constant rearranged analogues.

\begin{proposition}[Minimum entropy at fixed sharpness]
Let $s\in[0,1)$.

\smallskip
\noindent
\emph{Discrete case.}
The minimum of $H(P)$ over the level set $\{P:S(P)=s\}$ is attained. Every minimizer has ordered profile
\[
P^\uparrow=
(\underbrace{0,\dots,0}_{n-k},
 \underbrace{a,\dots,a}_{r},
 \underbrace{b,\dots,b}_{k-r}),
\]
for some $k\in\{2,\dots,n\}$, some $r\in\{1,\dots,k-1\}$, and some $0\le a\le b$, with
\[
ra+(k-r)b=1,
\]
and with the parameters chosen so that $S(P)=s$. The minimum is obtained by comparing the entropies over these finitely many two-level candidate families. At the grid values
\[
s=\frac{n-k}{n-1},
\]
the degenerate case $a=b=\frac1k$ yields the uniform distribution on $k$ outcomes.

\smallskip
\noindent
\emph{Continuous case.}
If $s=0$, the unique minimizer is the uniform density on $\Omega$. If $s\in(0,1)$, no minimizer exists and
\[
\inf\{H(f):S(f)=s\}=-\infty.
\]
\end{proposition}

\begin{proof}[Proof sketch]
In the discrete case, the feasible set
\[
\{P^\uparrow:\; S(P)=s\}
\]
is a closed and bounded convex subset of the ordered simplex. Since $H$ is continuous, the minimum is attained.
To characterize the minimizers, reparametrize the ordered profile by its successive increments:
\[
d_1=p_{(1)}, \qquad d_j=p_{(j)}-p_{(j-1)} \ \ (j\ge2).
\]
Then the ordering constraint $p_{(1)}\le \cdots \le p_{(n)}$ is equivalent to $d_j\ge0$ for all $j$, and normalization and the sharpness condition give two linear equalities. Since $H$ is strictly concave, its minimum over this convex set is attained at an extreme point. By the standard characterization of extreme points for such sets, at most $m$ coordinates are nonzero at any extreme point \citep[pp.~40-44]{luenberger2008}. Here $m = 2$, giving at most two nonzero increments $d_j$, which translates back to the ordered profiles with at most two distinct positive probability values. The minimum is therefore obtained by comparing the entropies over the finitely many two-level families parametrized above.

In the continuous case, let $L=|\Omega|$ and fix $s\in(0,1)$. Choose any $q\in(0,s)$ and, for sufficiently small $\varepsilon>0$, define a nondecreasing rearrangement $f_\varepsilon^\uparrow$ on $[0,L]$ by
\[
f_\varepsilon^\uparrow(t)=
\begin{cases}
0, & 0\le t<a_\varepsilon,\\[1mm]
c_\varepsilon, & a_\varepsilon\le t<L-\varepsilon,\\[1mm]
q/\varepsilon, & L-\varepsilon\le t\le L,
\end{cases}
\]
where $a_\varepsilon$ is chosen so that
\[
\int_0^L t\,f_\varepsilon^\uparrow(t)\,dt=\frac{L}{2}(1+s),
\]
and $c_\varepsilon$ is determined by normalization. Then $S(f_\varepsilon)=s$ by Eq.~(2.7), while
\[
H(f_\varepsilon)
=-(1-q)\log c_\varepsilon-q\log(q/\varepsilon)\longrightarrow -\infty
\qquad\text{as }\varepsilon\downarrow 0.
\]
Thus the entropy infimum is $-\infty$ and is not attained. The case $s=0$ is immediate from Proposition~2.2.
\end{proof}

\begin{proposition}[Minimum sharpness at fixed entropy]
Let $h_0$ lie in the range of the entropy functional.

\smallskip
\noindent
\emph{Discrete case.}
The minimum of $S(P)$ over the level set $\{P:H(P)=h_0\}$ is attained. Every minimizer has ordered profile
\[
P^\uparrow=
(\underbrace{0,\dots,0}_{n-k},
 \underbrace{a,\dots,a}_{r},
 \underbrace{b,\dots,b}_{k-r}),
\]
for some $k\in\{2,\dots,n\}$, some $r\in\{1,\dots,k-1\}$, and some $0\le a\le b$, with
\[
ra+(k-r)b=1,
\]
and with the parameters chosen so that
\[
-r\,a\log a-(k-r)\,b\log b = h_0.
\]

\noindent The minimum is obtained by comparing sharpness scores over these finitely many two-level candidate families. At the grid values
\[
h_0=\log k,
\]
the degenerate case $a=b=\frac1k$ yields the uniform distribution on $k$ outcomes.

\smallskip
\noindent
\emph{Continuous case.}
Let $L=|\Omega|$. To minimize $S(f)$ over the level set
\[
\{f:H(f)=h_0\},
\]
candidate minimizers are nondecreasing rearrangements of piecewise-constant two-level form
\[
f^\uparrow(t)=
\begin{cases}
0, & 0\le t<L-\ell,\\[1mm]
a, & L-\ell\le t<L-\eta,\\[1mm]
b, & L-\eta\le t\le L,
\end{cases}
\]
for some $0<\eta\le \ell\le L$ and some $0\le a\le b$, where
\[
(\ell-\eta)a+\eta b=1
\quad\text{and}\quad
-(\ell-\eta)a\log a-\eta b\log b=h_0.
\]
Whether a minimizer exists outside this class remains open.
\end{proposition}

\begin{proof}[Proof sketch]
In the discrete case, the entropy level set
\[
\{P^\uparrow:\; H(P)=h_0\}
\]
is a closed subset of the compact ordered simplex, and $S$ is continuous. Hence the minimum of $S$ on this level set is attained.

To identify minimizers, we first rule out interior points. Consider a region of the ordered simplex where the support size is fixed and the positive coordinates are strictly ordered. If a minimizer lay in the interior of such a region, then applying Lagrange multipliers to minimize $S$ subject to normalization and the entropy constraint would give an exponential profile
\[
p_{(j)} \propto e^{\beta w_j},
\]
which was shown above to be a maximizer of sharpness at fixed entropy. This implies that an interior point cannot be a minimizer. Therefore every minimizer must lie on the boundary, meaning that either a positive coordinate vanishes or two adjacent positive coordinates become equal. Iterating this boundary reduction implies that a minimizer must eventually lie on a lowest-dimensional boundary profile with at most two distinct positive values, as any profile with three or more distinct positive values still lies in the interior of some lower-dimensional face of the simplex. The minimum sharpness at entropy level $h_0$ is thus obtained by comparing sharpness scores over the finitely many two-level families parametrized above. At the grid values $h_0=\log k$, the degenerate case $a=b=1/k$ yields the uniform distribution on $k$ outcomes.

In the continuous case, write $g=f^\uparrow$ on $[0,L]$, where $L=|\Omega|$. The sharpness functional is linear in $g$, while the entropy constraint fixes the value of the concave functional $H(g)$. If a minimizer lay in an interior part of the admissible class, then the usual Lagrange-multiplier argument would again lead to the exponential family
\[
g(t)\propto e^{\lambda t},
\]
that is, the same family that maximizes sharpness at fixed entropy. This suggests that any minimizer should instead lie on the boundary of the admissible class. Motivated by the discrete two-level characterization, we therefore consider boundary profiles of the simplest piecewise-constant form, namely an initial zero segment followed by two positive plateaus:
\[
f^\uparrow(t)=
\begin{cases}
0, & 0\le t<L-\ell,\\[1mm]
a, & L-\ell\le t<L-\eta,\\[1mm]
b, & L-\eta\le t\le L.
\end{cases}
\]
The parameters are fixed by normalization and the entropy constraint. Whether a minimizer exists outside this two-level family remains open.
\end{proof}

\subsection{Visualizing Level Sets on the 2-Simplex and 3-Simplex}

We next examine level sets of sharpness, entropy, and variance by visualizing them on the 2-simplex and 3-simplex. The 2-simplex represents the space of all probability vectors $(p_1, p_2, p_3) \in \mathbb{R}^3$ satisfying $p_i \geq 0$ and $\sum_{i=1}^3 p_i = 1$. Each point on the simplex thus corresponds to a discrete distribution over three outcomes. On this space, the sharpness, entropy, and variance functionals partition the simplex into level sets, subsets of distributions that yield approximately the same value under the respective measure.

\begin{figure}
\centering{
\includegraphics[width=10cm,height=\textheight]{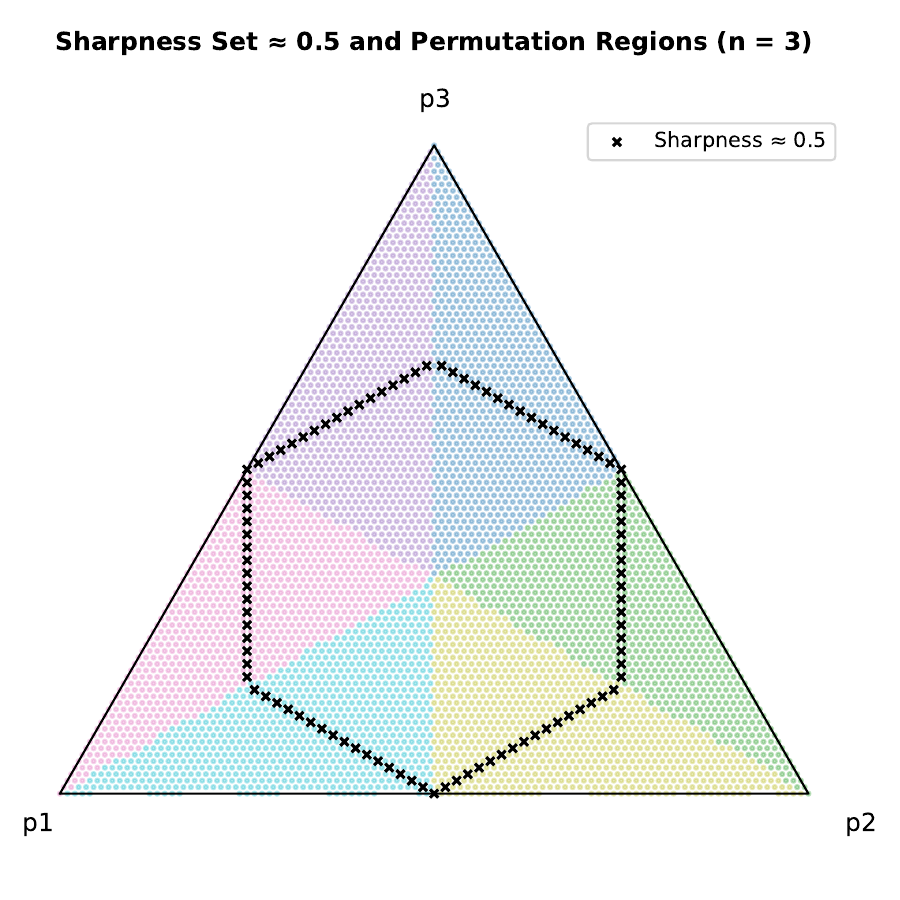}
}

\caption{\label{fig:ss05} The level set of distributions for which $S(P) \approx 0.5$ on the 2-simplex.}
\end{figure}

Figure~\ref{fig:ss05} displays the sharpness level set corresponding to $S(P) \approx 0.5$. To contextualize the visualizations, we partition the 2-simplex into six regions, each corresponding to a distinct permutation of the probabilities $(p_1, p_2, p_3)$. These regions are determined by the relative ordering of the three components and are delineated by boundary lines where two probabilities are equal. The vertices correspond to distributions where the respective outcome is given probability $1.0$. The uniform distribution $(1/3, 1/3, 1/3)$ lies at the center and is invariant under permutation.

The six regions are described as follows, moving clockwise from the top vertex:

\begin{enumerate}
    \item \textbf{Top right (blue):} \( p_1 < p_2 < p_3 \)
    \item \textbf{Right middle (green):} \( p_1 < p_3 < p_2 \)
    \item \textbf{Bottom right (yellow):} \( p_3 < p_1 < p_2 \)
    \item \textbf{Bottom left (light blue):} \( p_3 < p_2 < p_1 \)
    \item \textbf{Left middle (pink):} \( p_2 < p_3 < p_1 \)
    \item \textbf{Top left (violet):} \( p_2 < p_1 < p_3 \)
\end{enumerate}

\begin{figure}[h]
\centering
\begin{subfigure}[b]{0.32\textwidth}
    \includegraphics[width=\textwidth]{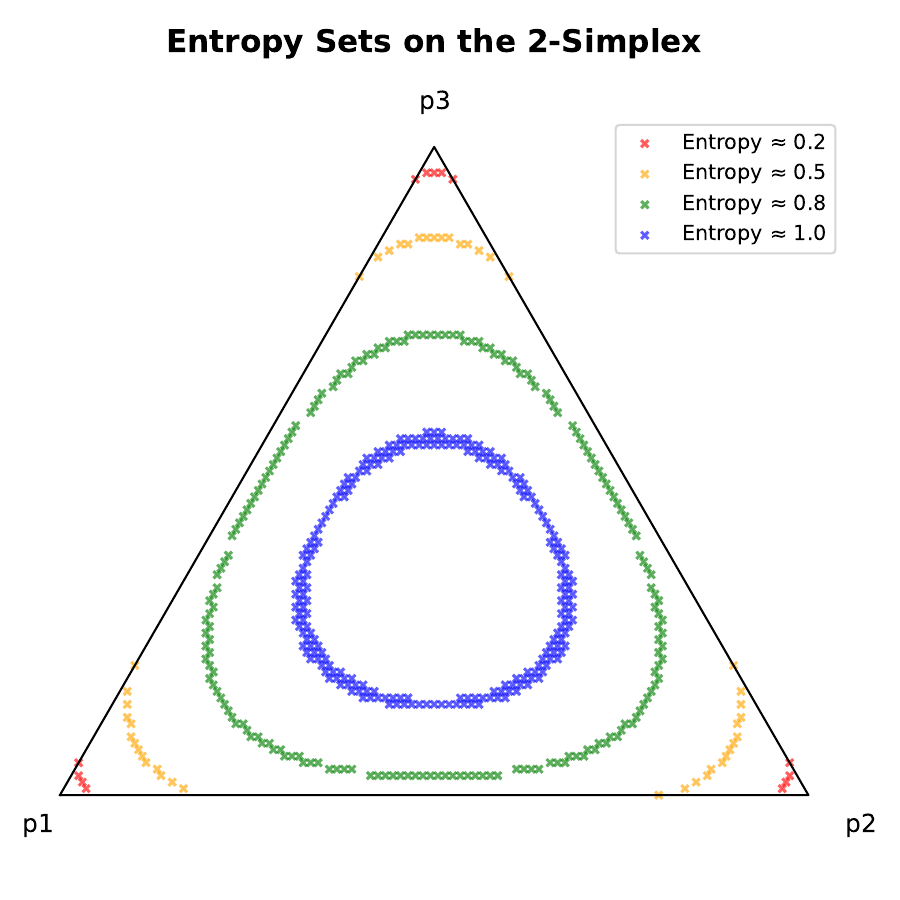}
    \caption{Entropy}
\end{subfigure}
\hfill
\begin{subfigure}[b]{0.32\textwidth}
    \includegraphics[width=\textwidth]{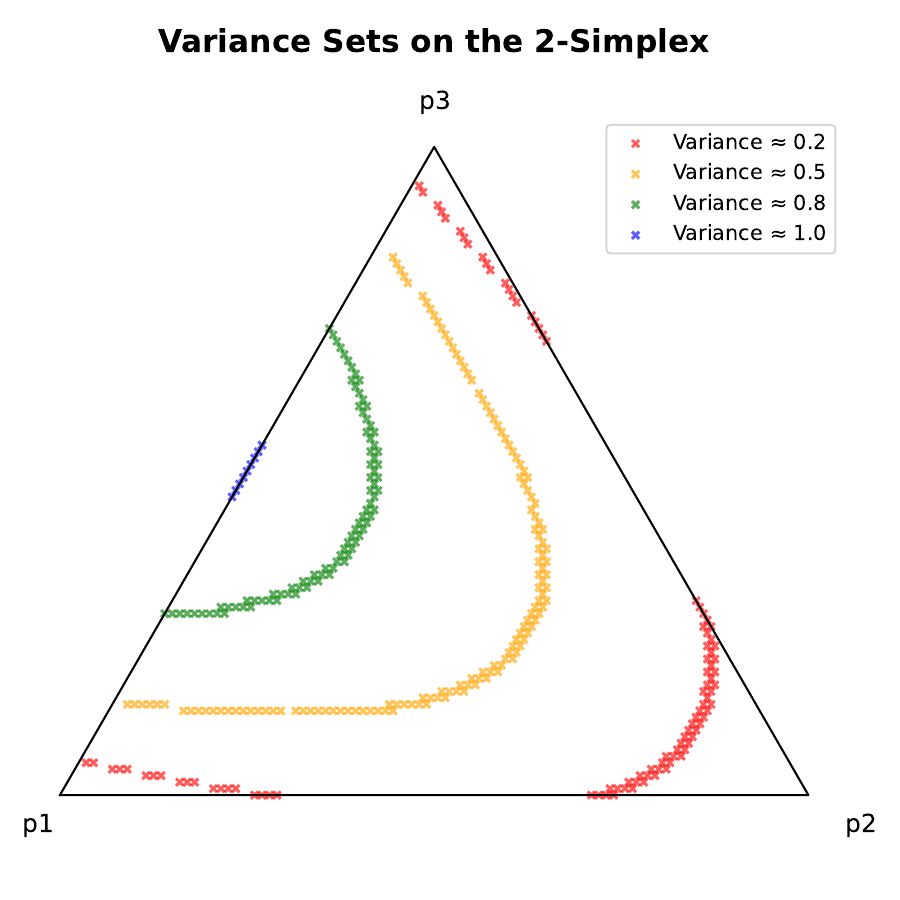}
    \caption{Variance}
\end{subfigure}
\hfill
\begin{subfigure}[b]{0.32\textwidth}
    \includegraphics[width=\textwidth]{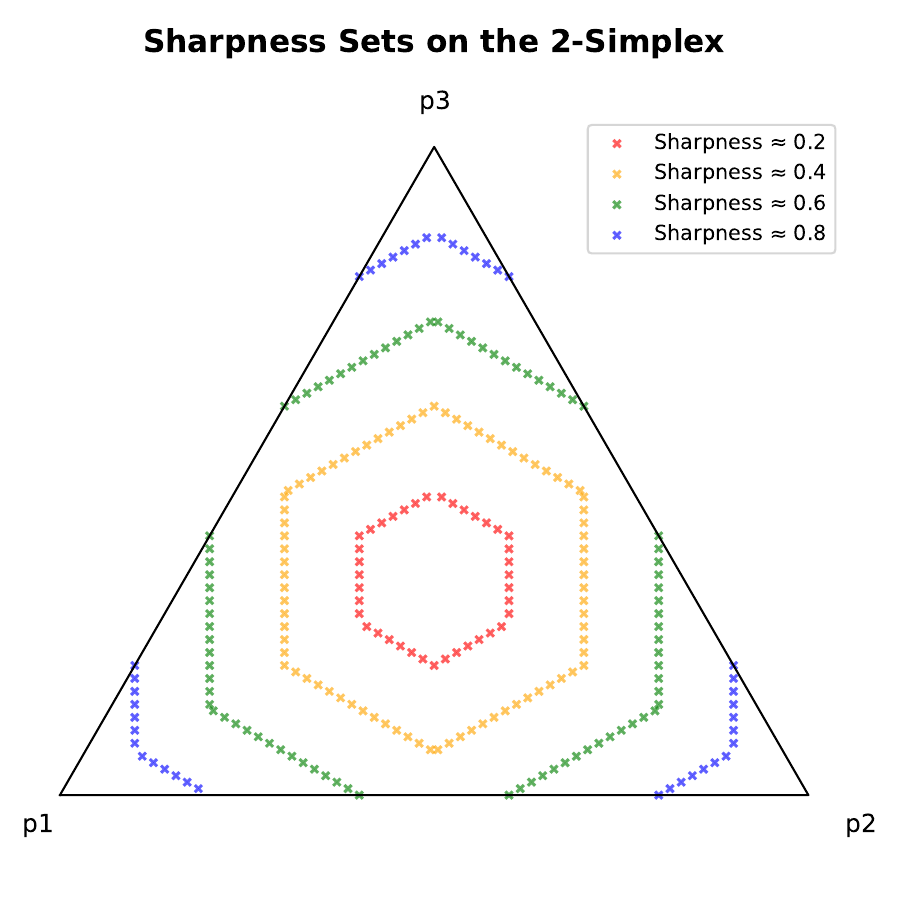}
    \caption{Sharpness}
\end{subfigure}
\caption{\label{fig:measure_sets}Level sets of entropy, variance, and sharpness on the 2-simplex.}
\end{figure}

To illustrate the behavior of the different measures, Figure~\ref{fig:measure_sets} displays level sets of entropy, variance, and sharpness on the 2-simplex. Throughout, for $P=(p_1,\dots,p_n)$ on $\{0,\dots,n-1\}$, we let $Y\sim P$ and write $\operatorname{Var}(Y)$ for its variance. Each measure partitions the space in a distinctive way, reflecting its particular sensitivity to different aspects of distributional structure.

Entropy reflects total uncertainty and it is thus maximized when each outcome is given equal probability and decreases when uncertainty is reduced. Uncertainty can be reduced effectively, for example, either by eliminating outcomes (by setting their probability to zero) or increasing top-1 mass. Sharpness tracks the concentration of mass within the distribution. It reflects the displacement of mass from low to high probability regions. Variance is sensitive to the spatial position of outcomes, so it is naturally minimized when mass is centered and maximized when mass is split between outcomes at opposite ends of the outcome space.

\begin{figure}[h]
\centering
\begin{subfigure}[b]{0.49\textwidth}
    \includegraphics[width=\textwidth]{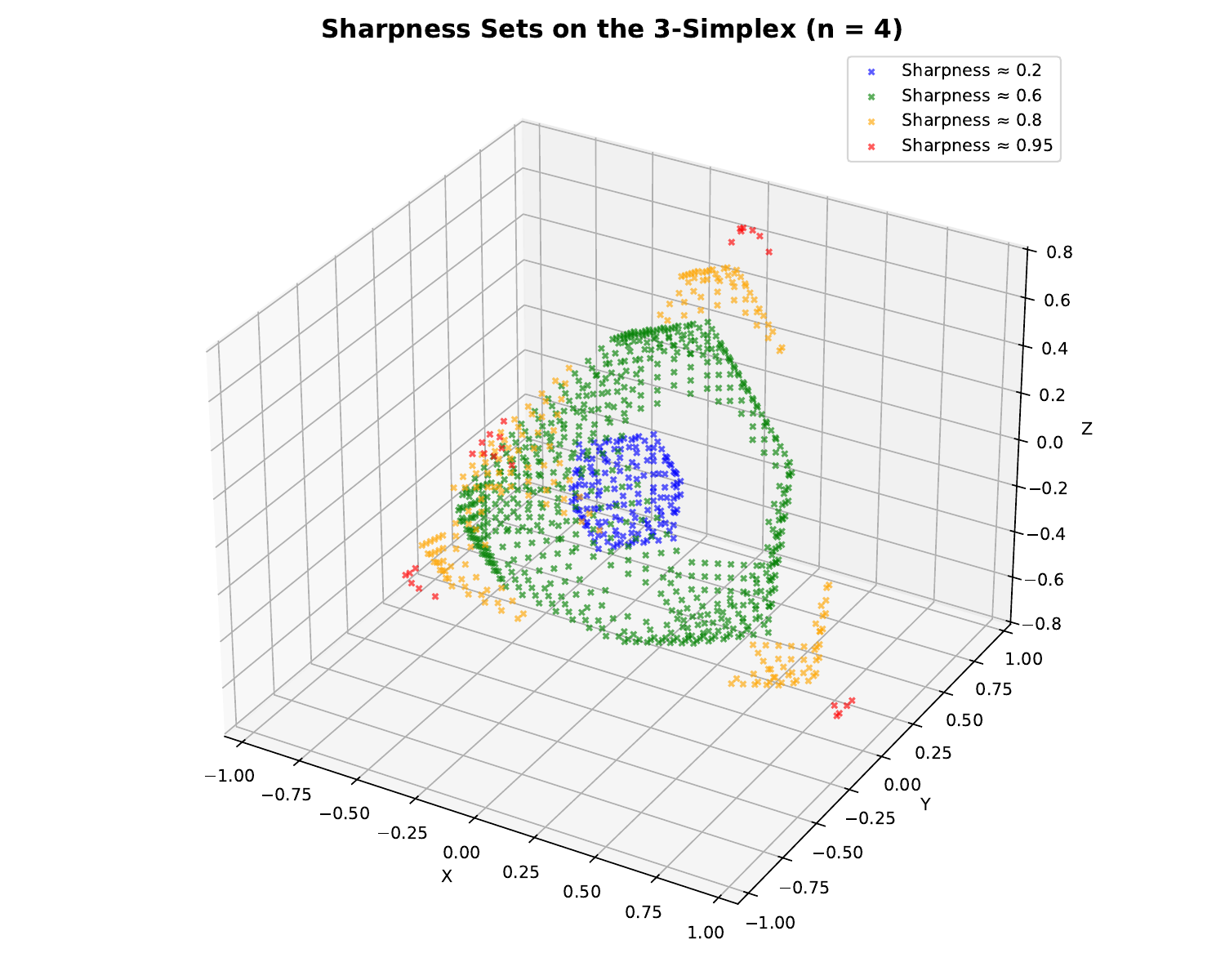}
    \caption{Sharpness}
\end{subfigure}
\hfill
\begin{subfigure}[b]{0.49\textwidth}
    \includegraphics[width=\textwidth]{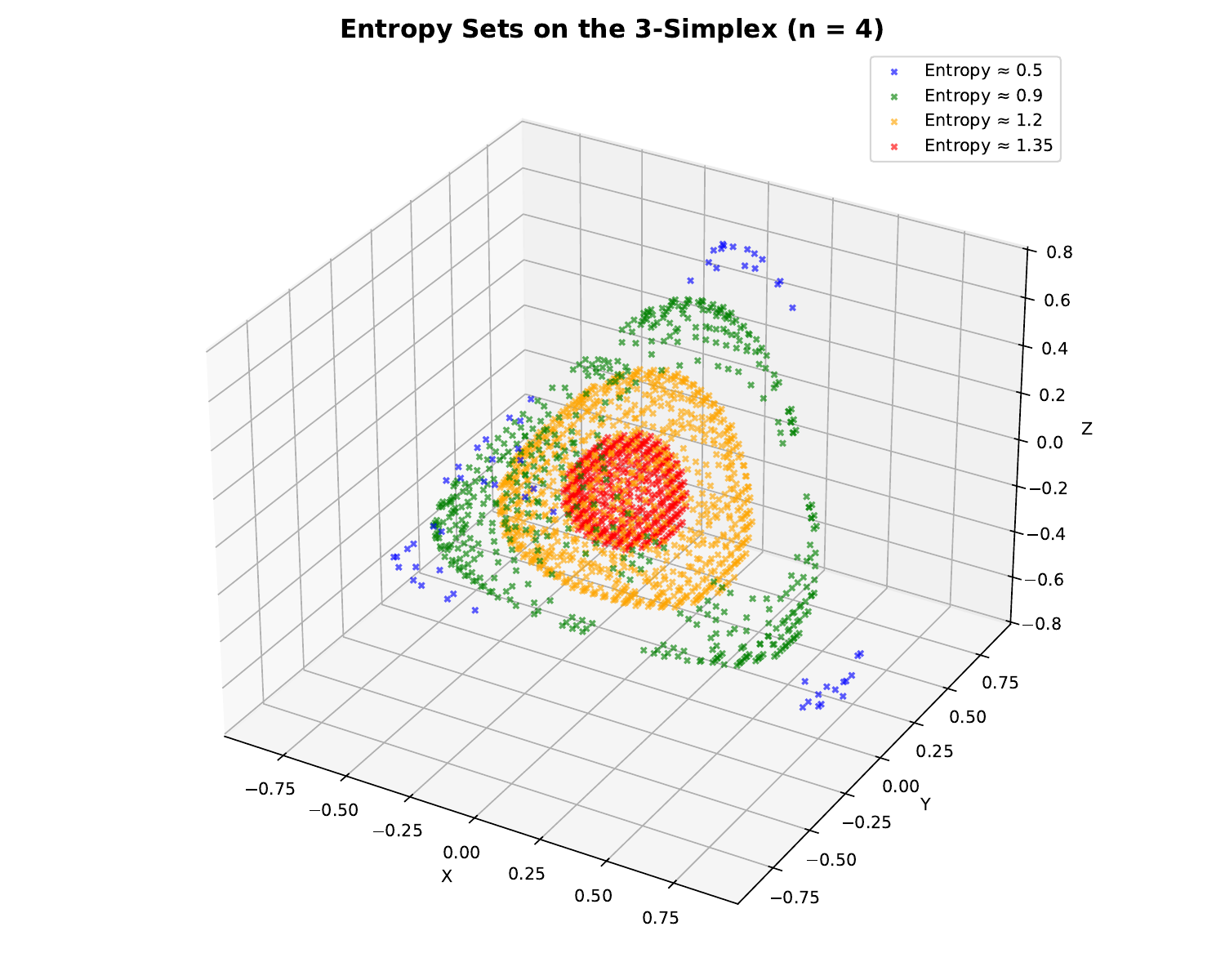}
    \caption{Entropy}
\end{subfigure}
\caption{\label{fig:measure_sets4}Level sets of sharpness and entropy on the 3-simplex.}
\end{figure}

The same behaviors can be observed when examining the space of discrete probability distributions over four outcomes. Distributions over four outcomes can be analyzed on the 3-simplex, which is defined by:
\[
\Delta^3 = \left\{ (p_1, p_2, p_3, p_4) \in \mathbb{R}^4 \;\middle|\; p_i \geq 0, \ \sum_{i=1}^4 p_i = 1 \right\}.
\]
The 3-simplex is visualized as a regular tetrahedron in $\mathbb{R}^3$, where each vertex again corresponds to a degenerate distribution concentrated entirely on one outcome. Here, the level sets of the measures trace out a thin surface embedded within the simplex, which is illustrated in Figure~\ref{fig:measure_sets4} for sharpness and entropy.

\begin{figure}
\centering{
\includegraphics[width=8.5cm,height=\textheight]{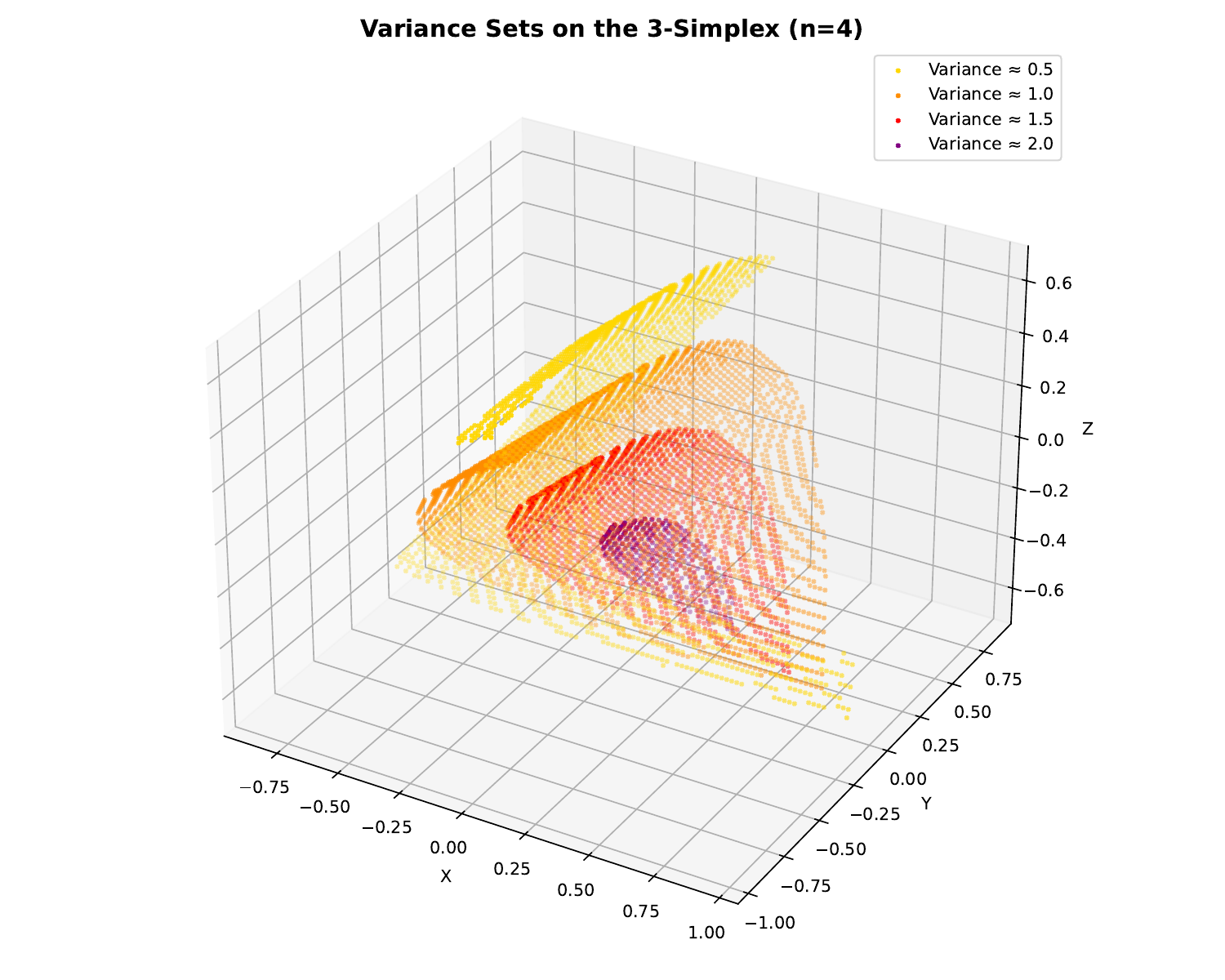}
}

\caption{\label{fig:vsets4}Variance level sets on the 3-simplex.}
\end{figure}

Figure~\ref{fig:vsets4} shows representative level sets of variance on the 3-simplex. Very different distributions can lie on the same level set if they spread mass across outcomes in such a way that the overall difference from the mean is preserved. Thus, the level sets are stretched along directions where mass is distributed more evenly across outcomes, and compressed along directions where the distribution becomes increasingly dominated by an outlier value.

\clearpage

\subsection{Variation Over Level Sets}
Given the distinctive behavior of each measure, they jointly provide a more detailed characterization of distributional shape. Sharpness and variance have a natural reciprocal relationship: variance measures spatial dispersion whereas sharpness measures concentration. Considered together, they distinguish between more or less sharp and more or less spatially spread out distributions within each other's level sets, revealing differences in distributional shape that either measure alone would miss. Sharpness and entropy, in contrast, have a certain dual relationship: when either measure is fixed, the other varies between a top-$n$ concentrated (minimum entropy and sharpness) and an exponential distribution (maximum entropy and sharpness).

\begin{figure}[h]
\centering
\begin{subfigure}[b]{0.49\textwidth}
    \includegraphics[width=\textwidth]{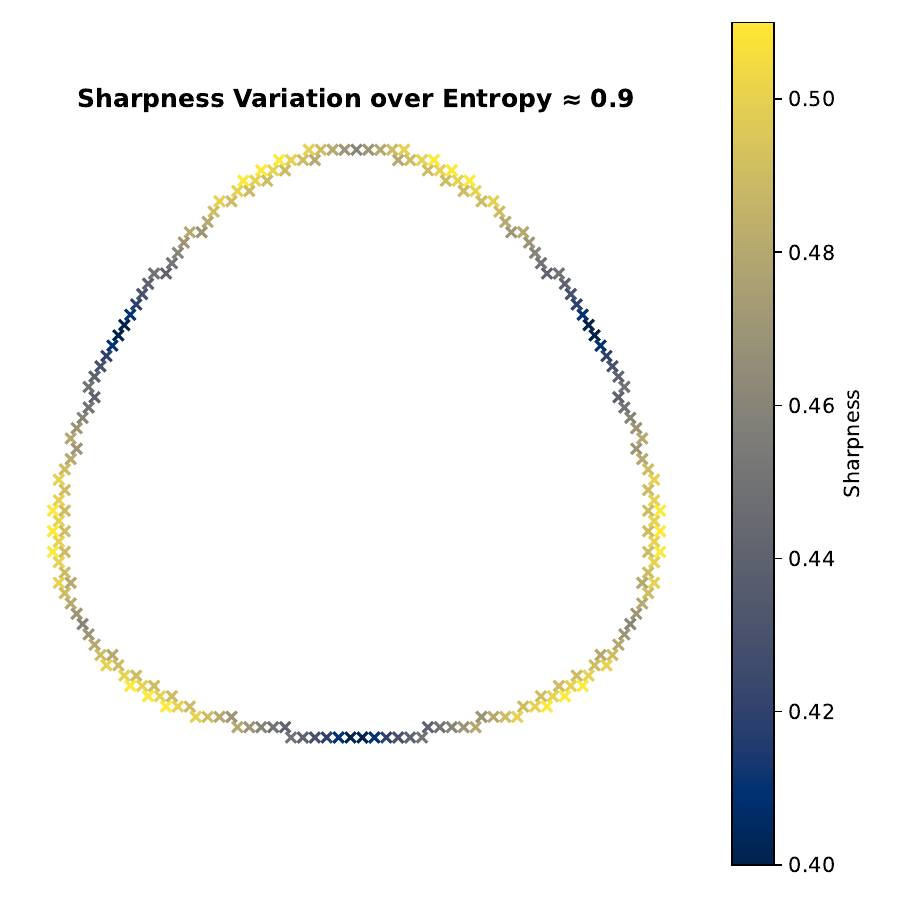}
    \caption{Sharpness Over Entropy}
\end{subfigure}
\hfill
\begin{subfigure}[b]{0.49\textwidth}
    \includegraphics[width=\textwidth]{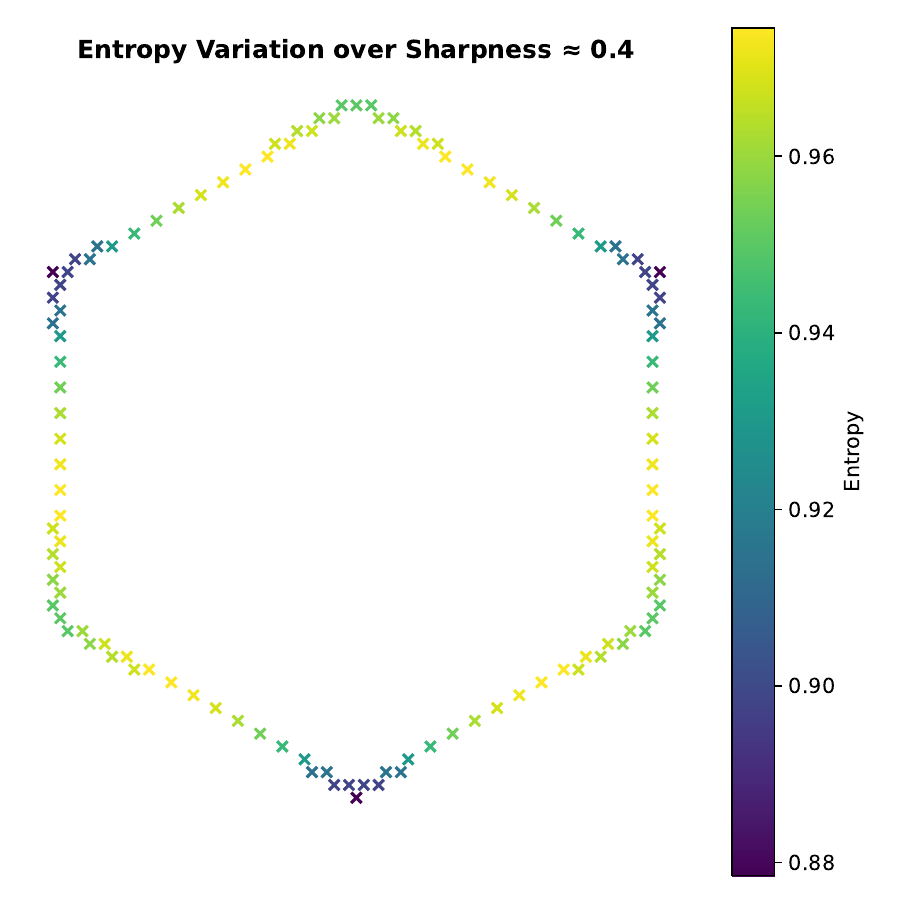}
    \caption{Entropy Over Sharpness}
\end{subfigure}
\caption{\label{fig:overlay1}Level sets of entropy and sharpness for n=3, with sharpness and entropy variation overlaid.}
\end{figure}

We begin by exploring the relationship between entropy and sharpness. Figure~\ref{fig:overlay1} illustrates how the two measures vary with respect to one another on the 2-simplex.

\clearpage

When sharpness is overlaid on entropy, its minimum values occur for the flattest distributions consistent with the entropy constraint. For example, for entropy level set $H(P) \approx 0.9$, distributions with minimal sharpness include distributions such as $\{0.07,\ 0.46,\ 0.47\}$. Within the same entropy level set, sharpness increases as the distribution becomes increasingly exponential in shape (e.g., here, $\{0.11,\ 0.27,\ 0.62\}$). Similarly, when entropy is overlaid on sharpness level sets, entropy is lowest when the distribution balances mass equally between a subset of the outcomes---for sharpness level set $S(P) \approx 0.4$, which corresponds roughly to the previous entropy level set, again when two of the three probabilities are equal (e.g., $\{0.47,\ 0.47,\ 0.06\}$). Within the same sharpness level set, entropy is maximized by an exponential shape (e.g., $\{0.55,\ 0.29,\ 0.15\}$).

\begin{figure}[h]
\centering
\begin{subfigure}[b]{0.49\textwidth}
    \includegraphics[width=\textwidth]{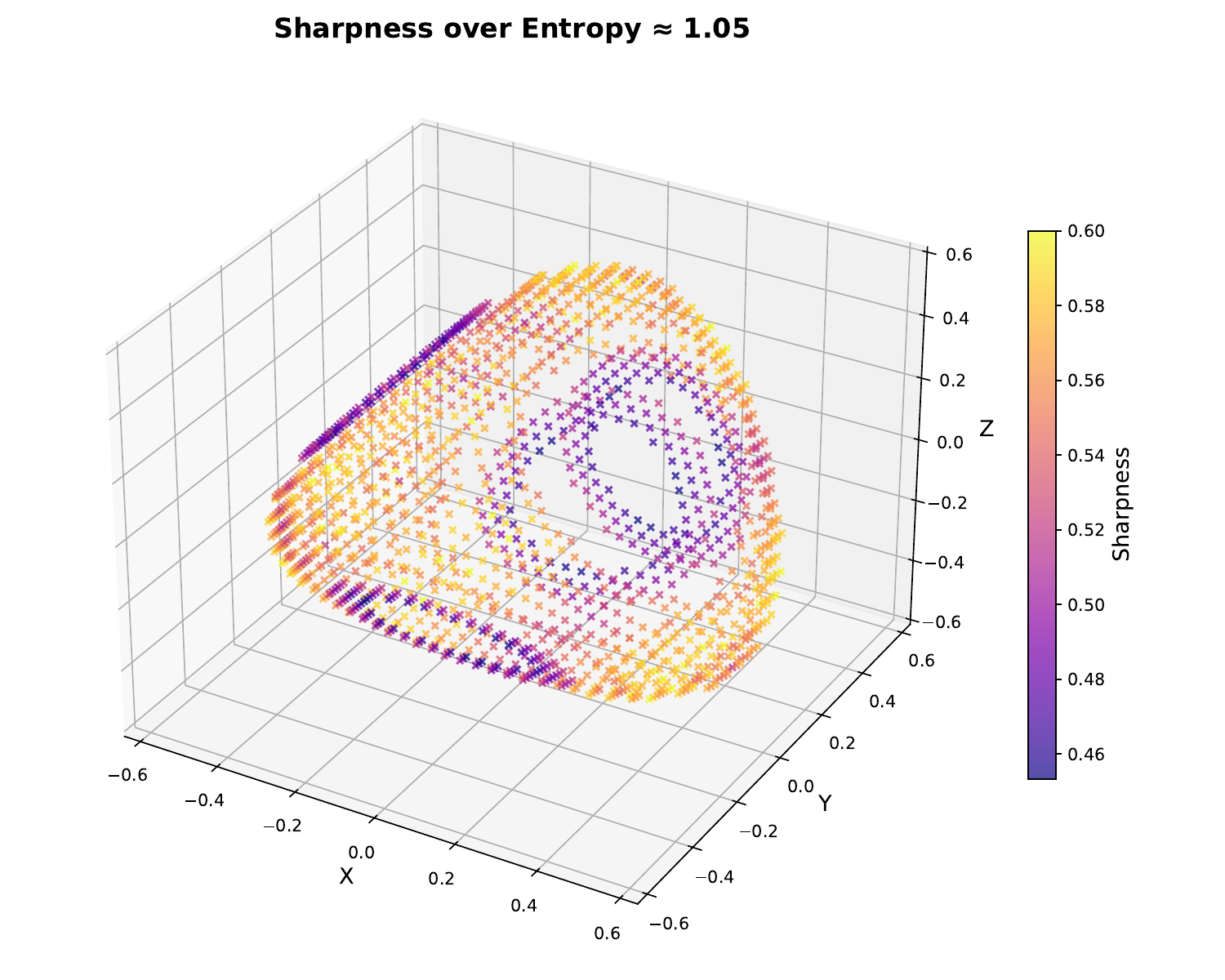}
    \caption{Sharpness Over Entropy}
\end{subfigure}
\hfill
\begin{subfigure}[b]{0.49\textwidth}
    \includegraphics[width=\textwidth]{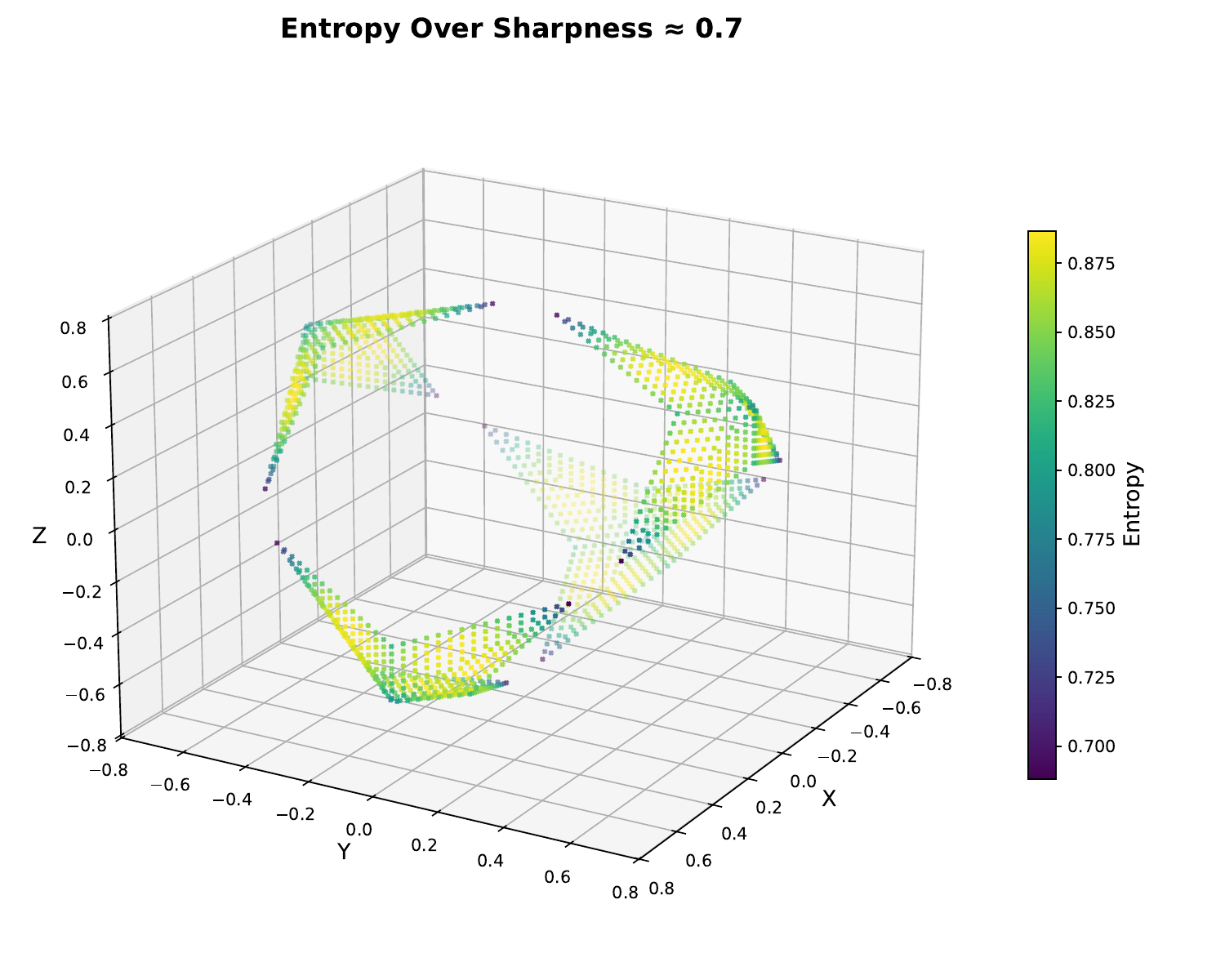}
    \caption{Entropy Over Sharpness}
\end{subfigure}
\caption{\label{fig:overlay3}Level sets of sharpness and entropy for n=4, with entropy and sharpness overlaid.}
\end{figure}

Figure~\ref{fig:overlay3} illustrates this behavior when n=4. Within the entropy level set, sharpness is minimized by flat distributions---for example, $\{0,\ 0.22,\ 0.38,\ 0.4\}$, with $S(P)$ $\approx 0.45$ for entropy $\approx 1.05$---and maximized by the exponential shape---for example, $\{0.6,\ 0.24,\ 0.04,\ 0.12\}$, with $S(P)$ = 0.6 for entropy $\approx 1.05$. For a moderately high sharpness score of 0.7, entropy is minimized by balancing mass between a subset of outcomes at the top---e.g., $\{0,\ 0,\ 0.45,\ 0.55\}$, with $H(P)$ $\approx 0.69$---and maximized by exponential distributions---e.g., $\{0.67,\ 0.07,\ 0.03,\ 0.23\}$, with $H(P)$ $\approx 0.89$.

\begin{figure}[h]
\centering
\begin{subfigure}[b]{0.49\textwidth}
    \includegraphics[width=\textwidth]{entropy_over_sharpness.pdf}
    \caption{Entropy Over Sharpness $\approx 0.70$}
\end{subfigure}
\hfill
\begin{subfigure}[b]{0.49\textwidth}
    \includegraphics[width=\textwidth]{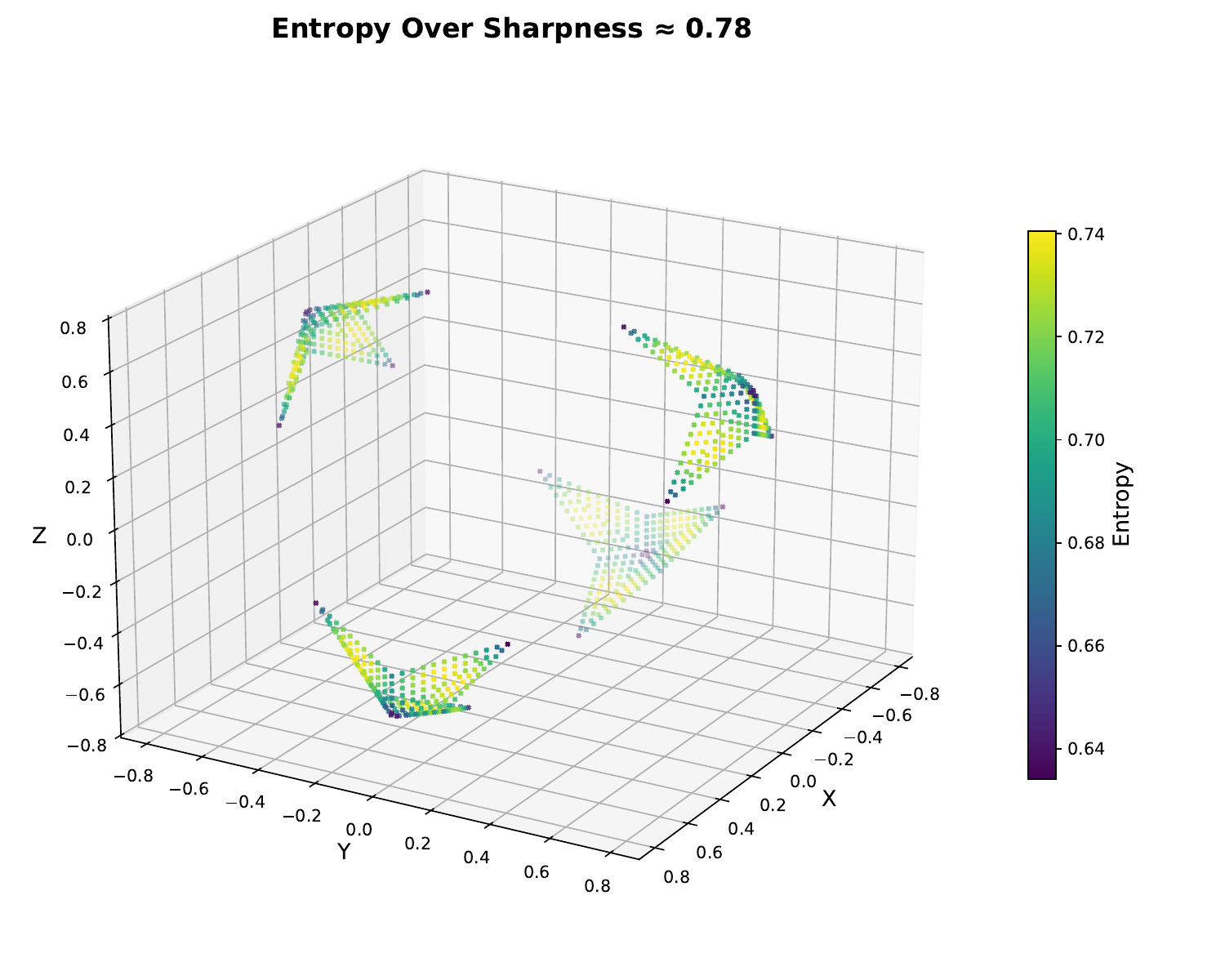}
    \caption{Entropy Over Sharpness $\approx 0.78$}
\end{subfigure}
\caption{\label{fig:overlay4}Sharpness level sets with entropy overlaid for n=4.}
\end{figure}

We may next consider what happens as mass becomes increasingly concentrated over the domain, meaning sharpness increases and entropy decreases. Within increasingly concentrated level sets, fewer potential distributional shapes remain viable that meet the strict sharpness or entropy constraint. Empirically, we find that the terminal viable shape appears to be of the form $\{a,\dots,a,b\}$, in which a single outcome dominates and the support is flat over the rest of the domain. We next examine when this shift occurs for the n=4 case, also illustrated in Figure~\ref{fig:overlay4}.

For $S(P)$ $\approx 0.70$, low entropy distributions are still dominated by distributions that spread mass more evenly between two top outcomes, and the minimizing shape is $\{0,0,a,b\}$. However, when mass can no longer be distributed between multiple outcomes to maintain the high sharpness constraint, the lowest entropy distributions shift to single-peaked distributions within the level set. For n=4, this occurs at $S(P)$ $\approx 0.78$, where the minimum entropy distributions now consist of single-peaked distributions such as $\{0.05,\ 0.05,\ 0.84,\ 0.06\}$ with $H(P)$ $\approx 0.63$, and the highest entropy is obtained by distributions such as $\{0.19,\ 0.75,\ 0.05,\ 0.01\}$, with $H(P)$ $\approx 0.75$.

Sharpness behaves in reciprocal fashion. For $H(P)$ $\approx 0.64$, minimum sharpness is obtained by distributions such as $\{0,\ 0,\ 0.34,\ 0.66\}$ with $S(P) = 0.77$ and maximum by distributions such as $\{0.80,\ 0.14,\ 0.05,\ 0.01\}$ with $S(P)$ $\approx 0.83$. At $H(P)$ $\approx 0.63$, a similar shift occurs mirroring that of entropy, where the lowest sharpness scores are obtained by distributions with a single peak and uniform flat tail (e.g., $\{0.05,\ 0.06,\ 0.06,\ 0.83\}$ with $S(P) \approx 0.78$), and the maximum is obtained by the exponential form (e.g., $\{0.01,\ 0.04,\ 0.14,\ 0.81\}$ with $S(P)$ $\approx 0.83$).

For growing values of n, similar behaviors are maintained: entropy and sharpness are minimized by evenly balanced subsets or concentration on a single outcome. However, for growing n, the switch to the final viable shape (top-1 dominance) occurs at lower sharpness levels. For example, in a sample of 5 000 000 distributions over n=10, and given a sharpness level set $S(P) \approx 0.3$, the minimum entropy values were achieved by more even distributions (e.g., distributions with 8 outcomes sharing roughly equal probability). By contrast, for sharpness level set $S(P) \approx 0.6$, distributions where one outcome has a disproportionate share of mass achieved the lowest entropy (top-1 dominance).

\begin{figure}[h]
\centering
\begin{subfigure}[b]{0.49\textwidth}
    \includegraphics[width=\textwidth]{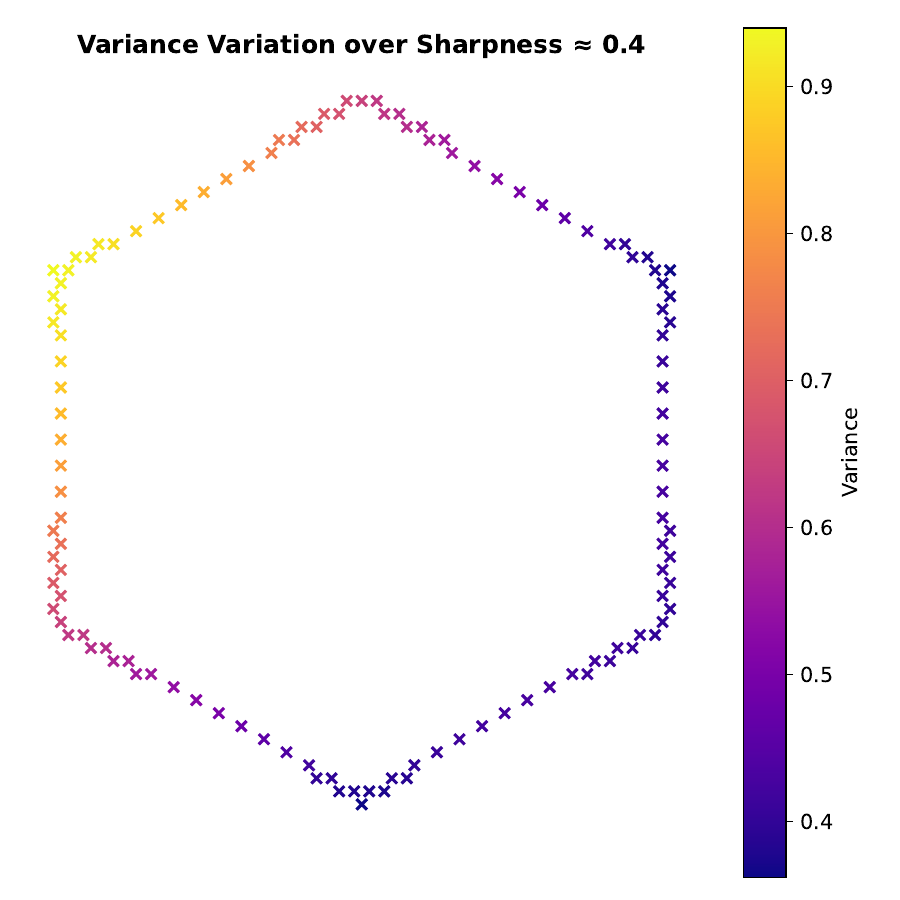}
    \caption{Variance Over Sharpness}
\end{subfigure}
\hfill
\begin{subfigure}[b]{0.49\textwidth}
    \includegraphics[width=\textwidth]{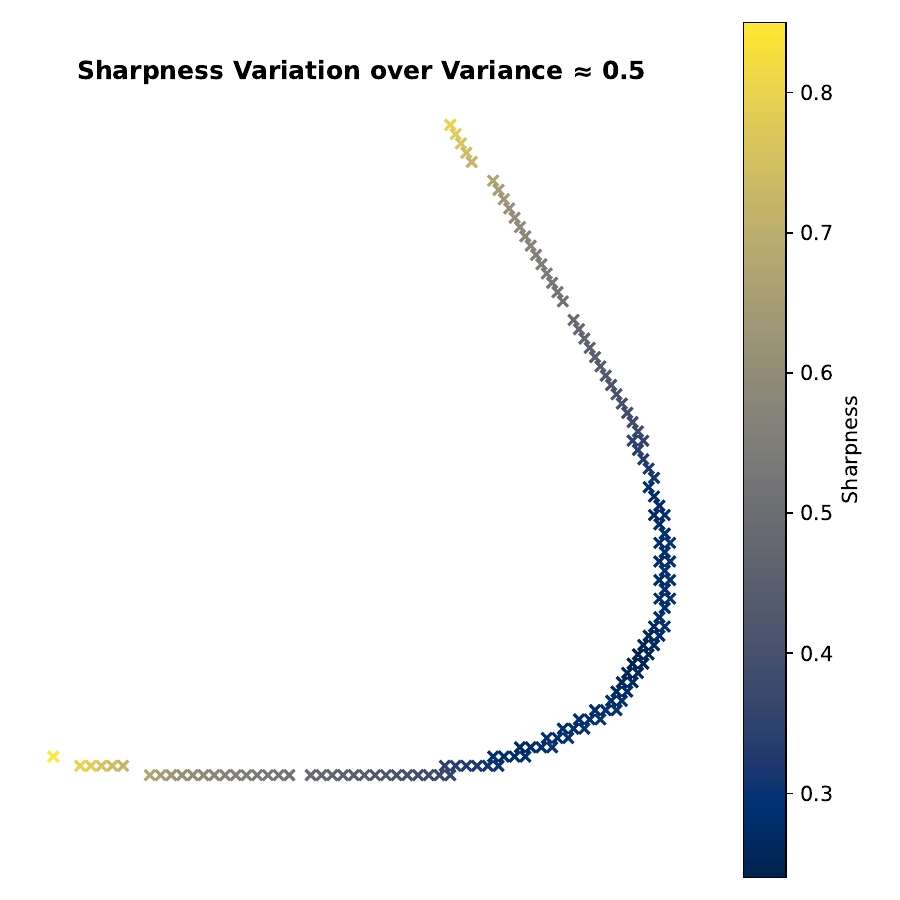}
    \caption{Sharpness Over Variance}
\end{subfigure}
\caption{\label{fig:overlay2}Level sets of sharpness and variance for n=3, with variance or sharpness overlaid.}
\end{figure}

In addition to the relationship between entropy and sharpness, variance provides additional diagnostic information by quantifying the spatial distribution of the probability mass. Variance can remain constant across distributions with vastly different sharpness scores, and vice versa. This complementary behavior is illustrated in Figure~\ref{fig:overlay2} on the 2-simplex (n=3).

For n=3 and sharpness level set $S(P) \approx 0.4$, distributions such as $\{0.47, \ 0.06, \ 0.47\}$ maximize variance, while variance is minimized by distributions that concentrate mass over adjacent positions (e.g., $\{0.47, \ 0.47, \ 0.06\}$). Conversely, sharpness score is lowest when the distribution is flat and highest when it is maximally concentrated within the variance level set. For example, for level set $Var(Y) \approx 0.5$, distributions such as $\{0.25, \ 0.49, \ 0.26\}$ yield the lowest sharpness scores (e.g., here, $S(P) = 0.24$), while distributions where one value dominates, such as $\{0.85, \ 0, \ 0.15\}$, yield a considerably higher score---here, $S(P) = 0.85$.

\begin{figure}
\centering{
\includegraphics[width=12cm,height=\textheight]{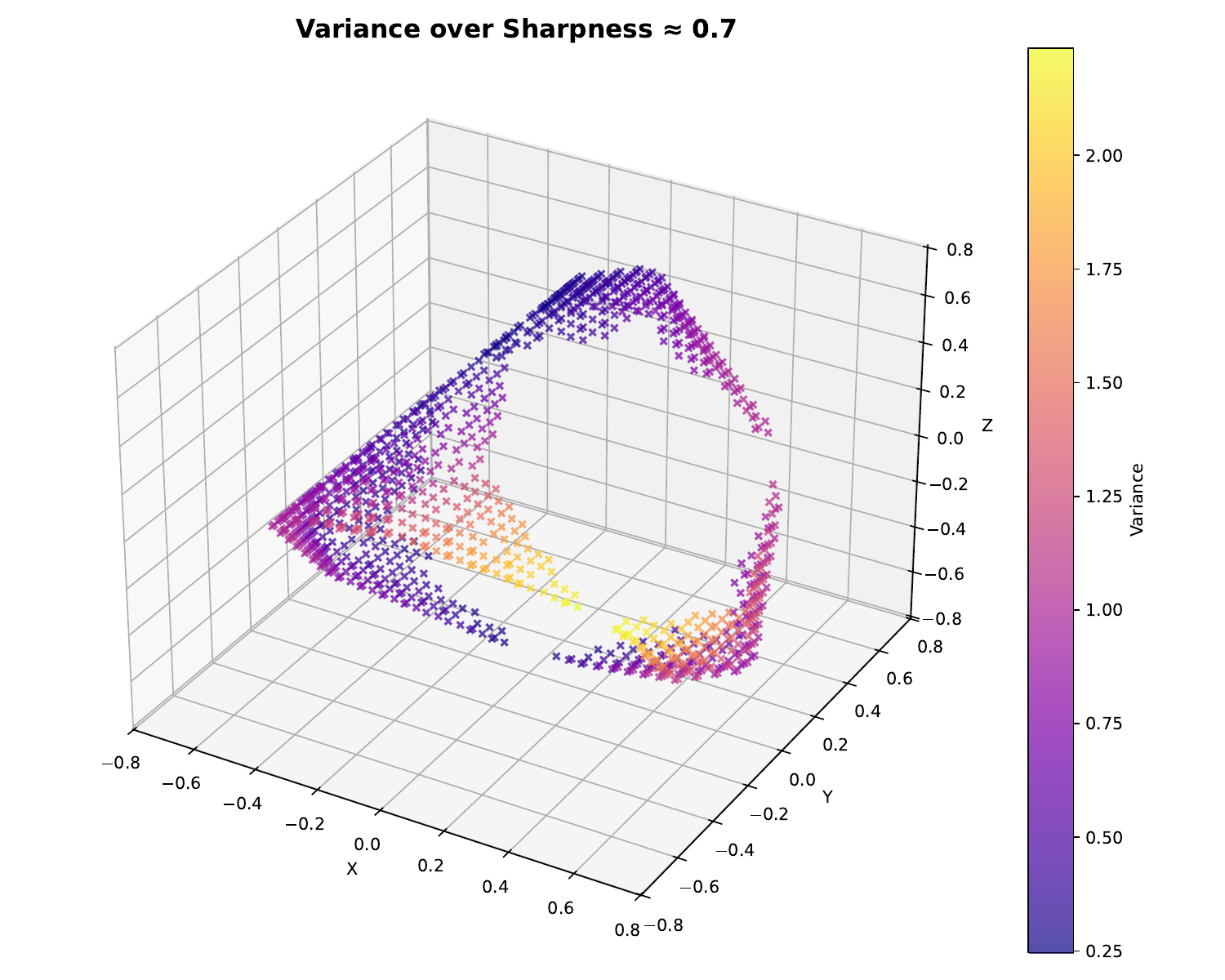}
}

\caption{\label{fig:varsh4}Sharpness level set $S(P) \approx 0.7$ on the 3-simplex, with variance overlaid.}
\end{figure}

Figure~\ref{fig:varsh4} illustrates variance over sharpness when n=4. The highest variance region corresponds to the distributions in which mass is concentrated at the opposite ends of the domain. In this case, where S(P) $\approx 0.70$, variance is maximized by distributions such as $\{0.44, \ 0, \ 0, \ 0.56\}$ and minimized by distributions such as $\{0, \ 0.44, \ 0.56, \ 0\}$.

\begin{figure}
\centering{
\includegraphics[width=12cm,height=\textheight]{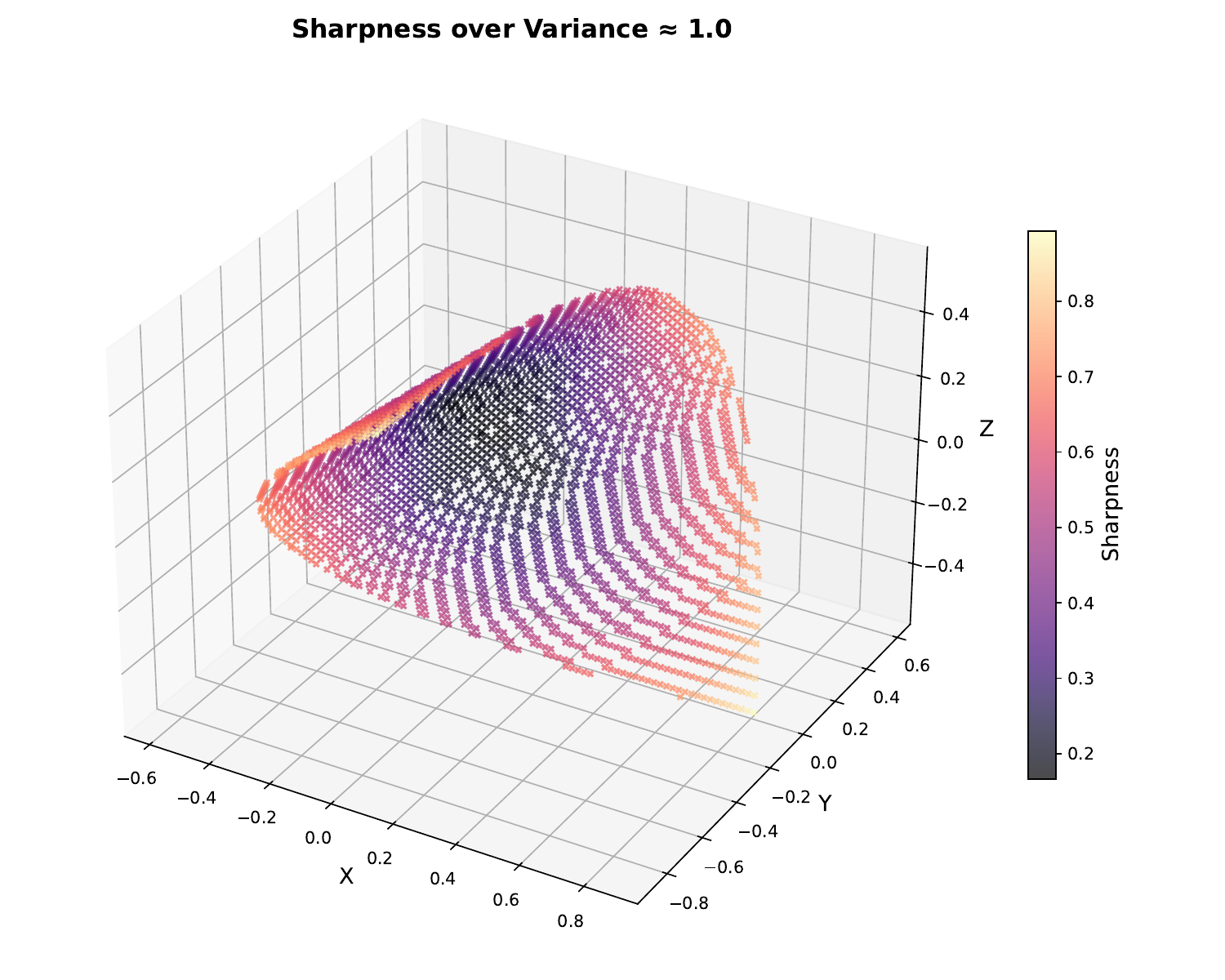}
}

\caption{\label{fig:shvar4}Variance level set $Var(Y) \approx 1.0$ on the 3-simplex, with sharpness overlaid.}
\end{figure}

Finally, Figure~\ref{fig:shvar4} illustrates sharpness over variance level set $Var(Y) \approx 1.0$. In this case, the range between the lowest and highest sharpness scores is extreme: a low of $S(P) \approx 0.17$ is achieved by $\{0.19, \ 0.32, \ 0.3, \ 0.19\}$, while a high of $S(P) \approx 0.89$ is achieved by $\{0.86, \ 0, \ 0.02, \ 0.12\}$. Given the behavior of each measure, the relationship between sharpness and variance is naturally maintained for higher values of n, where sharpness is highest for the most concentrated distributions whereas variance is maximized by packing mass at opposite ends of the domain.

For continuous cases, visualization becomes inherently more difficult; however, the basic relationships between the three measures are preserved. For example, in a sample of 100 000 unimodal and bimodal Gaussian distributions defined over a domain of measure 6, and given sharpness level set $S(f) \approx 0.6$, variance was minimized by a centered unimodal distribution and maximized by a bimodal distribution where the peaks occur at the ends of the domain. For bimodal distributions and a given sharpness set, entropy was minimized by distributions where the peaks are roughly equal, and maximized by unequal peaks. Conversely, sharpness scores increase when probability mass becomes more unequally concentrated. These relationships may be further explored by experimenting with different distributional shapes, investigating the respective minima and maxima.

\end{document}